\documentclass[pdflatex,sn-mathphys]{sn-jnl}
\usepackage{enumitem}
\usepackage{float}
\usepackage{blindtext}

\usepackage{color}
\usepackage{mathrsfs}
\usepackage{natbib}
\usepackage{geometry}
\usepackage{graphicx}
\usepackage{amsmath}
\usepackage{graphicx}
\usepackage{amsfonts}
\usepackage{amssymb}
\usepackage{bm}
\usepackage{dsfont}
\usepackage{epsfig,float,afterpage}
\usepackage{lineno}

\def\beq{\begin{equation}}
\def\eeq{\end{equation}}
\def\bea{\begin{eqnarray}}
\def\eea{\end{eqnarray}}
\def\nn{\nonumber}
\def\ba{\begin{array}}
\def\ea{\end{array}}

\newcommand{\dg}{\dagger}
\def\one{1\hskip -1mm{\rm l}}
\newlength{\sizeonefig}
\newlength{\sizetwofig}
\jyear{2021}

\begin{document}
\title[Nonequilibrium transport in open quantum systems of topological superconductors]{Nonequilibrium electrical, thermal and spin transport in open quantum systems of topological superconductors, semiconductors and metals}

\author[1]{\fnm{Nilanjan} \sur{Bondyopadhaya}}

\author[2]{\fnm{Dibyendu } \sur{Roy}}

\affil[1]{\orgdiv{Integrated Science Education and Research Centre}, \orgname{Visva-Bharati University}, \orgaddress{\city{Santiniketan}, \postcode{731235}, \country{India}}}

\affil[2]{ \orgname{Raman Research Institute}, \orgaddress{\city{Bangalore}, \postcode{560080}, \country{India}}}


\abstract{We study nonequilibrium transport in various open quantum systems whose systems
and leads/baths are made of topological superconductors (TSs), semiconductors, and
metals. Using quantum Langevin equations and Green’s function method, we derive
exact expressions for steady-state electrical, thermal, and spin current at the junctions
between a system and leads. We validate these current expressions by comparing them
with the results from direct time-evolution simulations. We then show how an electrical
current injected in TS wires divides into two parts carried by single electronic excitations
and Cooper pairs. We further show ballistic thermal transport in an open TS wire in
the topological phase under temperature or voltage bias. The thermal current values
grow significantly near the topological phase transition, where thermal conductance
displays a sharp quantized peak as predicted earlier. We relate the quantized thermal
conductance to the zero-frequency thermoelectric transmission coefficient of the open
TS wire. We also observe a large thermoelectric current near the topological transition
of the TS wires. We introduce a differential spin conductance which displays a quantized
zero-bias peak at zero temperature for a spinful TS wire in the topological phase. The
role of superconducting baths in transport is demonstrated by thoroughly examining the
features of zero-temperature differential electrical conductance and thermal conductance
in open systems with TS baths. Our new thermoelectric and spin transport findings in
various two-terminal geometries are beneficial to the present challenges in probing the
emergence of Majorana quasi-particles in experiments.}

\maketitle

\noindent
\section{Introduction}
\label{intro}

The study of quantum transport in superconducting materials has attracted much attention in recent years due to its applicability in detecting intriguing topological phases of superconductors \cite{Lutchyn2018, Beenakker2015}. Several recent experiments based on electrical transport measurements have strongly suggested possible evidence of Majorana bound states (MBSs), which are exotic quasiparticle excitations localized at the edges of one-dimensional (1D) topological superconductors (TSs) \cite{MourikScience2012, DasNature2012, NadjPergeScience2014, DengScience2016, FornieriNature2019}. While the electrical transport measurements can detect the emergence of protected zero modes of the MBSs, the thermal transport measurements are further useful in directly probing these chargeless Majorana quasiparticles \cite{AkhmerovPRL2011,FulgaPRB2011,Beenakker2015}. There has been a surge of interest in quantized thermal conductances for detecting fractionally charged and neutral modes \cite{Banerjee2017,Banerjee2018,Kasahara2018}. The MBSs in experimentally realized TSs made of hybrid nanowires combining spin-orbit coupled semiconductor and superconductor materials have a non-zero spin polarization \cite{Sticlet2012, Aligia2020}. Therefore, in such systems' spin transport is expected to provide valuable information about the systems' MBSs and topological properties \cite{Machado2017,Ohnishi2020,YangNanoLetter2020}. The thermal and spin transport measurements in probing the emergence of Majorana fermions can become essential tools in the present scenario when the detection of MBSs through electrical current peak signals is not unambiguous \cite{FrolovNatcomments2021, KayyalhaScience2020, YuNat2021, WangPRL2021, valentiniArxiv2020, saldaaaArxiv2021} because such electrical signals can also be produced in these devices due to other than Majorana fermions, such as, other quantum states that are not Majoranas and imperfections in the nanowire \cite{Kells2012,RoyPRB2013}. 

On the theoretical side, while there is a vast number of theoretical studies on electrical transport in systems with TSs \cite{AliceaReview2012, Stanescu2013,BolechPRL2007, LawPRL2009,Flensberg2010,Liu2012,Kells2012,DasSarma2012, RoyPRB2012,Zazunov2012,RoyPRB2013,LobosNJP2015,Yang2015,Zazunov2016,Sharma2016,Ioselevich2016,Bondyopadhaya2019,Bhat2020}, the thermal \cite{AkhmerovPRL2011,FulgaPRB2011,Nomura2012,Beenakker2015,Li_2017,SmirnovPRB2018,SmirnovPRB2019A} and mostly spin transport \cite{Tanaka2009,He2014} in these systems are much less explored. One of this paper's primary goals is to give a detailed and unified description of electrical, thermal, and spin transport in different devices made of TSs. The quantum transport in such devices are generally studied utilizing the generalized scattering theory \cite{Anantram1996, NilssonPRL2008, AkhmerovPRL2011, FulgaPRB2011}, the Keldysh  formalism \cite{cuevasPRB1996, BolechPRL2007, Flensberg2010, LobosNJP2015, SmirnovPRB2018}, and the quantum Langevin equations $\&$ Green's function (LEGF) method \cite{RoyPRB2012, RoyPRB2013, Bhat2020}. These theoretical techniques are mainly employed to calculate electrical currents and differential conductances, which are supposed to show non-trivial behaviour in superconductors' topological phases. We apply the LEGF method to develop a unified electrical, thermal, and spin transport description in 1D superconducting devices. The LEGF method is useful in the physical understanding of nonequilibrium processes as it provides a nice picture of the baths (generating the bias) as sources of thermal and quantum noise and dissipation. The last feature is also helpful in deriving fluctuations in the current (e.g., current-current correlators) applying this method.


The LEGF method is an open-quantum system formulation within the Heisenberg representation of quantum mechanics to study nonequilibrium quantum transport in mesoscale and nanoscale devices \cite{DharPRB2003, Segal2003, Kohler2005, DharPRB2006, DharRoy2006}. This method which is based on a direct solution of the Heisenberg equations of system and bath variables, has been applied extensively in calculating steady-state \cite{DharPRB2006, DharRoy2006} and time-dependent \cite{Kohler2005, KunduPRL2013} electrical, thermal, and optical transport properties in devices consisting of noninteracting baths (e.g., metals, harmonic oscillators) \cite{RoyDharPRB2007, RoyPRB2012, RoyPRB2013, RoyPRA2017, Manasi2018, Bhat2020}. An extension of this method for superconducting baths lacks to date. Such an extension is necessary for dealing with the Josephson effect within an open-quantum system framework \cite{Bondyopadhaya2019} and various Majorana braiding schemes proposed using TSs \cite{Lutchyn2018, AliceaReview2012}. In this paper, we consider hybrid devices of $X$-$Y$-$Z$ configuration where $X,Y$, and $Z$ can be made of TSs, normal metal (N), and spin-orbit coupled semiconductor (SM) in the presence of a magnetic field \cite{Bondyopadhaya2019}. Here, the $X$ and $Z$ wires act as leads/baths in thermal equilibrium, and the $Y$ wire is the system through which the transport is happening. Searching the signatures of TS wire leads in quantum transport when these TS wires are in the topological phase is another aspect of our study.
To this end, we consider both the Kitaev chain \cite{Kitaev2001} and the experimentally realizable 1D semiconductor-superconductor heterostructures \cite{LutchynPRL2010,OregPRL2010} as a TS lead.

It is worth noting that we have already performed a direct time-evolution study of electrical current at both the junctions of various hybrid devices of $X$-$Y$-$Z$ configuration in \cite{Bondyopadhaya2019}. We have detected a persistent and oscillating electrical current at both junctions of a TS-N-TS device, even in the absence of any phase or voltage or thermal bias when multiple MBSs and/or Andreev bound states (ABSs) within the bulk-gap are present near the junctions. Moreover, the amplitude and period of the oscillating current strongly depend on the middle N wire's initial conditions indicating the absence of thermalization.  Therefore, the presence of such bound states (both MBS and ABS) localized near the junctions prohibits the full hybrid devices from attaining a unique  nonequilibrium steady state (NESS) at a long time. It can be noted that the initial-condition dependence of supercurrents in a phase-biased superconducting nanojunction of topologically trivial BCS superconductors has also been investigated in Refs.~\cite{ZrirskiPRL2011,SoutoPRL2016,SoutoPRB2017,TarankoPRB2019}.

Interestingly, it has been demonstrated that the presence of MBSs amplifies the amplitude of zero-bias oscillating currents at the junctions compared to the amplitude of the same generated solely due to ABS. The generalized LEGF method can not be applied to study transport in those devices which do not have unique NESS. On the other hand, thermalization in tandem with a unique NESS can be achieved by tuning the system parameters in TS-N-$Z$ and TS-TS-$Z$ devices where TS is a Kitaev chain, and $Z$ is either an N wire or a Kitaev chain at the topological phase transition point (TP). A unique NESS is reached since the $Z$ lead's energy spectrum becomes gapless in such situations. There is no longer any bound state from the middle N/TS wire, and the middle wire gets equilibrated with the boundary wire(s). Similarly, we achieve thermalization and unique NESS in TS-SM-$Z$ and TS-TS-$Z$ devices with TS wires made of semiconductor-superconductor heterostructures when the $Z$ wire is either an SM or a TS at TP, and there is no mid-gap state in the spectrum of boundary wires. Hence one can apply generalized LEGF method to TS-N-Z, TS-SM-$Z$, and TS-TS-$Z$ devices once these systems attain steady-state. 

\subsection{Overview and new findings}
In Sec.~\ref{model}, we introduce the Hamiltonian of different models of TS, N, and SM wires and describe their statistical properties when they act a lead/bath. We particularly emphasize a general and detailed description of the TS wires as leads since this is the highlight of our present study. Next, we apply the generalized LEGF method in Sec.~\ref{formal} for deriving analytical expressions for the steady-state electrical, thermal and spin currents in various devices with TS leads. We also evaluate differential electrical and thermal conductances, which are incredibly convenient in identifying TS leads' role. This section is divided into two parts, (a) devices with Kitaev chains and (b) those with semiconductor-superconductor heterostructures. In Sec.~\ref{result}, we employ these expressions of currents and differential conductances to calculate several impressive results. We first validate our steady-state current formulas by comparing them to the long-time currents obtained from the direct time-evolution simulation. We then show how electrical current injected in one junction separates into different parts of charge current carried by single electronic excitations and Cooper pairs inside TS wires of an N-TS-N. 
 We further discuss several interesting features of thermal current and linear-response thermal conductance in the N-TS-N device. The properties of a sharp quantized peak of thermal conductance near the topological phase transition are especially highlighted.  We relate the quantized thermal conductance to the zero-frequency thermoelectric transmission coefficient of the open TS wire. We observe a large thermoelectric current near the topological transition of the TS wires, which might be potentially applicable. We then discuss an interesting electrical current asymmetry in a TS-N-N device with spatial asymmetry in tunneling rates. We  here introduce a differential spin conductance which displays a quantized zero-bias peak at zero temperature for a spinful TS wire in the topological phase. The role of superconducting leads/baths in transport is shown by thoroughly examining the zero-temperature differential electrical conductances (DECs) and thermal conductances in various devices. We conclude the paper's central part in Sec.~\ref{conc} by providing an outlook and problems of interest shortly. Eight appendices are further added to include analytical expressions and our method's various details for the interested readers.    
 
To summarize, we develop a generalized LEGF formalism which can be applied to study nonequilibrium electrical, thermal, and spin transport in an $X$-$Y$-$Z$ device made of superconducting leads as well as metallic leads; this is a technical advancement as the previous use of LEGF was only applicable to the devices with metallic leads. We note that the use of superconducting leads is of present interest as this can give better insight for Majorana detection \cite{Sharma2016, Yang2015}. On the application side of this generalized LEGF method, apart from verifying some of the already known results using this independent technique, we find several new features. (i) In a TS-TS-N device where TS is a Kitaev chain, we observe a quantized peak in thermal conductance near the TP of the middle TS wire only when the left superconducting lead is also at the TP. This feature of thermal conductance might have a potential application in detecting the topological phase of TS lead via thermal conductance measurement. (ii) We find a sizeable thermoelectric current in an N-TS-N device near the TP of the middle TS wire. Experimental measurements of such a large thermoelectric current or conductance in these devices would be much easier than the relatively small thermal conductance peak near the phase transition. Thus, the thermoelectric current or conductance might be a better probe to detect the TS wires' topological phase transition experimentally.  (iii) We show quantized zero-bias peaks of differential spin conductance in SM-TS-SM devices when the middle TS wire of semiconductor-superconductor heterostructures wire is in the topological phase. Further, we verify that the zero-bias peaks of differential spin conductance are robust against disorder. So, the detection of zero-temperature zero-bias peaks of differential spin conductance via spin tunneling spectroscopy may open up a new avenue to detect elusive Majorana fermions which emerge at the edges of TS wire in the topological phase. In the following sections, we will discuss this generalized LEGF method and its applications in detail.

\section{Topological superconductor leads and their statistical properties}
\label{model}
We consider a hybrid device of $X$-$Y$-$Z$ configuration consisting of a finite wire $Y$ whose left and right terminals (ends) are connected to the semi-infinite wire $X$ and $Z$ respectively. We here treat these semi-infinite $X$ and $Z$ wires as leads (baths) and impose canonical or grand-canonical equilibrium for their statistical properties from the beginning before they are connected to the middle $Y$ wire. Hereafter, we choose $X$, $Y$, and $Z$ wires to be an N or an SM or a TS wire. We are particularly interested in treating the $X$ and $Z$ wires made of TSs. Below, we introduce the mean-field Hamiltonians of two different 1D TS models and discuss their statistical properties in thermal equilibrium. These models are (a) the {\it Kitaev chain } of a spinless $p$-wave superconductor \cite{Kitaev2001} and (b) the {\it Majorana wire} of semiconductor-superconductor heterostructure. The latter one is a spinful TS engineered with a Rasbha spin-orbit coupled semiconductor nanowire proximity coupled to an s-wave superconductor in the presence of a magnetic field along the direction of the wire \cite{LutchynPRL2010,OregPRL2010}. We also study the Hamiltonians of N and SM wires as the limiting case of the Kitaev chain and Majorana wire, respectively, and their statistical properties in equilibrium. 

We here write the Hamiltonian of the TS wires and those of N and SM wires in a matrix format using a `double-fermion' basis, convenient for our nonequilibrium transport analysis with the quantum Langevin equations. Let us start by writing a most general 1D (mean-field) Hamiltonian \cite{PengPRB2017} of fermions as follows 
\beq
\frac{H}{\hbar}=\sum_{l'=1}^L \psi_{l'}^\dg U_{l'} \psi_{l'} +\sum_{l'=1}^{L-1}   (\psi_{l'+1}^\dg V_{l'} \psi_{l'}+\psi_{l'}^\dg V_{l'}^\dg \psi_{l'+1}),
\label{GHam}
\eeq
where $l'=1,\dots,L$ are the lattice sites along the wire, and $\psi_{l'}$ ($\psi_{l'}^\dg$) is a column (row) vector of fermion annihilation/creation operators at the $l'$-th lattice cite. This Hamiltonian is defined over an $LM$ dimensional Hilbert space $\mathbb{H}=h_1\otimes h_2 \otimes \dots \otimes h_L$ where $h_{l'}$ is an $M$ dimensional local Hilbert space defined at $l'$-th site. Clearly $\psi_{l'}$ is an $M$ dimensional column vector, whereas $U_{l'}$ and $V_{l'}$ are $M \times M$ dimensional matrices. In case of superconductors, $\psi_{l'}$ consists of both electron annihilation and creation operators. The generalized Hamiltonian (\ref{GHam}) can accommodate both the Kitaev chain and the Majorana wire for some particular choices of $\psi$, $U$ and $V$. 
\subsection{Kitaev chain}
\label{Kitaev}
To write the Kitaev chain in the above generalized form, we introduce the position space Nambu spinor $\psi_{l'}=(c_{l'},c_{l'}^\dg)^T$, which implies that the dimension of local Hilbert space ($h_{l'}$) is two ($M=2$). With the above form of $\psi_{l'}$ and the following choice of $U_{l'}$ and $V_{l'}$,
\[
U_{l'}=
\frac{1}{2}\begin{bmatrix}
-\epsilon &0\\
0& \epsilon  \\
\end{bmatrix},
\, \text{and} ~ ~
V_{l'}=
\frac{1}{2}\begin{bmatrix}
-\gamma & \Delta \\
-\Delta^* & \gamma \\
\end{bmatrix},
\]
the above Hamiltonian $H$ (\ref{GHam}) represents that of the Kitaev chain which is denoted by 
\bea
\frac{H_{\rm K}}{\hbar}=-\gamma \sum_{l'=1}^{L-1}(c^{\dg}_{l'}c_{l'+1}+c^{\dg}_{l'+1}c_{l'})-\epsilon \sum_{l'=1}^{L}(c^{\dg}_{l'}c_{l'}-\frac{1}{2}) - \sum_{l'=1}^{L-1}( \Delta c^{\dg}_{l'}c^{\dg}_{l'+1}+ \Delta^* c_{l'+1}c_{l'}), \nn \\
\label{Kham}
\eea
where, $\gamma$ is hopping, $\epsilon$ is the on-site energy, and $\Delta$ denotes superconducting pairing potential. The parameters $\gamma,\epsilon$ and $\Delta$ have dimension of frequency, and we assume them to be real hereafter. In order to study nonequilibrium transport using LEGF, we further introduce the following generalized basis:
\bea
 {\bf a} &\equiv& [ a_1, a_2, \dots, a_{2l'-1},a_{2l'},\dots, a_{2L-1}, a_{2L}]^T = [c_1, c^{\dg}_1, \dots,c_{l'},c_{l'}^\dg,\dots, c_{L}, c^{\dg}_{L}]^T. \nn \\
\label{bas1}
\eea
Clearly, $a_{2l'}=a^{\dg}_{2l'-1}$. In the above basis, $H_{\rm K}$ can be written in a quadratic form as follows, $H_{\rm K}=\frac{\hbar}{2}{\bf a}^{\dg}\mathcal{K}{\bf a}=\frac{\hbar}{2}\sum_{l,m}\mathcal{K}_{lm} a^{\dg}_la_m$, where $\mathcal{K}$ is an $2L \times 2L$ Hermitian matrix, and  $l,m =1,\dots,
2L $ \cite{Blaizot1986,Bondyopadhaya2019}. It can be noted that the index $l$ in $a_l$ (or $a_l^\dagger$) does not represent the actual physical site of the wire.  For a given $l$, one can define a map to the physical site $l'$ 
of spinless fermions as: $l'=(l+1)/2$ for odd values of $l$, and $l'=(l/2)$ for even values of $l$. In the 
presence of pairing ($\Delta \ne 0$), the superconducting wire undergoes a topological phase transition as $
\epsilon$ is tuned across $2\gamma$. The wire is in a topological phase for $\mid \epsilon \mid <2 \mid \gamma 
\mid$, and the TS wire hosts two spatially-localized MBSs at the opposite ends of the wire for a relatively long 
wire. The wire transits to a topologically trivial phase (non-topological phase) without the MBSs for $\mid
\epsilon \mid >2\mid \gamma \mid$. The topological phase transition near $ \mid \epsilon \mid =2 \mid \gamma \mid$ is also accompanied by a bulk-gap closing in its energy dispersion. The superconducting wire has a bulk-gap in its spectrum both in the topologically non-trivial and trivial phases, and the gap vanishes at the topological phase transition around $ \mid \epsilon \mid=2 \mid \gamma \mid $. The two phases of the Kitaev chain can be characterized unambiguously by the quantized value of the geometric phase, namely the Pancharatnam-Zak phase, which acts as a topological invariant for such 1D systems \cite{Zak1989, Vyas2019}. The Pancharatnam-Zak phase's values are $\pi$ and 0, respectively in the topological and non-topological phases. 

The Hamiltonian matrix $\mathcal{K}$ can be diagonalized by solving the Hermitian eigenvalue problem
\beq
\mathcal{K} U_r= \omega_r U_r,
\label{eigenv}
\eeq
where $\omega_r$ and $U_r$ are the $r$-th eigenvalue and corresponding eigenfunction of the Kitaev Hamiltonian. Since $\mathcal{K}$ is a Hermitian matrix in Nambu basis, it satisfies following property \cite{Blaizot1986} 
\beq
 \Sigma \, \mathcal{K} \, \Sigma=-\mathcal{K}^{ *}, 
 \label{hprop}
\eeq 
where, $\Sigma= \one_{L} \otimes \sigma_x$ and $\one_{L}$ is an identity matrix of size $L$. When $\mathcal{K}$ is real, all the components of $U_r$ are also real. From Eq. \ref{hprop}, it follows
\beq
\mathcal{K} V_r = -\omega_r \, V_r, \nn
\eeq
where, $V_r=\Sigma  U^{*}_r$.
Thus, the vector $V_r$ is also be an eigenvector with the eigenvalue $-\omega_r$. We group the eigenvalues of $\mathcal{K}$ into pairs ($\pm \,\omega_r $) with $\omega_r > 0$ .
The eigenvectors of $\mathcal{K} $ obey the completeness relation, 
\beq
\sum_{r>0}\left(  U_r U_r^{\dagger} + V_r V_r^{ \dagger}  \right) =\one_{2L},
\label{R1}
\eeq
where, the notation $r>0$ means that the sum is limited to the positive eigenvalues. From the properties of normalized eigenvectors, it follows
\beq
U_r^{\dagger}U_s=\delta_{rs}\,, ~~V_r^{\dagger}V_s=\delta_{rs}\, ,~~ U_r^{\dagger}V_r=0.
\label{R2}
\eeq
By using the aforesaid properties of eigenvectors and eigenvalues, one can express $\mathcal{K}$ in the following 
form
\beq
\mathcal{K}= \sum_{r>0} \omega_r (U_r U^{\dagger}_r- V_r V_r^{ \dagger}).
\label{Kmat}
\eeq
Sometimes it is more convenient to express (\ref{R1}) and (\ref{R2}) by the components of $U_r$ and $V_r$. Thus, denoting $U_r$ by
\[
 U_r= (
 \phi_r(1),
 \psi_r(1),
 \dots,
 \phi_r(L ),
 \psi_r(L )
)^T,
\label{XY}
\]
 one can rewrite (\ref{R1}) as 
\bea
&& \sum_{r>0}( \phi_r (i') \phi_r^{ *}(j')+\psi_r^{ *}(i') \psi_r(j'))=\delta_{i'j'}\,, \nn \\
&& \sum_{r>0}(\phi_r(i') \psi_r^{ *}(j')+\psi_r^{ *}(i') \phi_r(j'))=0.
\label{ortho}
\eea
Applying the  expressions (\ref{Kmat}) and  (\ref{ortho}), $H_{\rm K}$ (\ref{Kham}) can be expressed in a diagonal form: 
\bea
H_{\rm K}
&=& \sum_{r>0} \hbar\omega_r (q_r^{ \dg}\, q_r-q_r \,q_r^{\dg}) = \sum_{r>0} \hbar \omega_r (2q_r^{ \dg}\, q_r-1),\nn
\label{Khamq}
\eea
where, the fermionic {\it quasiparticle} destruction operators $q_r$ are defined as 
\bea
q_r  = U_r^\dg {\bf a}=  \sum_{i'=1}^{L}( \phi_r^{ *}(i')c_{i'} +\psi_r^{ *}(i') c_{i'}^\dagger).
\label{bogoliubov} 
\eea
The ground state energy of $H_{\rm K}$ is $E_g=-\sum_{r>0} \hbar\omega_r$, which corresponds to the quasiparticle vacuum $\mid \varnothing \rangle$. It can be shown that $q_r$ and $q_r^{ \dg}$ satisfy anticommutation relations (e.g., $\{q_r,q_s^{ \dg}\}=\delta_{r  s}$) that indicate fermionic nature of the quasiparticles. Evidently, this Hamiltonian can be diagonalized in terms of these Bogoliubov quasiparticles, which are linear superpositions of the excitations of negatively charged electrons and positively charged electron holes. These quasiparticle creation operators acting on $ \mid \varnothing \rangle$ create many-particle states. Moreover, any second-quantized fermionic operator defined on this Hilbert space automatically takes care of the Pauli exclusion principle. Using (\ref{ortho}), one can easily invert (\ref{bogoliubov}) to get back
\bea
c_{j'}=\sum_{r>0} ( \phi_r(j') q_r + \psi^{ *}_r(j') q_r^{\dg}), 
 \label{Klatticeop} 
\eea

We here use a semi-infinite Kitaev chain to model the TS leads/baths for $X$ or $Z$ wire. We assume that the wire is in thermal equilibrium at temperature $T$ and chemical potential $\mu$ before connecting it to the middle wire. It is now convenient for the superconducting leads to perform a gauge transformation such that the chemical potential does not explicitly appear in the leads' thermal density matrix. Under such gauge transformation, the chemical-potential differences  instead enter in our calculation through time-dependent phases in the tunnel couplings of the superconducting leads to the middle wire \cite{Zazunov2016}. Therefore, the quasiparticle modes of the superconducting leads/baths in thermal equilibrium satisfy the following relations:
\bea
 \langle q_r^{ \dg} \, q_s \rangle = f(\omega_r, T) \, \delta_{r s},   ~\langle q_r \, q_s^{ \dg} \rangle = f(-\omega_r, T) \, \delta_{r s}, 
 \label{exp}
\eea
where $\langle .. \rangle$ denotes equilibrium expectation with thermal density matrix. All other expectations like $\langle q_r^{ \dg}\,  q_s^{ \dg} \rangle$, $\langle q_r \, q_s \rangle $ are always zero. Here, $f(\omega,T)=1/(\exp[\hbar\omega/k_BT] +1)$ describes the equilibrium distribution of the fermionic quasiparticles of the baths. Here, we emphasize that the Fermi distribution does not capture the contribution of non-Abelian, zero-energy Majorana quasiparticles, which we assume being noninteracting, do not thermalize at temperature $T$. Nevertheless, the presence of such Majorana quasiparticles are expected to enter in our transport analysis through their contributions in the tunneling/scattering matrix of the transport coefficients. Using relations (\ref{Klatticeop}) and (\ref{exp}), one can readily evaluate the equilibrium correlations of particle creation and annihilation operators of the Kitaev chain. The detailed expressions of these correlation matrices are given in Appendix \ref{App3}. 

The Hamiltonian $H_{\rm K}$ (\ref{Kham}) reduces to that of a spinless N wire in the absence of pairing ($\Delta=0$). The derivation of a semi-infinite N lead's normal modes and their statistical properties are straight-forward \cite{DharPRB2006}, and are not reproduced here. In the N bath case, we explicitly include the chemical potential in its thermal density matrix, and the tunneling matrix between the N bath and middle wire is now time-independent. If an N bath is kept at temperature $T$ and chemical potential $\mu$, its thermal density matrix for normal modes can be derived from Eq.~\ref{exp} after substituting $(\hbar\omega_r-\mu)$ in the place of $\hbar\omega_r$. One can also find the equilibrium correlations in terms of particle creation and annihilation operators $c_{l'}^{\dg},c_{l'}$. 

\subsection{Majorana wire}
\label{Majorana}
The Hamiltonian of the Majorana wire can also be dealt in the same manner as the Kitaev chain. Nevertheless, the local Hilbert space now is four dimensional ($M=4$), and the Nambu spinor reads as $\psi_{l'}=(c_{l' \uparrow },c_{l' \downarrow },c_{l' \uparrow }^\dg,c_{l' \downarrow }^\dg)^T$, where $c_{l' \sigma}$ annihilates an electron with spin $\sigma =\uparrow,\downarrow$ at the lattice site $l'$. With the following forms of $U_{l'}$ and $V_{l'}$,
\[
U_{l'}=
\begin{bmatrix}
(\epsilon - \gamma) & B & 0 & -\Delta \\
B& -(\epsilon - \gamma) & \Delta & 0   \\
0 &  \Delta^* & (\epsilon - \gamma) & -B  \\
-\Delta^* & 0 & -B & -(\epsilon - \gamma) 
\end{bmatrix},~
\]
,
\[
V_{l'}=
\frac{1}{2}\begin{bmatrix}
-\gamma  & -\zeta & 0 &0 \\
\zeta & -\gamma & 0 & 0  \\
0 & 0 & \gamma & \zeta \\
0 & 0 & -\zeta &  \gamma
\end{bmatrix},
\]
 the generalized Hamiltonian, $H$ (\ref{GHam}) reduces to the Majorana wire Hamiltonian:
 \bea
 \frac{H_{\rm M}}{\hbar}&=&\sum_{l'=1}^{L-1}\big[-\gamma \sum_{\sigma=\uparrow, \downarrow}(c^{\dag}_{l',\sigma} c_{l'+1,\sigma}+ c^{\dag}_{l'+1,\sigma} c_{l',\sigma})+\zeta (c^{\dag}_{l'+1,\uparrow}c_{l',\downarrow}-c^{\dag}_{l'+1,\downarrow}c_{l',\uparrow} \nn \\
 && +c^{\dag}_{l',\downarrow}c_{l'+1,\uparrow}-c^{\dag}_{l',\uparrow}c_{l'+1,\downarrow})\big] +2\sum_{l'=1}^{L}\big[(\epsilon-\gamma)\sum_{\sigma=\uparrow, \downarrow}(c^{\dag}_{l',\sigma}c_{l',\sigma}-\frac{1}{2}) \nn \\
 && +B (c^{\dag}_{l',\uparrow}c_{l',\downarrow}+c^{\dag}_{l',\downarrow}c_{l',\uparrow})- (\Delta c^{\dag}_{l',\uparrow}c^{\dag}_{l',\downarrow}+\Delta^* c_{l',\downarrow} c_{l',\uparrow}) \big], \nn \\
\label{Mham}
\eea
where, $\epsilon$ represents the on-site energy, $\Delta$ is the proximity induced $s$-wave superconducting pairing potential, $B$ is the magnetic field applied along the direction of wire (say, x-axis), $\gamma$ is the hopping, and $\zeta$ is the Rashba spin-orbit coupling strength. We again assume all the parameters,  which have dimension of frequency, to be real. We define $x,y,z$ components of total spin by $\sigma_x^T=\frac{\hbar}{2}\sum_{l'=1}^{L}(c^{\dag}_{l',\uparrow}c_{l',\downarrow}+c^{\dag}_{l',\downarrow}c_{l',\uparrow}), \sigma_y^T=\frac{\hbar}{2} \sum_{l'=1}^{L}i(c^{\dag}_{l',\downarrow}c_{l',\uparrow}-c^{\dag}_{l',\uparrow}c_{l',\downarrow})$, and $\sigma_z^T=\frac{\hbar}{2} \sum_{l'=1}^{L}(c^{\dag}_{l',\uparrow}c_{l',\uparrow}-c^{\dag}_{l',\downarrow}c_{l',\downarrow})$, respectively. We have $[\sigma_x^T,H_{\rm M}] \ne 0$ for $\zeta \ne 0$, $[\sigma_y^T,H_{\rm M}] \ne 0$ for $B \ne 0$, and $[\sigma_z^T,H_{\rm M}] \ne 0$ for $\zeta$ or $B\ne0$. We later discuss consequences of the above commutation relations (conservation laws) on spin transport in Majorana wires.

Further, we introduce a generalized basis for the Majorana wire:
\bea
{\bf b} & \equiv & [ b_1, b_2, b_3, b_4, \dots, b_{4L-3}, b_{4L-2}, b_{4L-1}, b_{4L}]^T \nn \\
 &=&[c_{1 \uparrow}, c_{1 \downarrow}, c^{\dg}_{1 \uparrow}, c^{\dg}_{1 \downarrow} , \dots, c_{L \uparrow}, c_{L \downarrow}, c^{\dg}_{L \uparrow}, c^{\dg}_{L \downarrow}]^T.
\label{bas2}
\eea
In this basis, the Majorana wire Hamiltonian can also be written in a quadratic form as follows, $H_{\rm M}=\frac{\hbar}{2}{\bf b}^{\dg}\mathcal{M}{\bf b}=\frac{\hbar}{2}\sum_{l,m}\mathcal{M}_{lm}b^{\dg}_lb_m$, where  $\mathcal{M}_{lm}$ is an $4L \times 4L $ square matrix and $l,m =1,\dots,4L  $. These $b_l$ operators are mutually related by following relations: $b_{4l'-3}^\dg=b_{4l'-1}$ and $b_{4l'-2}^\dg=b_{4l'}$. For this case, the index $l$ of operator $b_l$ can be related to the actual physical lattice site $l'$ applying a set of rules: (i) $l'=l/4$ if $l$ is even and divisible by $4$, (ii) $l'=(l+2)/4$ if $l$ is even but not divisible by $4$, (iii) $l'=(l+1)/4$ if $l$ is odd and $(l+1)$ is divisible by $4$, (iv) $l'=(l+3)/4$ if $l$ is odd but $(l+1)$ is not divisible by $4$. The Majorana wire undergoes a topological phase transition at a certain critical magnetic field, $B_c = \sqrt{\Delta^2+\epsilon^2}$ \cite{LutchynPRL2010,OregPRL2010, AliceaReview2012}. For an applied magnetic field $B>B_c$, this heterostructure is driven into a chiral $p$-wave topological superconducting phase supporting two zero-energy MBSs at the two ends of the nanowire. However, in the opposite limit $B<B_c$, this system remains in a non-topological phase, and does not host MBS at the edges. 

Like the Kitaev chain, the Majorana wire Hamiltonian matrix $\mathcal{M}$ can also be diagonalized using the Bogoliubov quasiparticles after solving the Hermitian eigenvalue problem:
\beq
\mathcal{M} U_r= \omega_r U_r,
\label{Meigenv}
\eeq
where, $\omega_r$ the $r$-th eigenvalue and $U_r$ is the corresponding eigenfunction of the Majorana wire Hamiltonian. In the Nambu spin basis, $U_r$ is represented by an $4L$ dimensional column vector. The eigenvalues of $\mathcal{M}$ can also be grouped into pairs ($\pm \,\omega_r $) with $\omega_r > 0$. If $V_r$ represents an eigenvector with an eigenvalue $-\omega_r$, then $U_r$ and $V_r$ satisfy completeness relation:
\beq
\sum_{r>0}\left(  U_r U_r^{\dagger} + V_r V_r^{ \dagger}  \right) =\one_{4L}\, ,
\label{CM1}
\eeq
and orthogonality relations similar to (\ref{R2}). Hence, we write $\mathcal{M}$ as
\beq
\mathcal{M}= \sum_{r>0} \omega_r (U_r U^{\dagger}_r- V_r V_r^{ \dagger})\, .
\label{Mmat}
\eeq
To define the quasiparticle excitations for the Majorana wire ($ H_{\rm M}$), we first express the eigenvectors in terms of their components: 
\bea
U_r&=&\left[\Phi_{r }(1), \Psi_{r}(1), \dots,  \Phi_{r }(L),  \Psi_{r }(L)\right]^T \, ,\nn \\
V_r &=& \left[\Psi_{r }^*(1), \Phi_{r}^*(1), \dots,  \Psi_{r }^*(L),  \Phi_{r }^*(L)\right]^T\, ,
\eea
where $\Phi_r(l')$, and $\Psi_r(l')$ are two component objects defined as $\Phi_r(l')=(\phi_{r \uparrow}(l'), \phi_{r \downarrow}(l'))$, and $\Psi_r(l')=(\psi_{r \uparrow}(l'), \psi_{r \downarrow}(l'))$.
Now, we define the fermionic Bogoliubov quasiparticles as 
\bea
q_r &=&U_r^\dg {\bf b}= \sum_{i'=1}^{L} \, \sum_{\sigma= \uparrow, \downarrow} ( \phi_{r \sigma }^{ *}(i')c_{i' \sigma}+\psi_{r \sigma} ^{ *}(i') c_{i' \sigma}^\dagger)\, , \nn \\
q_r^{\dagger}  &=& {\bf b}^\dg U_r = \sum_{i'=1}^{L} \sum_{\sigma= \uparrow, \downarrow} ( \phi_{r \sigma }(i')c_{i' \sigma}^\dagger+\psi_{r \sigma}(i') c_{i' \sigma})\, .\nn \\
\label{Mbogoliubov} 
\eea
We can also express the electron operators using the Bogoliubov quasiparticles by inverting the above relations (\ref{Mbogoliubov}) while utilizing orthonormality relations of eigenvectors:
\bea
c_{j' \sigma }&=&\sum_{r>0} ( \phi_{r \sigma} (j') q_r + \psi^{ *}_{r \sigma} (j') q_r^{\dg}), \nn \\
c_{j' \sigma }^{\dg}&=&\sum_{r>0} ( \phi^{ *}_{r \sigma} (j') q_r^{\dg} + \psi_{r \sigma }(j')q_r),
 \label{Mlatticeop} 
\eea
where $\sigma=\uparrow,\downarrow$. Applying the anti-commutation relations of quasiparticles, it is easy to express the Majorana wire Hamiltonian in a diagonal form:
\bea
H_{\rm M}
&= \sum_{r>0} \hbar\omega_r (2q_r^{ \dg}\, q_r-1).\nn
\label{Mhamq}
\eea
Like the Kitaev chain leads, we include the chemical-potential differences as a time-dependent phase in the Majorana wire lead's tunnel couplings to the middle wire. The thermal density matrix of quasiparticle modes of a semi-infinite Majorana wire bath kept at a temperature $T$ takes exactly similar forms as Eq.~\ref{exp}. In this case, it is assumed that the chemical potentials for up and down spins are the same. This density matrix can be used to calculate the equilibrium correlations in terms of electrons' creation and annihilation operators $c_{j' \sigma }^{\dg}, c_{j' \sigma }$. The detailed expressions of equilibrium correlations are given in Appendix \ref{App4b}. 

The Hamiltonian of an SM in the presence of a magnetic field used in our study can be obtained from $H_{\rm M}$ (\ref{Mham}) by dropping the superconducting pairing $\Delta=0$. Like an N bath, we explicitly include the chemical potential ($\mu$) in the thermal density matrix for an SM bath. The thermal density matrix for its quasiparticle modes can also be derived from Eq. \ref{exp} by substituting $(\hbar\omega_r-\mu)$ in the place of $\hbar\omega_r$.

\section{Quantum Langevin equations and steady-state transport}
\label{formal}
\begin{figure}
\includegraphics[width=0.99\linewidth]{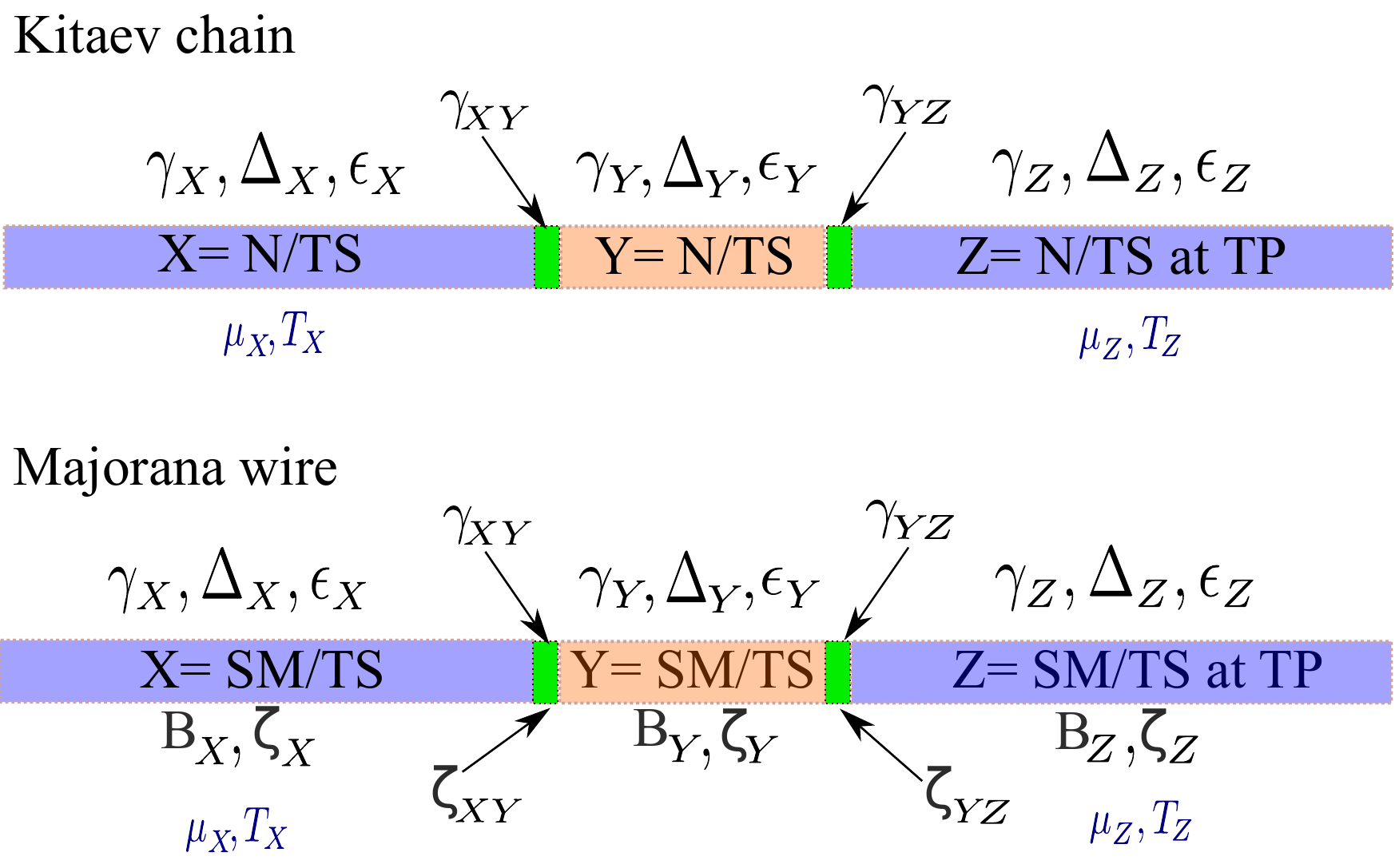}
\centering
\caption{Schematic of open quantum systems of $X$-$Y$-$Z$ configuration where $X$, $Y$, $Z$ are made of topological superconductor (TS) of Kitaev chain and normal metal (N) [top], and TS of Majorana wire and semiconductor (SM) [bottom]. The finite middle $Y$ wire acts a system, and the semi-infinite  boundary $X,Z$ wires are leads/baths, which are kept at some chemical potential $(\mu_{\rm X},\mu_{\rm Z})$ and temperature $(T_{\rm X}, T_{\rm Z})$ from the beginning. The individual components of the devices are coupled via tunneling $\gamma_{\rm XY},\gamma_{\rm YZ}$ and spin-orbit coupling $\zeta_{\rm XY},\zeta_{\rm YZ}$.}
\label{cartoon}
\end{figure}

In this section, we discuss the procedure of calculating steady-state nonequilibrium transport properties using a generalized LEGF method. As mentioned before, the hybrid device $X$-$Y$-$Z$ consists of three separate wires $X$, $Y$, and $Z$. Total length of the device is $L=L_{\rm X}+L_{\rm Y}+L_{\rm Z}$, and $L_{\alpha'}$ is the length of $\alpha'$ wire for $\alpha'= X,Y,Z$. We assume that both $L_{\rm X}$ and $L_{\rm Z}$ are much greater than $L_{\rm Y}$, thus both $X$ and $Z$ wires can be treated as baths connected to $Y$. The first site of the $Y$ wire is connected to the $L_{\rm X}$-th site of left bath, and the $L_{\rm Y}$-th site of $Y$ wire is connected to the first site of right bath (see Fig.~\ref{cartoon}).
The Hamiltonian of the full hybrid device, which is made of $X,Y,Z$ wires and two contacts, is given by
\beq
H^{\rm F}=H^{\rm X}+H^{\rm Y}+H^{\rm Z}+H^{\rm XY}+H^{\rm YZ},
\label{H1}
\eeq
where $H^{\alpha'}$ is the Hamiltonian corresponding to the $\alpha'$ wire, and the contact Hamiltonians for $X$-$Y$ and $Y$-$Z$ junctions are $H^{\rm XY}$ and $H^{\rm YZ}$, respectively. The exact form of $H^{\alpha'}$ and $H^{\alpha' \beta'}$ should be chosen according to the type of wire and junction. In our study, $H^{\alpha'}$ can be Hamiltonian of a Kitaev or a Majorana chain of TS or an N or an SM, which we have introduced in the previous section. We choose $H^{\alpha' \beta'}$  to be either an N junction or an SM junction respectively for our study with the Kitaev chain and the Majorana wire. As discussed in the previous section, $\alpha'$ wire Hamiltonian ($H^{\alpha'}$) can be written in the matrix form, which is denoted by $\mathcal{H}^{\alpha'}$.

Here, we generalize the LEGF method so that it can also be applied to the devices of our interest, i.e., a hybrid junction with superconducting (topological/non-topological) leads. Let us briefly indicate the steps leading to generalized quantum Langevin equations of motion. We assume that the leads/baths ($X$ and $Z$ wires) are disconnected from the middle wire ($Y$) at time $t \leq t_0$. Each bath is assumed to be in thermal equilibrium characterized by its temperature $T_\alpha$ and chemical potential $\mu_\alpha$ for $\alpha={X,Z}$. It should be noted that we do not explicitly include chemical potentials in the thermal density matrix of superconducting baths. Nevertheless, we do have chemical potentials for an N or an SM bath. Here we make some critical approximations to make analytical progress. We assume the mean-field superconducting pair potential, $\Delta$ of the TS wires remains fixed during time-evolution.  We have here avoided any self-consistent evolution of $\Delta$ for weak nonequilibrium boundary conditions. This approximation also implies no time-evolution of the superconducting phase of the middle TS wire \footnote{The time evolution of phase of the middle TS wire can be investigated using the first-principle/direct time-evolution numerics employed later (see Appendix \ref{App7})}. Further we assume that the temperature scale of the superconductor ($T$) is much smaller than the critical temperature of the superconductor ($T_c$), so we can approximate $\Delta(T) \sim \Delta (T=0)$ within this low temperature regime. Effect of higher temperature will be manifested as a suppression of conductance peak height \cite{Sharma2016}.  In principle, any nonequilibrium boundary condition such as a voltage or a temperature bias would influence the pairing of the middle TS wire, which is then needed to be determined within a self-consistent mean-field approach in nonequilibrium. However, the qualitative features of our results are not affected as the existence of superconducting gap and the gap closing phenomenon are crucial for such features \cite{LobosNJP2015}. It will be an interesting problem to model such nonequilibrium scenario self-consistently and study the feedback mechanism that stabilizes all the mean-field parameters.

In the following, we are mostly interested in studying time-independent transport in hybrid junctions featuring unique nonequilibrium steady-state.  Thus, we do not consider transport in a device with two superconducting leads kept at a non-zero chemical potential difference ($\mu_{\rm X} -\mu_{\rm Z} \neq 0 $), which results in time-dependent steady-state. Nevertheless, for a system with one TS lead and another N/SM lead, we choose to bias the N/SM lead with a non-zero chemical potential while keeping the superconducting one at zero chemical potential  to avoid unnecessary complication that may arise from the shifting of quasiparticle energy levels due to the chemical potential. In all the devices of type TS-TS-N/TS-TS-SM, we are mainly interested in calculating different types differential conductances at $Y$-$Z$ junction. These are local quantities which are not affected by the supercurrent that may develop at $X$-$Y$ junction due to the phase difference, $\delta \phi=\phi_{\rm X} -\phi_{\rm Y}$ when the complex pairing potentials of two superconductors are $\Delta_{\rm X}=\mid \Delta_{\rm X} \mid e^{i \phi_{\rm X}}$ and $\Delta_{\rm Y}=\mid \Delta_{\rm Y} \mid e^{i \phi_{\rm Y}}$ respectively. So, for simplicity we choose the pair potentials of both the superconductors to be real as the phase difference has no role in the differential conductances calculated for right TS-N/SM junction. However for TS-TS-TS at TP, relative phases of pairing potentials play some important role, but we avoided that complicacy by choosing them to be real as this particular case is used only to prove the applicability of NEGF for such junction.

In the subsequent subsections, we study charge, energy, and spin transport in the devices of $X$-$Y$-$Z$ configuration in which at least one component is either the Kitaev chain or the Majorana wire. We first discuss the LEGF method and find expressions for steady-state electrical and thermal current for the Kitaev chains. Later, we sketch the approach and derive expressions for electrical and spin current in steady-state for the Majorana wire. We highlight the important steps and provide relatively concise formulas in the main text, and include lengthy expressions and details in the appendices. Due to the complexity of spin-orbit coupled Majorana wire, we do not repeat the tedious calculation of energy current for such a system. However, our method for energy current can readily be applied for the Majorana wire.

\subsection{Electrical and thermal current : Kitaev chain}
\label{kita}

For a hybrid device with spinless TS wires, we choose normal metallic contacts with tunneling rates $\gamma_{\alpha' \beta'}$, whose Hamiltonians read as
\beq
H^{\rm \alpha' \beta'}=-\hbar\gamma_{\rm \alpha' \beta'}(c_{l'}^{\dg}c_{l'+1}+c_{l'+1}^{\dg}c_{l'}),
\label{Kcontact}
\eeq
where $\alpha' \beta'={\rm XY}$, $l'=L_{\rm X}$, and $\alpha' \beta'={\rm YZ}$, $l'=L_{\rm X}+L_{\rm Y}$ respectively for $X$-$Y$ and $Y$-$Z$ junction. At time $t=t_0$, we connect the baths to the opposite ends of the middle $Y$ wire and look for the steady-state properties of the $Y$ wire at later time $t$ ($t>> t_0+\tau$; $\tau $ is some characteristic time scale for reaching the steady state). As we need to consider the time evolution of total $X$-$Y$-$Z$ device, it is convenient to use the full generalized basis: ${\bf a} \equiv [a_1, a_2, \dots, a_{2L-1}, a_{2L}]^T$. For $t > t_0$, the Heisenberg equations of motion for the $Y$ wire variables are given by
\bea
 \dot{a}_l &=& -i \sum_{m=2L_{\rm X}+1}^{2L_{\rm XY}} \mathcal{H}^{\rm Y}_{lm} a_m -i\gamma_{\rm XY}\sum_{k= 2L_{\rm X}-1}^{2L_{\rm X}} (-1)^k a_l \, \delta_{l,k+2}-i\gamma_{\rm YZ}\sum_{k= 2L_{\rm X}+1}^{2L_{\rm X}+2} (-1)^k a_l \, \delta_{l,k-2} \, , \nn \\
\label{eomwire}
\eea
where, $l = 2L_{\rm X}+1,\dots,2L_{\rm XY} $ and $L_{\rm XY}=L_{\rm X}+L_{\rm Y}$.  Similarly, the Heisenberg equations for the creation and annihilation operators of $X$ and $Z$ wires read as
\bea
 \dot{a}_l  = -i \sum_{m=1}^{2L_{\rm X}} \mathcal{H}^{\rm X}_{lm} a_m +i\gamma_{\rm XY}\sum_{k= 2L_{\rm X}+1}^{2L_{\rm X}+2} (-1)^{k+1} a_l \, \delta_{l,k-2} \, ,
\label{eombathl}
 \eea
 for $l = 1, \dots, 2L_{\rm X}$, and
 \bea
 \dot{a}_l = -i \sum_{m=2L_{\rm XY}+1}^{2L} \mathcal{H}^{\rm Z}_{lm} a_m 
 +i\gamma_{\rm YZ} \sum_{k= 2L_{\rm XY}-1}^{2L_{\rm XY}} (-1)^{k+1} a_l \, \delta_{l,k+2} \, ,
 \label{eombathr}
 \eea
for $l = 2L_{\rm XY}+1,\dots,2L $. The Eqs.~\ref{eombathl}, \ref{eombathr} are coupled, inhomogeneous, first-order differential equations that can be formally solved for the boundary bath operators by using the retarded Green's function. Substituting these solutions for the bath operators into Eq.~\ref{eomwire}, one can rewrite Eq.~\ref{eomwire} as Eq.~\ref{eom4}, a generalized quantum Langevin equation (see Appendix \ref{App1} for a detailed discussion). The quantum Langevin equations of $Y$ wire variables can be solved in the frequency domain using the Fourier transformation. However, the application of Fourier transformation is reliable only for the systems with unique NESS such that a memory of the initial state of the middle wire is irrelevant. It is worth mentioning here that the Fourier transform method is not applicable in the presence of bound states which prevent equilibration, and one needs to solve Eq.~\ref{eom4} numerically to examine the time evolution in such a case \cite{DharPRB2006, Bondyopadhaya2019}. Therefore, we solve the quantum Langevin equations of the middle wire by Fourier transformation only when a unique NESS is reached. To this end, we first consider the limit $t_0 \rightarrow -\infty$. The Fourier transform of the wire variables are defined as $\tilde{a}_l (\omega)=\frac{1}{2\pi}\int_{-\infty}^{\infty} dt \, a_l(t)\, e^{i \omega t}$. We get the following steady-state solutions for $\tilde{a}_l(\omega)$ after taking Fourier transform of the quantum Langevin equations of the $Y$ wire variables, $a_l(t)$ (\ref{eom4}):
\bea
 \tilde{a}_l(\omega)&=&\sum_{m=2L_{\rm X}+1}^{2L_{\rm XY}} \tilde{G}^+_{l,m}(\omega)\left( \sum_{k=1,2} \tilde{\eta}_{k}^{\rm X} (\omega)\, \delta_{m,2L_{\rm X}+k} +\sum_{k=1,2} \tilde{\eta}_{k}^{\rm Z} (\omega)\, \delta_{m,2L_{\rm XY}+k -2} \right) , \nn \\
\label{a1}
\eea 
where $l=2L_{\rm X}+1,\dots,2L_{\rm XY} $. Here, $\tilde{\eta}^{\rm X}_{1,2}(\omega) $ and $\tilde{\eta}^{\rm Z}_{1,2}(\omega) $ are the noise terms arising in the process of integrating out the variables of $X$ and $Z$ bath, respectively (see Appendix \ref{App1} for definition). These noise terms keep track of the nonequilibrium boundary conditions across the middle wire, which we impose in the beginning through the boundary wires. The retarded Green's function $\tilde{G}^+(\omega)$ of the full system in the Fourier domain is defined as
\bea
\tilde{G}^+(\omega)={(\omega \one_{2L_{\rm Y}} - \mathcal{H}^{\rm Y}- \tilde{\Sigma}_{\rm X}^+(\omega) -\tilde{\Sigma}^+_{\rm Z}(\omega))^{-1}} ={(\omega \one_{2L_{\rm Y}} - \tilde{\mathcal{H}}^{\rm Y})}^{-1}\,,
\label{FG0}
 \eea
where $\tilde{\Sigma}^+_{\rm X,Z}$ are the self-energy corrections to the $Y$ wire Hamiltonian originated from its interactions to the respective baths. The effective Hamiltonian matrix of the $Y$ wire which can be non-Hermitian, is given by $\tilde{\mathcal{H}}^{\rm Y}=\mathcal{H}^{\rm Y}+ \tilde{\Sigma}^+_{\rm X}(\omega) +\tilde{\Sigma}^+_{\rm Z}(\omega)$. These $\tilde{\Sigma}_{\rm X,Z}^+$ are square matrices of dimension $2L_{\rm Y} \times 2L_{\rm Y}$. The components of the self-energy terms $\tilde{\Sigma}^+_{\rm X,Z}$ are as following:
\bea
 [\tilde{\Sigma}_{\rm X}^+(\omega)]_{lm}& =& \gamma_{\rm XY} ^2 \sum_{k=2L_{\rm X}-1}^{2L_{\rm X}} {\tilde{G}^{\rm  X +}_{k,k}(\omega)}\, \delta_{l,k+2}\, \delta_{l,m} \nn \\
 &&~~~~~~~~~~~~~~~~~~ -\gamma_{\rm XY} ^2 \sum_{k,k'=2L_{\rm X}-1 \atop k \neq k'}^{2L_{\rm X}}{\tilde{G}^{\rm  X +}_{k,k'}(\omega)}\, \delta_{l,k+2}\, \delta_{m,k'+2} \, ,\nn \\
 {[\tilde{\Sigma}^+_{\rm Z}(\omega)]}_{lm}& =& \gamma_{\rm YZ}^2 \sum_{k=1}^{2}{\tilde{G}^{\rm Z+}_{k,k}(\omega)}\,\delta_{l,2L_{\rm XY}-2+k}\, \delta_{l,m} \nn \\
 &&~~~~~~~~~~~~~~~~~~ -\gamma_{\rm YZ} ^2 \sum_{k,k'=1 \atop k \neq k'}^{2}{\tilde{G}^{\rm  Z +}_{k,k'}(\omega)}\, \delta_{l,2L_{\rm XY}-2+k}\, \delta_{m,2L_{\rm XY}-2+k'} \, ,\nn \\
\eea
where $l,m= 2L_{\rm X}+1,\dots,2L_{\rm XY} $. Here, $\tilde{G}^{\alpha +}_{l,m}(\omega)$ is the retarded Green's function of isolated bath wires ($\alpha ={X,Z}$). For $\tilde{G}^{\rm X +}_{l,m}(\omega)$, $l,m =2L_{\rm X}-1,2L_{\rm X}$ correspond to the right most site (i.e. $L_{\rm X}$-th site) of the $X$ reservoir, whereas in case of $\tilde{G}^{\rm Z +}_{l,m}(\omega)$, $l,m=1,2$ correspond to the left most site (i.e. $L_{\rm XY}+1$-th site) of the $Z$ reservoir in the full $X$-$Y$-$Z$ device. The detailed definitions and expressions of $\tilde{G}^{\alpha +}_{l,m}(\omega)$ are given in Appendix \ref{App2}. Since $\tilde{\mathcal{H}}^{\rm Y}$ is a block diagonal matrix, numerical values of $\tilde{G}^+(\omega)$ can be calculated by inverting $(\omega \one_{2L_{\rm Y}} -\tilde{\mathcal{H}}^{\rm Y})$.

 We further write steady-state solutions for some of the bath variables $\tilde{a}_l(\omega)$ (\ref{solres}) defined at the edges of the baths. For example, these $\tilde{a}_{2L_{\rm X}-1}(\omega )$ and $\tilde{a}_{2L_{\rm X}}(\omega )$ for the left $X$ bath read as 
\bea
\tilde{a}_{2L_{\rm X}-1}(\omega )\gamma_{\rm XY} &=&  -\tilde{\eta}^{\rm X}_1(\omega)-
\sum_{m=2L_{\rm X}+1}^{2L_{\rm X}+2}{[\tilde{\Sigma}^+_{\rm X}(\omega)]}_{2L_{\rm X}+1,m} \, \tilde{a}_{m}(\omega),\nn \\
 \tilde{a}_{2L_{\rm X}}(\omega )\gamma_{\rm XY} &=&  \tilde{\eta}^{\rm X}_2(\omega)+
\sum_{m=2L_{\rm X}+1}^{2L_{\rm X}+2}{[\tilde{\Sigma}^+_{\rm X}(\omega)]}_{2L_{\rm X}+2,m} \, \tilde{a}_{m}(\omega) \, .\nn
\label{bathvariable}
\eea
These boundary variables of the baths would be useful in evaluating the transport coefficients through the middle wire, which we discuss below.

 For these hybrid devices with spinless particle, the transport coefficients of interest are electrical (charge) 
and thermal (energy) conductance. The electrical conductance measurements in such devices are sensitive to the 
emergence of the Majorana zero modes at the edges of the Kitaev chains. However, the charge neutrality of the 
Majorana quasiparticles poses a challenge to unambiguous detection of such topologically protected modes through 
electrical conductance. It is rather interesting to probe these charge-neutral modes through the thermal 
transport, which we also evaluate here \cite{Banerjee2017, Banerjee2018, Kasahara2018, AkhmerovPRL2011, Li_2017, 
SmirnovPRB2018}. Since the total number of particles (spinless electrons) is not conserved for the TS wires, the 
particle current of spinless electrons is, in general, not well-defined inside the TS wires. Nevertheless, the 
total particle number (of spinless electrons) is conserved for N wires or N junctions, and we mostly define 
charge currents carried by spinless electrons at those segments. Using the conservation of particles (of spinless 
electrons) at the junctions, we describe the charge/electrical current across the links after multiplying 
particle current by electron's charge $e$:

\beq
J^e_{\rm \alpha' \beta'}=ie\gamma_{\rm \alpha' \beta'}\langle (c_{l'}^{\dg}c_{l'+1}-c_{l'+1}^{\dg}c_{l'})\rangle,\label{Kcurrent}
\eeq
where again $\alpha' \beta'={\rm XY}$, $l'=L_{\rm X}$, and $\alpha' \beta'={\rm YZ}$, $l'=L_{\rm XY}$ respectively for the $X$-$Y$ and $Y$-$Z$ junction. The expectation $\langle .. \rangle$ denotes averaging over the initial density matrix of the baths. To find the electrical current using Eq.~\ref{Kcurrent}, we first evaluate the noise-noise correlations for the baths. In  Appendix \ref{App3}, we outline the procedure of finding the noise-noise correlations, and we also list there all the noise-noise correlations relevant for the calculation of electrical current at the junctions. Next, we rewrite the currents in Eq.~\ref{Kcurrent} in the following compact form:
\bea
 J^e_{\rm XY} &=&- 2e \gamma_{\rm XY} \, \text{Im}[\langle a_{2L_{\rm X}+1}^\dagger (t) a_{2L_{\rm X}-1} (t) \rangle]\, ,
\label{jxy} \\
 J^e_{\rm YZ } 
&=&- 2e \gamma_{\rm YZ} \, \text{Im}[\langle a_{2L_{\rm XY}+1}^\dagger (t) a_{2L_{\rm XY}-1} (t) \rangle] \, .
\label{jyz} 
\eea
First we take the Fourier transformation of the bath and wire variables in Eqs.~\ref{jxy} and \ref{jyz}, then those variables are substituted with Eq.~\ref{a1} and the Fourier transformed version of Eq.~\ref{solres}. Using the noise-noise correlations (\ref{NN1a}-\ref{NN2}), we finally obtain the analytical expressions for $J^e_{\rm XY}$ and $J^e_{\rm YZ}$ (see Appendix \ref{App5}). Due to the absence of conservation of total number of particles inside a superconductor, we observe that $J^e_{\rm XY} \neq J^e_{\rm YZ}$ for N-TS-N devices for arbitrary nonequilibrium boundary conditions \cite{RoyPRB2012}. In Sec.~\ref{result}, we show an interesting conversion of a part of the injected electrical current to a Cooper pair current inside the middle TS wires. Nevertheless, $J^e_{\rm XY}=J^e_{\rm YZ}$ for a symmetric bias, $\mu_{\rm X}=-\mu_{\rm Z}$, in an N-TS-N device. The emergence of a zero-energy MBS is expected to manifest a quantized zero-bias peak of height $2e^2/h$ in the zero-temperature DEC. For N-TS-N devices placed under a symmetric bias, we define DEC as $2\left[\frac{dJ^e_{\rm XY}}{dV}\right]$ or $2\left[\frac{dJ^e_{\rm YZ}}{dV}\right]$, where $\mu_{\rm X}-\mu_{\rm Z}=eV$ \footnote{Note that \cite{RoyPRB2012} seems to have missed the above 2 factor in the definition of DEC for a symmetric bias.}. Writing $\mu_{\rm X}=-\mu_{\rm Z}=\mu$, we can also express the DEC in such a device as $e\left[\frac{dJ^e_{\rm XY}}{d\mu }\right]$ or $e\left[\frac{dJ^e_{\rm YZ}}{d \mu}\right]$. For TS-N-N ans TS-TS-N devices, we apply an asymmetrical bias by setting $\mu_{\rm X}=0$ and $\mu_{\rm Z}= eV$, and change $eV$ to find DEC. For such systems, we are only interested in the DEC at the $Y$-$Z$ junction, which is defined as $\left[\frac{dJ^e_{\rm YZ}}{d V}\right]$. 

The expressions for $J^e_{\rm XY}$ and $J^e_{\rm YZ}$ can be written in a simple and neat Landauer current form in the presence of a temperature bias and zero chemical potentials,  $\mu_{\rm X}=\mu_{\rm Z}=0$. There, $J^e_{\rm XY}$ (\ref{cuL}) and $J^e_{\rm YZ}$ (\ref{cuR}) are simplified as a product of a frequency-dependent transmission coefficient $\mathcal{T}(\omega)$ and a difference between the Fermi functions of the boundary leads. We note that generally $J^e_{\rm XY} \neq J^e_{\rm YZ}$ for a TS-TS-N and a TS-TS-TS at TP in the above scenario. However, the expression \ref{cuR1} implies that the $J^e_{\rm XY}=J^e_{\rm YZ}$ for an N-TS-N device in the above limit of bias. Thus, we get the following expression for the thermoelectric current generated by a temperature bias from the $X$ and $Z$ bath with temperature $T_{\rm X}$ and $T_{\rm Z}$, respectively: 
\bea
J^e_{\rm XY}=J^e_{\rm YZ}=\int_{- \infty}^{ \infty}d \omega \,e\, \mathcal{T}(\omega) \left(f(\omega,T_{\rm X}) -f(\omega,T_{\rm Z})\right) =\frac{\pi^2 k_B^2 eT}{3 \hbar^2}\left[\frac{ \partial \mathcal{T}}{\partial \omega}  \right]_{\omega=0}   \Delta T, \nn \\
\label{pcur} 
\eea
where $\mathcal{T}(\omega)~=\mathcal{T}^1_{\rm XY}(\omega)+\mathcal{T}^2_{\rm XY}(\omega)$, whose explicit forms are given in  Appendix~\ref{App5}. We obtain the last expression in the above equation by making a linear response expansion for small temperature differences $\Delta T \ll T$, where $T_{\rm X}=T+\Delta T/2$ and $T_{\rm Z}=T-\Delta T/2$. As discussed above, all over the calculation we assume $T$ itself is also small so that we can ignore the effect of $T$ on the superconducting pair potential. We show later that $\left[\frac{ \partial \mathcal{T}}{\partial \omega}  \right]_{\omega=0}$ shows a sharp dip near the topological phase transition of the middle TS wires. We here propose to experimentally probe the topological phase transition in TS wires by measuring such a large dip in the thermoelectric current.

While electrical currents and differential conductances are extensively explored in an N-TS junction for the search of elusive Majorana fermions, the thermal/energy currents are relatively less studied in such junction \cite{AkhmerovPRL2011, Li_2017, SmirnovPRB2018, SmirnovPRB2019A}. However, recent experiments \cite{Banerjee2017, Banerjee2018, Kasahara2018} have suggested it might be possible to probe such very small thermal conductances arising from a few conducting channels rather accurately. Motivated by these developments, we derive the expressions of energy currents and linear-response thermal conductance in hybrid devices made of Kitaev chains. Further motivation stems from the fact that while electrical current is not the same across such mean-field models of TS wires, the energy current remains the same across the TS wires, which we explicitly demonstrate in Appendix~\ref{App7}. Using the continuity equation for the conserved energy across the junction between two wires, we derive the following  expressions for energy current at $X$-$Y$ and $Y$-$Z$ junctions:
\bea
 J^{u}_{\rm XY}&=&\bar{J}^q_{\rm XY}+\bar{J}^p_{\rm XY}  \nn \\ 
&& = 2\hbar \gamma_{\rm XY} \left( \gamma_{\rm Y} {\rm Im}[\langle c^\dg_{L_{\rm X}+2}(t) c_{L_{\rm X}} (t)\rangle ]+ \Delta_{\rm Y}{\rm Im}[\langle c^\dg_{L_{\rm X}+2}(t) c^\dg_{L_{\rm X}} (t)\rangle ] \right) \nn \\ 
&&~~~~~~~~~~~~~~~~~~~~~~~~~~~~~~~~~~~~~~~~ +    2 \hbar \gamma_{\rm XY} \epsilon_{\rm Y}
{\rm Im}[\langle c^\dg_{L_{\rm X}+1}(t) c_{L_{\rm X}} (t)\rangle ]   \, ,\nn \\ 
\label{ecuurent1} \\ \nn \\ \nn \\
 J^u_{\rm YZ}&=&\bar{J}^q_{\rm YZ}+\bar{J}^p_{\rm YZ} \nn \\
&& = 2 \hbar\gamma_{\rm YZ} \left( \gamma_{\rm Y} {\rm Im}[\langle c^\dg_{L_{\rm XY}+1}(t) c_{L_{{\rm XY}-1}} (t)\rangle ]-\Delta_{\rm Y} {\rm Im}[\langle c^\dg_{L_{\rm XY}+1}(t) c^\dg_{L_{{\rm XY}-1}} (t)\rangle ]\right) \nn \\
&&~~~~~~~~~~~~~~~~~~~~~~~~~~~~~~~~~~~~~~~~~ + 2\hbar \gamma_{\rm YZ} \epsilon_{\rm Y}{\rm Im}[\langle c^\dg_{L_{\rm XY}+1}(t) c_{L_{\rm XY}} (t)\rangle ]  \, ,\nn \\
\label{ecuurent2}
\eea
assuming the pairing potential $\Delta_{\rm Y}$ is real. The first parts within the big parenthesis in both (\ref{ecuurent1}) and (\ref{ecuurent2}) denote $\bar{J}^q_{\alpha' \beta'}$, and the other parts of the expressions represent $\bar{J}^p_{\alpha' \beta'}$, which is zero when the on-site energy of the $Y$ wire is zero. As shown in the above expressions, $\bar{J}^q_{\alpha' \beta'}$ can be separated into two parts; one is electronic part which has explicit dependence on the hopping parameter, and another is Cooper pair part which is explicitly related to the pairing potential. Like the electrical currents, we can again find explicit expressions of the energy currents in the steady state by using the steady-state solutions of the variables appearing in Eqs.~\ref{ecuurent1} and \ref{ecuurent2}. Due to the conservation of total energy in the middle wire in our all studied models, the energy current remains the same across the middle wire including for a TS wire in an N-TS-N device. The detailed expression of the energy current is given in  Appendix \ref{App6}. When the boundary lead wires are kept at a finite temperature bias and at zero chemical potential ($\mu_{\rm X}=\mu_{\rm Z}=0$), the expression of $J^{u}_{\rm XY}$ and $J^{u}_{\rm YZ}$ (\ref{encfull}) can be expressed in a simple Landauer current form as a multiplication of a frequency-dependent transmission coefficient with a difference between the Fermi functions of the boundary baths \cite{DharPRB2006, DharRoy2006, RoyDharPRB2007}.

In the linear response regime, we can further simplify the Landauer form of energy current by assuming $T_{\rm X}=T+\Delta T/2$ and $T_{\rm Z}=T-\Delta T/2$ along with $\mu_{\rm X}=\mu_{\rm Z}=0$, where the temperature difference $\Delta T=T_{\rm X}-T_{\rm Z}$ is much small compared to the mean temperature $T$. By expanding the Fermi functions about $T$, we then write for the energy current:
\bea
J^{u}_{\rm XY}= J^{u}_{\rm YZ} &=&\int_{- \infty}^{ \infty}  d \omega \, \hbar \mathcal{A}(\omega) \left(f(\omega,T_{\rm X}) -f(\omega,T_{\rm Z})\right) \nn \\
&=& \int_{- \infty}^{ \infty}d \omega \,\hbar \left( \mathcal{A}(0)+ \left[\frac{ \partial \mathcal{A}}{\partial \omega}  \right]_{\omega=0}\omega \right) \left[\frac{\partial f}{\partial T}\right]_{T}\Delta T \nn \\
&=& \frac{\pi^2 k_B^2 }{3 \hbar}\left[\frac{ \partial \mathcal{A}}{\partial \omega}  \right]_{\omega=0}  T \Delta T,
\label{heatcur}
\eea
where $\mathcal{A}(\omega)=\mathcal{A}^1(\omega)+\mathcal{A}^2(\omega)$ (Appendix \ref{App6}) is the frequency-dependent transmission coefficient of energy across the middle $Y$ wire due to a temperature bias. We now define linear-response thermal conductance $G_{\rm T}$ as
\bea
 G_{\rm T}=\frac{J^{u}_{\rm XY}}{\Delta T}=\frac{\pi^2 k_B^2T }{3 \hbar}\left[\frac{ \partial \mathcal{A}}{\partial \omega}  \right]_{\omega=0},
\label{lrDTC}
\eea
where we can identify $[\partial \mathcal{A}/\partial \omega]_{\omega=0} =\mathcal{T}(\omega=0)$ in an N-TS-N device and other devices with TS leads. We also notice that $\mathcal{A}(\omega)= \omega \mathcal{T}(\omega)$ (\ref{pcur}) for all values of $\omega$ when $\epsilon_{\rm Y}=0$ in an N-TS-N device. The last result is probably due to the existence of two independent Majorana (conduction) channels without any scattering between them when $\epsilon_{\rm Y}=0$ \cite{RoyPRB2012}.  $G_{\rm T}$ shows some interesting features across the TP in an N-TS-N device. We discuss the properties of steady-state energy current and $G_{\rm T}$ for different hybrid devices in Sec.~\ref{thermal}.

\subsection{Electrical and spin current: Majorana wire}
\label{TransMajorana}
Next, we explore steady-state quantum transport in spinful models of superconductors and semiconductors. Apart from the electrical currents, we intend to find the features of spin current in such devices, which have at least one part made of Majorana wires. Motivated by the experimental set-ups in \cite{MourikScience2012, DengScience2016}, we now consider that the tunneling Hamiltonians also include spin-orbit couplings. Thus, the tunneling Hamiltonians for the $X$-$Y$ and $Y$-$Z$ junctions are the following:
\bea
 H^{\alpha' \beta'}&=&-\hbar\gamma_{\alpha' \beta'} \sum_{\sigma=\uparrow, \downarrow} (c^{\dag}_{l',\sigma} c_{l'+1,\sigma} + c^{\dag}_{l'+1,\sigma} c_{l',\sigma} ) \nn \\
 && ~~~~~~~~~~~~~~~~~~~~~~~~~~+ \hbar \zeta_{\alpha' \beta' }(c^{\dag}_{l'+1,\uparrow}c_{l',\downarrow}-c^{\dag}_{l'+1,\downarrow}c_{l',\uparrow}+h.c.)\, ,\nn \\
\label{tunham}
\eea
where $\alpha' \beta' ={\rm XY}$, $l'=L_{\rm X}$, and $\alpha' \beta' = {\rm YZ}$, $l'=L_{\rm XY}$ respectively for the X-Y and Y-Z junction. Here, $\gamma_{\alpha' \beta'}$ represents tunneling rate, and $\zeta_{\alpha' \beta' }$ represents the strength of Rashba spin-orbit coupling at the tunnel junction. 

The basic structure of the generalized LEGF is same for Kitaev and Majorana wire lead. However, the calculation of steady-state electrical and spin currents at the junctions of these devices made of Majorana/SM wire is a bit cumbersome due to the presence of spin degrees of freedom and spin-orbit coupling in the Majorana and SM wires. So without going into much details, we highlight some of the main steps to find the steady-state solutions of the Heisenberg's equations of motion in Appendix \ref{App4a}. We also explain the method of calculating noise-noise correlations for Majorana wire leads in Appendix \ref{App4b}.

We define the total electrical current of both spin components of electrons at the junctions between the wires. Again, the total particle density of electrons at the junctions is conserved, and we use the continuity equations to write the charge currents of electrons across the junctions. We immediately get the following expressions for electrical current from the $X$ bath to the $Y$ wire and from the $Y$ wire to the $Z$ bath, respectively:
\bea
\mathcal{J}^e_{\rm XY}&=& -2e\gamma_{\rm XY} \sum_{\sigma= \uparrow, \downarrow }{\rm Im}[\langle c^{\dg}_{L_{\rm X}+1, \sigma}(t)\, c_{L_{\rm X}, \sigma }(t) \rangle] \nn \\
&&~~~~~-2e \zeta_{\rm XY} \sum_{\sigma,\sigma'=\uparrow,\downarrow \atop \sigma \neq \sigma'}(-1)^{r(\sigma)}{\rm Im}[\langle c^{\dg}_{L_{\rm X}+1, \sigma}(t)\, c_{L_{\rm X}, \sigma' }(t) \rangle],\label{currSOXY} \nn \\
\\
\mathcal{J}^e_{\rm YZ}&=& -2e\gamma_{\rm YZ} \sum_{\sigma= \uparrow, \downarrow }{\rm Im}[\langle c^{\dg}_{L_{\rm XY}+1, \sigma}(t)\, c_{L_{\rm XY}, \sigma }(t) \rangle] \nn \\
&&~~~~~-2e \zeta_{\rm YZ} \sum_{\sigma,\sigma'=\uparrow,\downarrow \atop \sigma \neq \sigma'}(-1)^{r(\sigma)}{\rm Im}[\langle c^{\dg}_{L_{\rm XY}+1, \sigma}(t)\, c_{L_{\rm XY}, \sigma' }(t) \rangle], \label{currSOYZ} \nn \\
\eea
where, $r(\uparrow) (r(\downarrow))=1(2) $. These expressions can be directly employed to study time evolution of electrical currents at the junctions after numerically solving the equations of motion for annihilation and creation operators of the full device. To calculate junction currents at NESS using the generalized LEGF, we first take the Fourier transformation of the expressions (\ref{currSOXY}) and (\ref{currSOYZ}), then substitute the variables ($\tilde{b}_l(\omega)$) of $Y$ wire (\ref{b1}) and the baths into it. Finally, plugging the noise-noise correlations, we can derive the electrical currents at the junctions in the steady state. Like the TS-N-N and TS-TS-N devices, we set $\mu_{\rm X}=0$ and $\mu_{\rm Z}= eV$, and tune $eV$ to find DEC in TS-SM-SM and TS-TS-SM devices. For such systems, we are again interested in the DEC at $Y$-$Z$ junction, which is defined as $\left[\frac{d\mathcal{J}^e_{\rm YZ}}{d V}\right]$.

Recently, the spin transport has been experimentally explored in semiconductor -superconductor hybrid devices \cite{YangNanoLetter2020}. In order to calculate spin transport in devices with the Majorana wires, we need to define spin currents at the junctions. The expressions for spin currents at the junctions can be obtained by applying the continuity equation for the local spin density around the junctions. We define local spin density operator for the $x$-component of spin at site $l'$ as $\sigma_x^{l'}=\frac{\hbar}{2}(c^{\dag}_{l',\uparrow}c_{l',\downarrow}+c^{\dag}_{l',\downarrow}c_{l',\uparrow})$. Employing the continuity equations at the terminal sites of the $Y$ wire ($l'=L_{\rm X}+1,L_{\rm XY}$), we derive below the expressions of $x$-component of spin current from the $X$ bath to the $Y$ wire and from the $Y$ wire to the $Z$ bath, respectively:
\bea
 \mathcal{J}^{s_x}_{\rm XY}&=& -\hbar\gamma_{\rm XY} \sum_{\sigma,\sigma'=\uparrow,\downarrow \atop \sigma \neq \sigma'} {\rm Im}[\langle c^{\dg}_{L_{\rm X}+1, \sigma}(t)\, c_{L_{\rm X}, \sigma'}(t) \rangle] \nn \\
 &&~~~~~~~~~~~~~~~~~~~~~~~~~+ \hbar \zeta_{\rm XY} \sum_{\sigma=\uparrow,\downarrow} (-1)^{r(\sigma)}{\rm Im}[\langle c^{\dg}_{L_{\rm X}+1, \sigma}(t)\, c_{L_{\rm X}, \sigma }(t) \rangle],\nn \\
\label{currXYSX} 
\eea
\bea
 \mathcal{J}^{s_x}_{\rm YZ}&=& -\hbar\gamma_{\rm YZ} \sum_{\sigma,\sigma'=\uparrow,\downarrow \atop \sigma \neq \sigma'} {\rm Im}[\langle c^{\dg}_{L_{\rm XY}+1, \sigma}(t)\, c_{L_{\rm XY}, \sigma'}(t) \rangle] \nn \\ 
 &&~~~~~~~~~~~~~~~~~~~~~~~+ \hbar \zeta_{\rm YZ} \sum_{\sigma=\uparrow,\downarrow} (-1)^{r(\sigma)}{\rm Im}[\langle c^{\dg}_{L_{\rm XY}+1, \sigma}(t)\, c_{L_{\rm XY}, \sigma }(t) \rangle],\nn \\
\label{currYZSX} 
\eea
where, $r(\uparrow) (r(\downarrow))=1(2) $.
We later apply the above expressions (\ref{currXYSX},\ref{currYZSX}) to find the expectation value of spin currents in different devices made of Majorana wires. Inspired by the extensive applications of DEC in probing the topological state of superconductors in experiments, we here introduce differential spin conductance (DSC), which we write as $\left[\frac{d\mathcal{J}^{s_x}_{\rm YZ}}{d V}\right]$ for the $Y$-$Z$ junction. In Sec.~\ref{spincur}, we discuss some novel features of zero-temperature DSC in various spinful TS wire devices by tuning $\mu_{\rm Z}$ from $-eV$ to $eV$ while keeping $\mu_{\rm X}=0$. 

\section{Results and discussion}
\label{result}
In this section, we discuss many new results that we obtained from our steady-state electrical, thermal, and spin current formulas which have been derived using the generalized LEGF method in the previous section. We have earlier argued that the generalized LEGF technique is applicable to those systems which reach a unique NESS in the long-time limit $t \rightarrow \infty$. For example, the LEGF can easily be applied to an N-TS-N device because such a junction satisfies the requirements for the unique NESS \cite{RoyPRB2012, Bhat2020}. However, an $X$-$Y$-$Z$ configuration with a superconducting wire for $X$ lead does not attain a unique NESS unless the $Z$ wire is an N/SM or a topological superconductor at TP. The basic criterion for achieving a unique NESS in these devices is the absence of a bound state with energy within the full system's bulk energy gap. There is no bound state in the hybrid device when the energy spectrum of $Z$ wire is gapless for an N/SM or a TS at TP, thus the currents at the junctions in the long-time limit ($t \rightarrow \infty$) do not depend on the initial density matrix of the middle wire indicating a unique NESS \cite{Bondyopadhaya2019}. Below, we validate our steady-state current expressions by comparing them with the currents' long-time values calculated directly through numerically solving the Heisenberg equations of the full device using the time-dependent Green's function techniques \cite{DharPRB2006,Bondyopadhaya2019}.

It should be noted that we choose some arbitrary initial conditions (e.g., local density) for the $Y$ wire in our direct time-evolution numerics \cite{Bondyopadhaya2019}. For example, we can choose $n_{l'}$ as the initial number of spinless fermions at $l'$-th site of the $Y$ wire for spinless models. Since every site of a spinful system like an SM/Majorana wire can be filled by two spins; $n_{l',\sigma}$ represents the number of fermions with spin $\sigma$ for spinful models. In our first principle calculation, we numerically evaluate some dynamical quantities like electric and thermal currents. We infer that the system has reached a unique NESS if these dynamical quantities become independent of time and the initial densities of $Y$ wire in the long-time limit. In Appendix \ref{App7}, for the shake of completeness, we briefly discuss the aforesaid numerical method for studying the time evolution of the currents in an $X$-$Y$-$Z$ device made of Kitaev chain and N wires. 
 
The rest of this section is divided in the four subsections to discuss (1) charge current carried by Cooper pairs in an N-TS-N device where the TS is a Kitaev chain and in an SM-TS-SM device with a Majorana wire, (2) thermal and thermoelectric currents and conductances, (3) electrical current and DEC in hybrid devices with Kitaev chain leads, and  (4) spin current, DEC and DSC in hybrid devices with Majorana wire leads.

\subsection{Charge current carried by Cooper pairs in N-TS-N $\&$ SM-TS-SM}
\label{pairCur}
\begin{figure}
\includegraphics[width=0.99\linewidth]{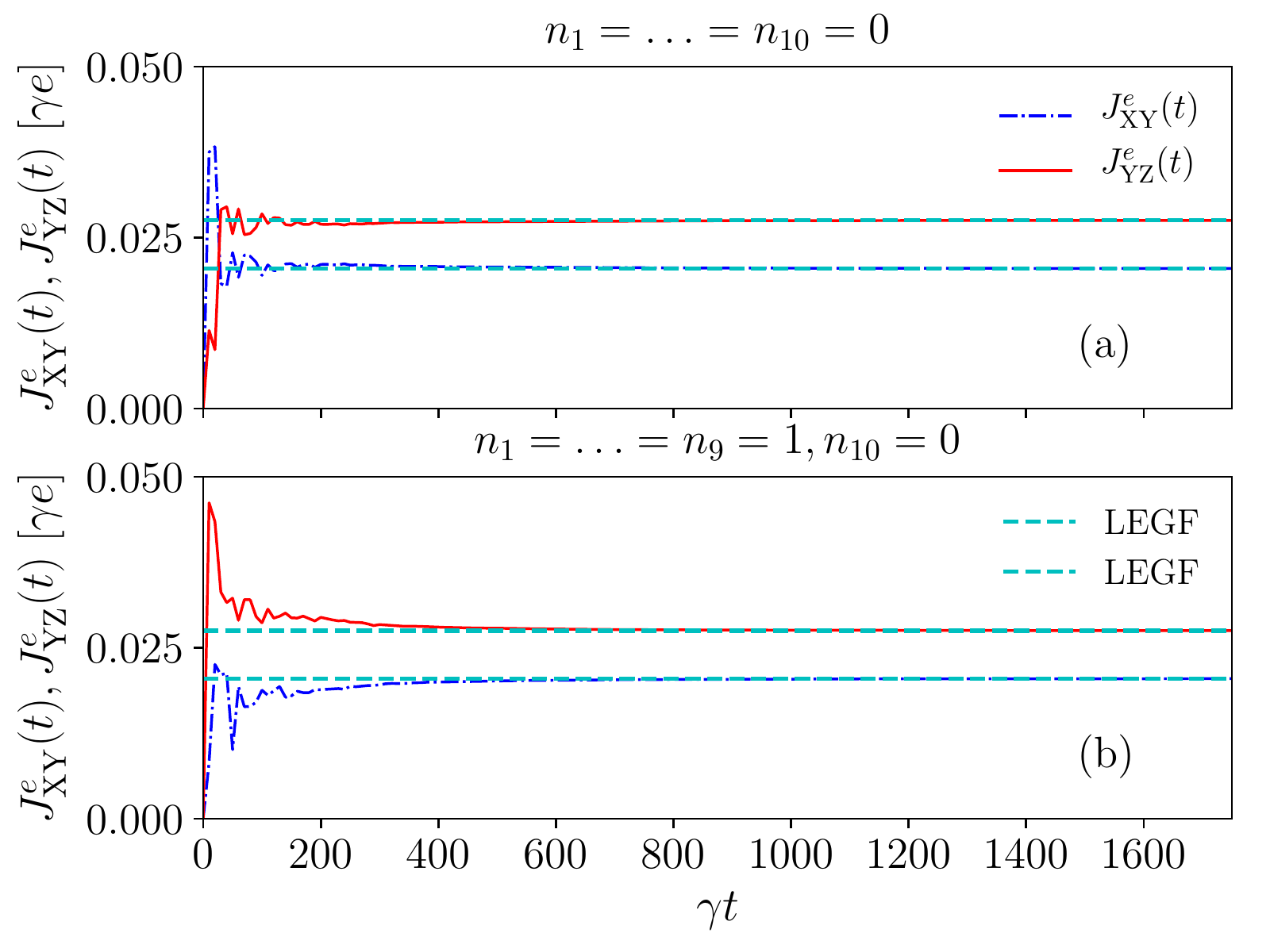}
\centering
\caption{Comparison of electrical currents $J^e_{\rm XY}(t), J^e_{\rm YZ}(t)$ obtained from the first-principle/direct time-evolution numerics and the generalized LEGF method (dashed lines) for an N-TS-N device made of a Kitaev chain. The initial numbers of spinless electrons $(n_{l'})$ at the middle TS wire used for the time-evolution numerics  are indicated on the headings. In both panels, $L_{\rm X}=L_{\rm Z}=2000, L_{\rm Y}=10$, $\gamma_{\rm X}=\gamma_{\rm Z}=\gamma=1$, $\gamma_{\rm Y}=0.5$, $\Delta_{\rm X}=\Delta_{\rm Z}=0,\Delta_{\rm Y}=0.15$, $\epsilon_{\rm X}=\epsilon_{\rm Z}=0$, $\epsilon_{\rm Y}=0.01$, $\gamma_{\rm XY}=\gamma_{\rm YZ}=0.25$, $T_{\rm X}=0.02,T_{\rm Z}=0.02$ and $\mu_{\rm X}=0.2,\mu_{\rm Z}=-0.4$. All above parameters except lengths are in units of $\gamma$.}
\label{pcNTSN}
\end{figure}

\begin{figure}
\includegraphics[width=0.99\linewidth]{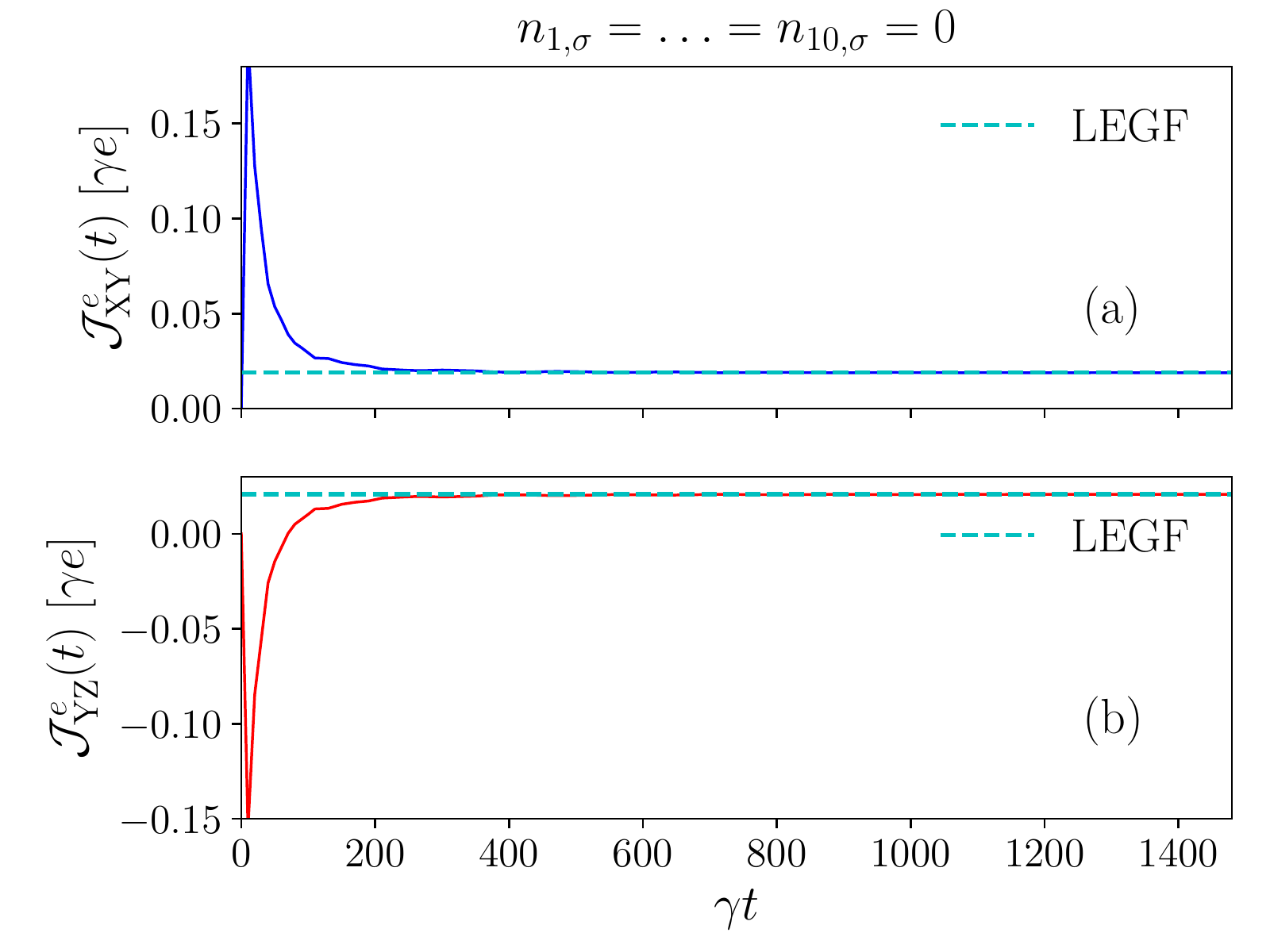}
\centering
\caption{Comparison of electrical currents $\mathcal{J}^e_{\rm XY}(t), \mathcal{J}^e_{\rm YZ}(t)$ obtained from the first-principle/direct time-evolution numerics and the generalized LEGF method (dashed lines) for an SM-TS-SM device of a Majorana wire. The initial density $n_{l',\sigma}=c^{\dg}_{l',\sigma}c_{l',\sigma}$ with $\sigma=\uparrow,\downarrow$ of the middle TS wire for the direct time-evolution numerics is indicated on the heading of the top panel. In both panels, $L_{\rm X}=L_{\rm Z}=1500, L_{\rm Y}=10$, $\gamma_{\rm X}=\gamma_{\rm Z}=\gamma=1$, $\gamma_{\rm Y}=0.5$, $\Delta_{\rm Y}=0.15,\Delta_{\rm Y}=\Delta_{\rm Z}=0, \zeta_{\rm X}=\zeta_{\rm Y}=\zeta_{\rm Z}=\zeta_{\rm XY}=\zeta_{\rm YZ}=0.2, B_{\rm X}=B_{\rm Y}=B_{\rm Z}=0.3$, $\epsilon_{\rm X}=\epsilon_{\rm Z}=0, \epsilon_{\rm Y}=0.01$, $\gamma_{\rm XY}=\gamma_{\rm YZ}=0.25$, $T_{\rm X}=T_{\rm Z}=0.02$ and $\mu_{\rm X}=0.2,\mu_{\rm Z}=-0.4$. All above parameters except lengths are in units of $\gamma$. }
\label{pcSMTSSM}
\end{figure}

First, we validate the generalized LEGF method for an N-TS-N and an SM-TS-SM device by comparing the steady-state currents with the results obtained by numerically evaluating the time-evolution of the full devices. The first-principle/direct numerics shows that the long-time transport in the N-TS-N (SM-TS-SM) device is independent of the initial conditions for the finite TS wire when the energy band of the N (SM) wire is wider than that of the TS wire. Therefore, the results obtained from the time-evolution numerics and the generalized LEGF should match with each other in such scenarios. In Fig.~\ref{pcNTSN}, we compare the LEGF formulas with the time-evolution numerics. We find very good agreement between the two values at a long time. Moreover, the values of electrical currents at a long time do not depend on the initial density of the middle TS wire as shown in panels of Fig.~\ref{pcNTSN}(a,b). Here, our time-evolution numerics shows that the currents at $X$-$Y$ and $Y$-$Z$ junctions reach constant non-zero values respectively, $0.02049$ and $0.02751$ for both the initial conditions of the $Y$ wire. On the other hand, the steady-state calculation yields $J^e_{\rm XY}=0.020492$ and $J^e_{\rm YZ}=0.02751$. Thus, both the calculations match excellently. We further observe  such a good matching for other sets of parameters as well as for an SM-TS-SM device where the LEGF values of $\mathcal{J}^e_{\rm XY}=0.018851$ and $\mathcal{J}^e_{\rm YZ}=0.020826$ do agree with the time-evolution numerics as shown in Fig.~\ref{pcSMTSSM}. Here the numerical values of the junction currents are in units of $\gamma e$. These results confirm the validity of generalized LEGF approach for steady-state electrical transport in N-TS-N and SM-TS-SM devices. 

\begin{figure}
\includegraphics[width=0.99\linewidth]{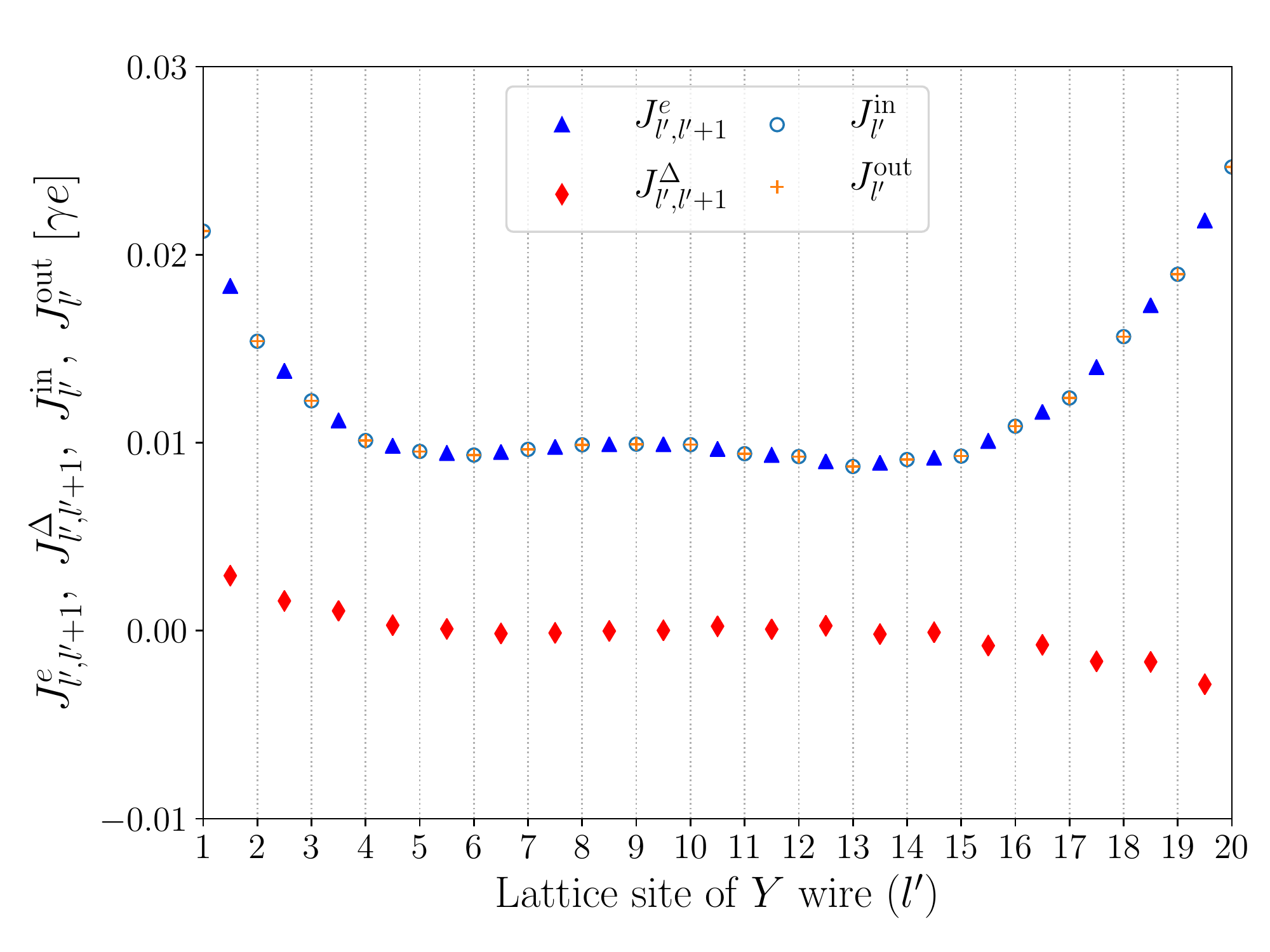}
\centering
\caption{Dependence of different components of the electrical currents on position of the lattice site inside the middle superconducting wire of a N-TS-N device made of a Kitaev chain. The parameters are $L_{\rm Y}=20$, $\gamma_{\rm X}=\gamma_{\rm Z}=\gamma=1$, $\gamma_{\rm Y}=0.5$, $\Delta_{\rm Y}=0.1$, $\epsilon_{\rm X}=\epsilon_{\rm Z}=0,\epsilon_{\rm Y}=0.05$, $\gamma_{\rm XY}=\gamma_{\rm YZ}=0.25$, $T_{\rm X}=T_{\rm Z}=0.02$ and $\mu_{\rm X}=0.2,\mu_{\rm Z}=-0.4$. All above parameters except lengths are in units of $\gamma$.}
\label{NTSN_BondC}
\end{figure}

We find from Fig.~\ref{pcNTSN} (Fig.~\ref{pcSMTSSM}) that $J^e_{\rm XY} \neq J^e_{\rm YZ}$ for an N-TS-N device \cite{RoyPRB2012} (or $\mathcal{J}^e_{\rm XY} \neq \mathcal{J}^e_{\rm YZ}$ for an SM-TS-SM device \cite{RoyPRB2013}), since the total and local electron number are not conserved inside a superconductor modeled by a mean-field Hamiltonian as ours in this paper. The violation of local and global electron number conservation can be expressed, respectively, as $[c^{\dagger}_{l'}c_{l'},H^{\rm Y}] \neq 0$ and $[\sum_{l'=L_{\rm X}+1}^{L_{\rm X}+L_{\rm Y}}c^{\dagger}_{l'}c_{l'},H^{\rm Y}] \neq 0$ 
 when $\Delta_{\rm Y} \neq 0$ for a spinless $Y$ wire. Due to the local as well global non-conservation of electron number inside a superconductor the electron current density is same neither locally nor globally, i.e., the incoming electron current density is not equal to the outgoing electron current density in a two-terminal geometry. We can rather say that such superconductor modelled by a mean-field Hamiltonian acts as a reservoir for its electrons. Thus, the single-electron charge current entering into the superconductor from the left edge may not be equal to the single-electron charge current coming out from the right edge of the superconductor. We note that the tunneling Hamiltonians (\ref{Kcontact},\ref{tunham}) do not contain any pairing term; thus, the junction current, that is passing between the boundary wire (lead) and the middle wire, is solely carried by single electrons. Nevertheless, Cooper pairs play a significant role in the electrical, thermal, and spin transport inside the superconductors. We discuss below the contribution of the Cooper pairs to the electrical current inside a superconductor.  

For devices like N-TS-N and SM-TS-SM, some interesting phenomena can be revealed when we investigate the charge currents inside the superconductors made of either a Kitaev chain or a Majorana wire. Inside a Kitaev chain, the total charge current can be written as a sum of currents carried by (i) single electrons ($J^e$) and (ii) pairs of electrons or Cooper pairs ($J^\Delta$). The latter is the charge current generated due to the motion of the Cooper pairs. It can be demonstrated using the continuity equation for charge density at $l'$-th site of the $Y$ wire made of a Kitaev chain that the charge current going into $l'$-th site is $J^{\rm in}_{l'}=J^e_{l'-1,l'}-J^\Delta_{l'-1,l'}$, and the charge current coming out from $l'$-th site is $J^{\rm out}_{l'}=J^e_{l',l'+1}+J^\Delta_{l',l'+1}$, where $J^e_{l',l'+1}$ is the single-electron current from site $l'$ to $l'+1$, and $J^\Delta_{l',l'+1}$ is the Copper pair contribution to the charge current between site $l'$ to $l'+1$. Considering the continuity equation for charge density at $l'$-th site ($\rho_{l'}=c_{l'}^\dg c_{l'}$), expressions for $J^e_{l',l'+1}$ and $J^\Delta_{l',l'+1}$ are obtained as:
\bea
&&J^e_{l',l'+1}=-2e\gamma_{\rm Y} {\rm Im}[ \langle c^\dg_{l'+1}(t) c_{l'} (t)\rangle ]\, , \nn \\
&&J^\Delta_{l',l'+1}=2e\Delta_{\rm Y} {\rm Im}[ \langle c^\dg_{l'}(t) c^\dg_{l'+1} (t)\rangle ]\,,
\label{bondcurrent}
\eea
where $\gamma_{\rm Y}$ and $\Delta_{\rm Y}$ are the hopping and the real pairing potential of the $Y$ wire respectively. From the conservation of total electrical charges at any (e.g., $l'$-th) site inside the TS wire, we have $J^{\rm in}_{l'}=J^{\rm out}_{l'}$ for $l'=2,\dots,L_{\rm Y}-1$, which implies 
\beq
J^e_{l'-1,l'}-J^\Delta_{l'-1,l'}=J^e_{l',l'+1}+J^\Delta_{l',l'+1}.
\label{bondcurcon}
\eeq
Now, $J^e_{\rm XY}$ is the incoming electrical current for the first site ($l'=1$), where as $J^e_{\rm YZ}$ is the outgoing electrical current for the last site ($l'=L_{\rm Y}$). Thus, we further have $J^e_{\rm XY}=J^{\rm out}_{1}$ and $J^{\rm in}_{L_{\rm Y}}=J^e_{\rm YZ}$. For notational convenience, we use $J^{\rm in}_{1}$ and $J^{\rm out}_{L_{\rm Y}}$ to represent $J^e_{\rm XY}$ and $J^e_{\rm YZ}$ respectively. Now, the conservation of electrical charge yields 
\beq
J^e_{\rm XY}=J^e_{\rm YZ}+2\sum_{l'=1}^{L_{\rm Y}-1}J^\Delta_{l',l'+1},
\label{nonceleccur}
\eeq
which relates the electrical currents at the left and right junction. In Fig.~\ref{NTSN_BondC}, we plot different components of the electrical currents which are carried by single electrons ($J^e_{l',l'+1}$) and Cooper pairs ($J^\Delta_{l',l'+1}$) for each bond $(l',l'+1)$, along with the incoming current ($J^{\rm in}_{l'}$) and the outgoing current ($J^{\rm out}_{l'}$) at each lattice sites of the middle wire. We depict $J^e_{l',l'+1}$ and $J^\Delta_{l',l'+1}$ by triangle and diamond symbols respectively, and these are placed at the middle of $l'$-th and $(l'+1)$-th sites as these are associated with $(l',l'+1)$ bond. On the other hand, $J^{\rm in}_{l'}$ and $J^{\rm out}_{l'}$ which are respectively represented by circle and `+' symbols, are placed just at the position of $l'$-th lattice site. 

\subsection{Thermal $\&$ thermoelectric currents and conductances}
\label{thermal}

We first discuss the thermal current and thermal conductance of a Kitaev chain in an N-TS-N device. In Fig.~\ref{EcNTSN}, we show that the values of $J^{u}_{\rm XY}(t)$ and $J^{u}_{\rm YZ}(t)$ due to a voltage bias become equal at long times for any initialization of the middle wire which confirms our prediction in Sec.~\ref{kita} based on the conservation of total energy for a TS or an N middle wire. Both for a voltage bias and a thermal bias, the energy transport seems to be ballistic (i.e., $J^{u}_{\rm XY}$ and $J^{u}_{\rm YZ}$ are independent of $L_{\rm Y}$) in an N-TS-N device for the TS wire in a topological phase \footnote{We note that thermal transport is also ballistic for a middle wire made of an N or a SM wire.}. In Table.~\ref{table1}, we show the values of $J^{u}_{\rm XY}~(=J^{u}_{\rm YZ})$ for different $L_{\rm Y}$ in an N-TS-N device with a finite temperature bias (second column) and a finite voltage bias (third column) when the middle TS wire is in a topological phase. The values of $J^{u}_{\rm XY}~(=J^{u}_{\rm YZ})$ do not seem to vary with $L_{\rm Y}$ for longer $L_{\rm Y}$ within the numerical precision in our study. We note that we take a relatively large bias to achieve better numerical precision in showing the ballistic thermal transport in the middle TS wire's topological regime. Such a large bias includes the contributions of the above-gap modes in the thermal transport. We also find a ballistic thermal transport near the TS wire's topological phase transition, as shown in the last column of Table.~\ref{table1}. In comparison to the topological phase, the values of thermal current  are relatively higher near the TP (even for a smaller bias), where the thermal conductance rapidly changes with $\epsilon_{\rm Y}$. We discuss it below.

\begin{table}
\begin{center}
 \begin{tabular}{c|c|c|c}
 \multicolumn{1}{c|}{} &     \multicolumn{1}{c|}{$\epsilon_{\rm Y}=0.2\gamma_{\rm Y}, \Delta T \ne 0 $} & \multicolumn{1}{c|}{$\epsilon_{\rm Y}=0.2\gamma_{\rm Y}, \Delta \mu \ne 0$} & \multicolumn{1}{c}{$\epsilon_{\rm Y}=2\gamma_{\rm Y}, \Delta \mu \ne 0$}\\
      \hline
      $L_{\rm Y}$ & $J^{u}_{\rm XY}=J^{u}_{\rm YZ}$ & $J^{u}_{\rm XY}=J^{u}_{\rm YZ}$& $J^{u}_{\rm XY}=J^{u}_{\rm YZ}$\\
      \hline
      10 &  0.00093305  & -0.00077447 &  -0.00077616\\
      20 &  0.00106418  & -0.00033649 &  -0.00082534\\ 
      40 &  0.00094839  & -0.00039224 &  -0.00085083\\
      80 &  0.00104907  & -0.00043694 &  -0.00085033\\
     120 &  0.00100268  & -0.00041746 &  -0.00085032 \\
    \end{tabular}
    \caption{Length dependence of the energy current $J^{u}_{\rm XY}, J^{u}_{\rm YZ}$ in an N-TS-N device made of a Kitaev chain. Here, $\gamma_{\rm X}=\gamma_{\rm Z}=\gamma=1$, $\gamma_{\rm Y}=0.5$, $\Delta_{\rm X}=\Delta_{\rm Z}=0,\Delta_{\rm Y}=0.15$, $\epsilon_{\rm X}=\epsilon_{\rm Z}=0$, $\gamma_{\rm XY}=\gamma_{\rm YZ}=0.25$ in all columns. We keep $\Delta T=T_{\rm X}-T_{\rm Z}=0.18, \mu_{\rm X}=\mu_{\rm Z}=0$ in the second column, $T_{\rm X}=T_{\rm Z}=0.02, \Delta \mu=\mu_{\rm X}-\mu_{\rm Z}= 1.2~ (\mu_{\rm X}=-\mu_{\rm Z}=0.6)$ in the third column, and $T_{\rm X}=T_{\rm Z}=0.02, \Delta \mu= 0.6~ (\mu_{\rm X}=-\mu_{\rm Z}=0.3)$ in the fourth column. All above parameters except lengths are in units of $\gamma$.}
    \label{table1}
  \end{center}
\end{table}

\begin{figure}
\includegraphics[width=0.99\linewidth]{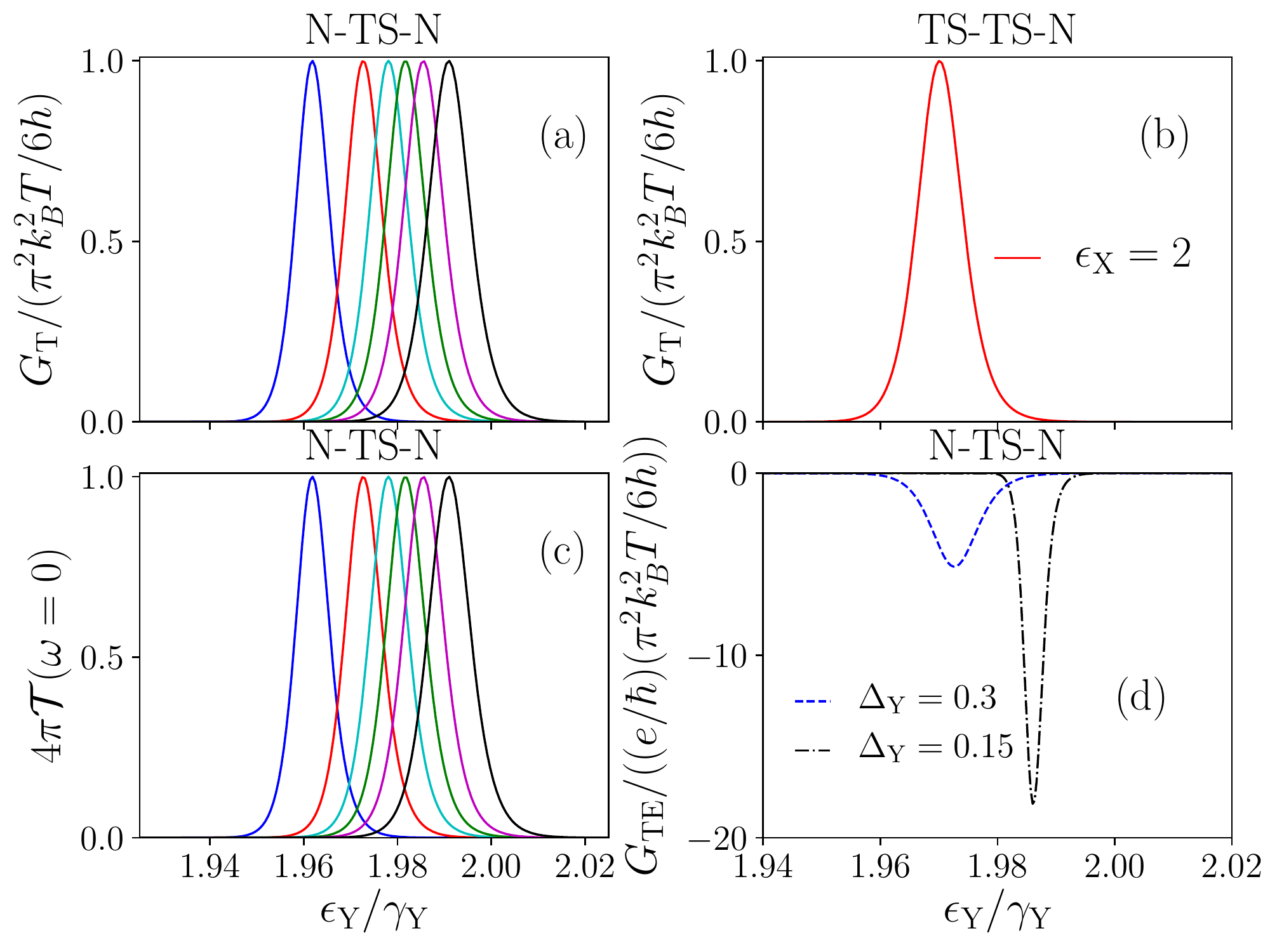}
\centering
\caption{Thermal conductance $G_{\rm T}$, zero-frequency thermoelectric transmission $\mathcal{T}(\omega=0)$ and thermoelectric conductance $G_{\rm TE}$ of the middle TS wire as a function of onsite energy $\epsilon_{\rm Y}$ in an N-TS-N and a TS-TS-N device made of a Kitaev chain. In all panels, $L_{\rm Y}=100$, $\gamma_{\rm X}=\gamma_{\rm Y}=\gamma_{\rm Z}=\gamma=1$, $\epsilon_{\rm Z}=0, \Delta_{\rm Z}=0$. Further, $\epsilon_{\rm X}=\Delta_{\rm X}=0,  \Delta_{\rm Y}=0.3, \gamma_{\rm YZ}=\gamma_{\rm XY}=[0.05,0.15,0.25,0.35,0.5,1]$ from left to right curves in panels (a,c), $\Delta_{\rm X}= \Delta_{\rm Y}=0.3, \gamma_{\rm YZ}=\gamma_{\rm XY}=0.15$ in panel (b), and $\epsilon_{\rm X}=\Delta_{\rm X}=0, \gamma_{\rm YZ}=\gamma_{\rm XY}=0.15$ in panel (d). All above parameters except lengths are in units of $\gamma$.}
\label{DTC}
\end{figure}

We notice that the linear-response thermal conductance $G_{\rm T}$ shows a sharp  peak near the TP as $\epsilon_{\rm Y}$ is tuned through a topological phase transition in the middle TS wire. The value of the sharp peak in $G_{\rm T}$ at the transition is quantized, and its value is $\pi^2k_B^2T/6h$, which was predicted earlier in \cite{AkhmerovPRL2011,FulgaPRB2011,Beenakker2015}. Such a behavior of $G_{\rm T}$ across the TP has been proposed for the detection of the topological phase transition in TS wires. Our open quantum-system description of thermal transport confirms the quantized height, which we depict in Fig.~\ref{DTC}(a). Nevertheless, we further observe that the height of sharp peak in $G_{\rm T}$ at the transition is sensitive to the tunneling rates $\gamma_{\rm XY}$ and $\gamma_{\rm YZ}$ for the open TS wires as shown in Fig.~\ref{DTC}(a). The value of $\epsilon_{\rm Y}$, where the sharp peak in $G_{\rm T}$ appears in the open TS wire, seems to move towards $2\gamma_{\rm Y}$ (the TP for an isolated TS wire) as $\gamma_{\rm XY},\gamma_{\rm YZ}$ approach $\gamma_{\rm X}=\gamma_{\rm Y}=\gamma_{\rm Z}$.

We can also relate the quantized value of $G_{\rm T}$ to the zero-frequency thermoelectric transmission coefficient ($\mathcal{T}(\omega=0)$) across the open TS wire under the temperature bias as described in Sec.~\ref{kita}. In Fig.~\ref{DTC}(c), we display $G_{\rm T}/(\pi^2k_B^2T/6h)=4\pi \mathcal{T}(\omega=0)$ for different tunneling rates. We further observe a large thermoelectric current and a huge dip in the thermoelectric conductance $G_{\rm TE}=J^{e}_{\rm XY}/\Delta T=(\pi^2k_B^2eT/3\hbar^2)\left[\frac{ \partial \mathcal{T}}{\partial \omega}  \right]_{\omega=0}$ near the topological phase transition of the middle TS wire in an N-TS-N device. The height of the dip in $G_{\rm TE}$ seems to depend on the values of pairing amplitude of the TS wires, and the height increases with a decrease in $\Delta_{\rm Y}$ as shown in Fig.~\ref{DTC}(d). The large values of thermoelectricity may generate potential applications of these TS wires. Experimental measurements of such a large dip in the thermoelectric current or conductance in these devices would be much easier than that of the relatively small thermal conductance peak near the phase transition. Therefore, the thermoelectric current or conductance might be a better probe to detect the TS wires' topological phase transition experimentally. 

We inspect the features of $G_{\rm T}$ in various devices with TS leads to unveil the role of TS leads in transport. In Fig.~\ref{DTC}(b), we show $G_{\rm T}$ from the $Y$-$Z$ junction with $\epsilon_{\rm Y}$ in a TS-TS-N device by varying $\epsilon_{\rm X}$ of the TS lead. Surprisingly, we observe a quantized peak in $G_{\rm T}$ near the TP of the middle TS wire only when the TS lead is also at the TP. When the TS lead is away from the TP, the peak in $G_{\rm T}$ near the TP of the middle TS wire disappears. Therefore in Fig.~\ref{DTC}(b), we have plotted just one quantized peak in $G_{\rm T}$ corresponding to $\epsilon_{\rm X}=2\gamma_{\rm X}$. To our opinion, a topological transition of the TS leads can also be probed by the thermal conductance measurements. 

\subsection{Electrical current $\&$ differential conductance in devices with Kitaev chain leads}

Different hybrid systems with one or multiple TS leads have been investigated in the recent years to probe emergence of Majorana quasiparticles as well as for efficient quantum devices \cite{AliceaReview2012,Rokhinson2012,Yang2015,Zazunov2016,Sharma2016,Ioselevich2016,Bondyopadhaya2019,Rokhinson2012}. In Appendix~\ref{App7}, we show validity of steady-state transport in TS-N-Z and TS-TS-Z with Z=N and TS at TP by comparing the LEGF results with the direct time-evolution numerics at long time. 
We observe an interesting electrical current asymmetry in a TS-N-N device with spatial asymmetry (broken parity). The spatial asymmetry in such devices can be engineered by creating different tunneling rates at the $X$-$Y$ and $Y$-$Z$ junctions. It is generally expected to have different forward and backward currents under the reversal of bias when there is spatial asymmetry in nonlinear models. Such a difference in currents (rectification) between the forward and reversed bias is generated due to a variation in the distributions of inelastically scattered modes under bias reversal, which is intrinsically related to the nonlinearity. However, the quantum transport for our noninteracting model of N wires and mean-field model of TS wires is expected to be linear. Therefore, a rectification in electrical current generally is not expected in our TS-N-N devices even in the presence of different tunneling rates. Nevertheless, we find a large change in the steady-state electrical currents when we reverse the tunneling rates keeping all other parameters including the bias unaltered. For example, we find $J^e_{\rm XY}=J^e_{\rm YZ}=-0.014245$ (in units of $\gamma e$) for $\gamma_{\rm XY}=0.125, \gamma_{\rm YZ}=0.25$, and $J^e_{\rm XY}=J^e_{\rm YZ}=-0.006628$ (in units of $\gamma e$) for $\gamma_{\rm XY}=0.25, \gamma_{\rm YZ}=0.125$, where $L_{\rm Y}=3 $, and the other relevant parameters in the units of $\gamma$ for both cases are $\gamma_{\rm X}=\gamma_{\rm Z}=\gamma=1, \gamma_{\rm Y}=0.5, \Delta_{\rm X}=0.3,\Delta_{\rm Y}=\Delta_{\rm Z}=0, \epsilon_{\rm X}=\epsilon_{\rm Z}=0,\epsilon_{\rm Y}=0.05, T_{\rm X}=T_{\rm Z}=0.02$ and $\mu_{\rm X}=0,\mu_{\rm Z}=0.5$. Such a change in electrical current  is related to different hybridization of the Majorana quasiparticle at the TS wires near $X$-$Y$ junction for different strength of the tunneling rate at that junction. The electrical currents at the both junctions are the same for each set of tunneling rates as expected for a middle N wire. Nevertheless, the electrical currents at the junctions do not change when the bias, e.g., the temperature of the leads, is reversed keeping the tunneling rates fixed at the junctions. This clarifies no true rectification in these hybrid devices with a mean-field model of TS wires. 

\begin{figure}
\includegraphics[width=0.99\linewidth]{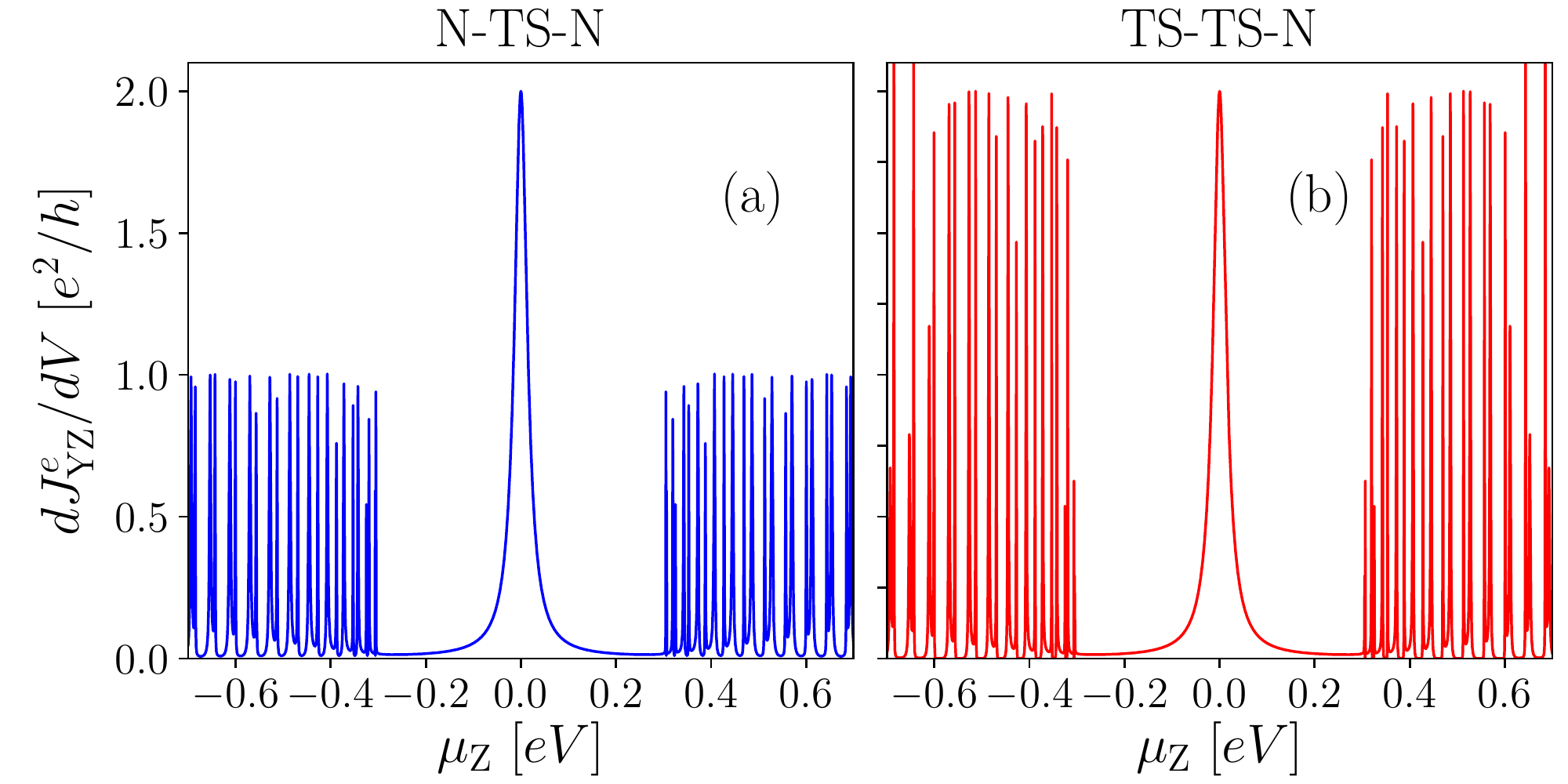}
\centering
\caption{Zero-temperature differential electrical conductance $dJ^e_{\rm YZ}/dV$ at the $Y$-$Z$ junction of an N-TS-N device (a) and a TS-TS-N device (b), where the TS wires are made of a Kitaev chain. In both panels, $L_{\rm Y}=50$, $\gamma_{\rm X}=\gamma_{\rm Z}=1.0,\gamma_{\rm Y}=0.5$, $\Delta_{\rm Y}=0.15, \Delta_{\rm Z}=0$, $\epsilon_{\rm X}=\epsilon_{\rm Y}=\epsilon_{\rm Z}=0$, $\gamma_{\rm XY}=\gamma_{\rm YZ}=0.15$, and $\mu_{\rm X}=0$. Also, $\Delta_{\rm X}=0$ in (a), and $\Delta_{\rm X}=0.3$ in (b). All above parameters except lengths are in units of $\gamma$.}
\label{DECKitaev1}
\end{figure}

\begin{figure}
\includegraphics[width=0.99\linewidth]{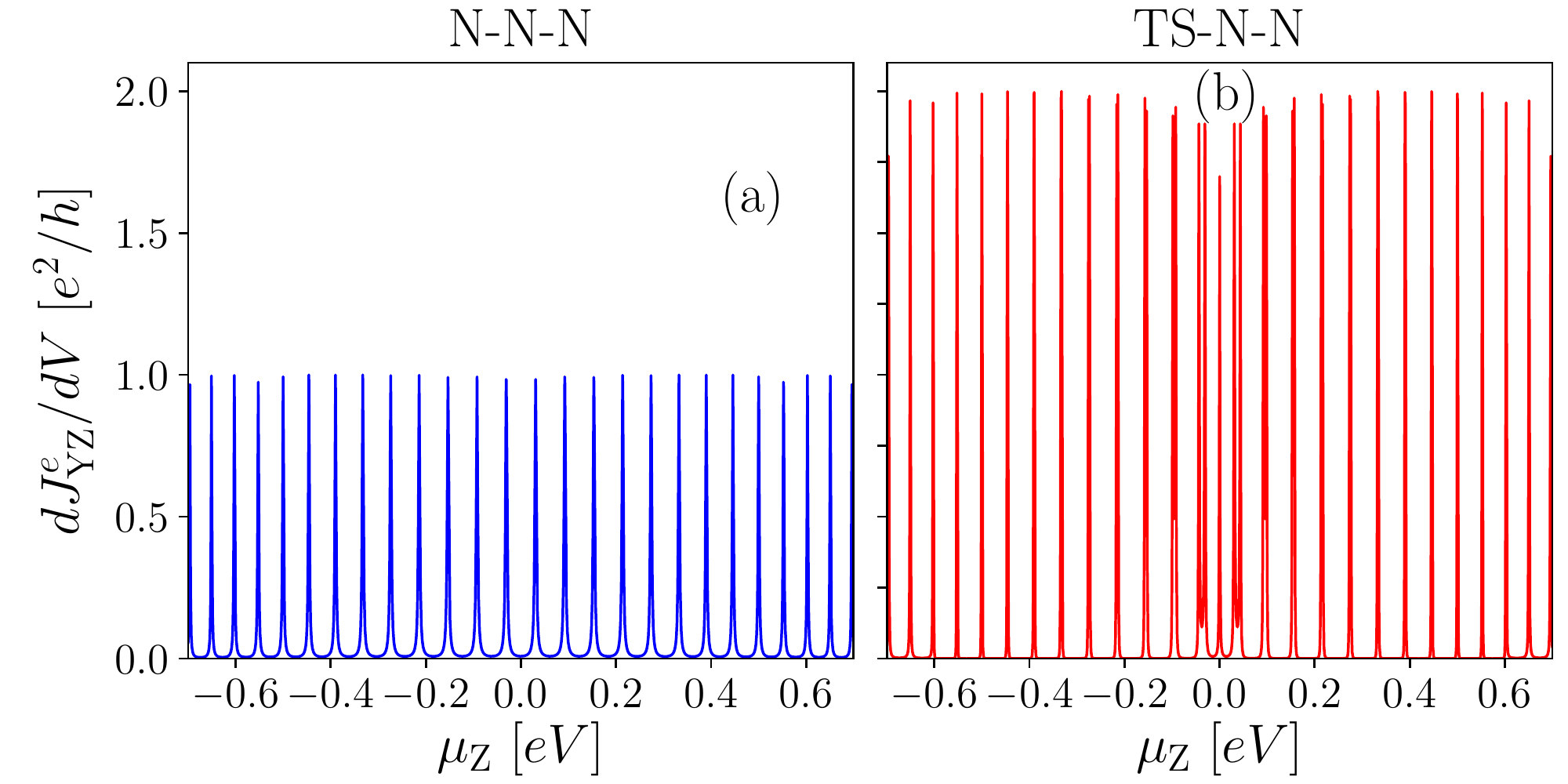}
\centering
\caption{Zero-temperature differential electrical conductance $dJ^e_{\rm YZ}/dV$ at the $Y$-$Z$ junction of an N-N-N device (a) and a TS-N-N device made of a Kitaev chain (b). In both panels, $L_{\rm Y}=50$, $\gamma_{\rm X}=\gamma_{\rm Z}=1.0,\gamma_{\rm Y}=0.5$, $\Delta_{\rm Y}=\Delta_{\rm Z}=0$, $\epsilon_{\rm X}=\epsilon_{\rm Y}=\epsilon_{\rm Z}=0$, $\gamma_{\rm XY}=\gamma_{\rm YZ}=0.15$, and $\mu_{\rm X}=0$. Also, $\Delta_{\rm X}=0$ in (a), and $\Delta_{\rm X}=0.3$ in (b). All above parameters except lengths are in units of $\gamma$.}
\label{DECKitaev2}
\end{figure}

To identify the unique role of a TS lead compared to an N lead, we also investigate zero-temperature DEC in different devices. Such a zero-temperature DEC has been earlier proposed and applied to detect the emergence of MBS in various junctions of TS wires such as an N-TS (or SM-TS) and an N-TS-N (or SM-TS-SM) devices  \cite{DasNature2012, MourikScience2012, NadjPergeScience2014}.  A quantized zero-bias peak of height $2e^2/h$ within the superconducting pairing gap appears in the zero-temperature DEC when the middle TS wire is in a perfect topological phase. The zero-bias peak in the zero-temperature DEC disappears in the topologically trivial phase of the TS wires. Here, we further apply the zero-temperature DEC to quantify TS leads' role in our different hybrid devices. In Fig.~\ref{DECKitaev1}, we compare the zero-temperature DEC at the right TS-N junction of a TS-TS-N device to that of an N-TS-N device. While the height of the zero-bias DEC peak is the same in both cases, the height of the DEC peaks above the superconducting pairing gap is $2e^2/h$ for a TS-TS-N device in contrast to $e^2/h$ for an N-TS-N device. This is an intriguing feature as the DEC properties at the $Y$-$Z$ junction are mostly expected to depend on local properties (e.g., the density of states) of the $Y$ and $Z$ wires for a TS middle wire, which does not conserve the particle number. Therefore, our results indicate that while the features of zero-bias DEC are mostly determined by the local properties of the $Y$ and $Z$ wires, the above-gap DEC peaks are controlled by the properties of both leads ($X$ and $Y$ wires) nonlocally \cite{AkhmerovPRL2011}. The height $2e^2/h$ of the above-gap DEC peaks is mainly due to the superconducting pairing of the $X$ wire, which we confirm by keeping the $X$ wire in a topologically trivial phase. For superconducting leads, both electron-type and hole-type quasiparticle excitations contribute to the density of states for the energy spectrum above the superconducting gap. In contrast, for metallic leads, only electrons contribute to the density of states. Moreover, for the energy range ($E$), which is much greater than the pairing gap of the superconducting leads ($E >>\Delta$), the quasiparticle density of states for the superconductor is almost double of the density of states for the normal metal at that energy $E$ \cite{Timm}. A larger value of density of states for superconducting lead, which is almost two times that of the metallic bath, is the reason for observing above-gap DEC with an approximate height of around $2e^2/h$. Further, the width of the finite-voltage above-gap DEC peaks is mainly controlled by the couplings $\gamma_{\rm XY},\gamma_{\rm YZ}$. 

In Fig.~\ref{DECKitaev2}, we further compare the zero-temperature DEC at the right N-N junction from a TS-N-N device to that of an N-N-N device. We again find that the DEC peaks' height is $2e^2/h$ for a TS-N-N device in contrast to $e^2/h$ for an N-N-N device. We also notice a weak zero-bias DEC peak in a TS-N-N device whose height is almost $2e^2/h$. The zero-bias peak emerges due to the MBS in the TS wires across the left junction. Therefore, the TS lead's MBS has a signature in the right link of a coherent device. To clarify the role of the topological phase of the TS leads, we check the zero-temperature DEC in a TS-N-N device when the TS lead is in a topologically trivial phase. We observe that most of the zero-temperature DEC peaks disappear within the left TS lead's bulk-gap in a trivial phase. The height of the zero-bias peak is also less than $e^2/h$ for the TS lead in a trivial phase.

\subsection{Spin current, differential electrical $\&$ spin conductances in devices with Majorana wires}
\label{spincur}
\begin{figure}
\includegraphics[width=0.99\linewidth]{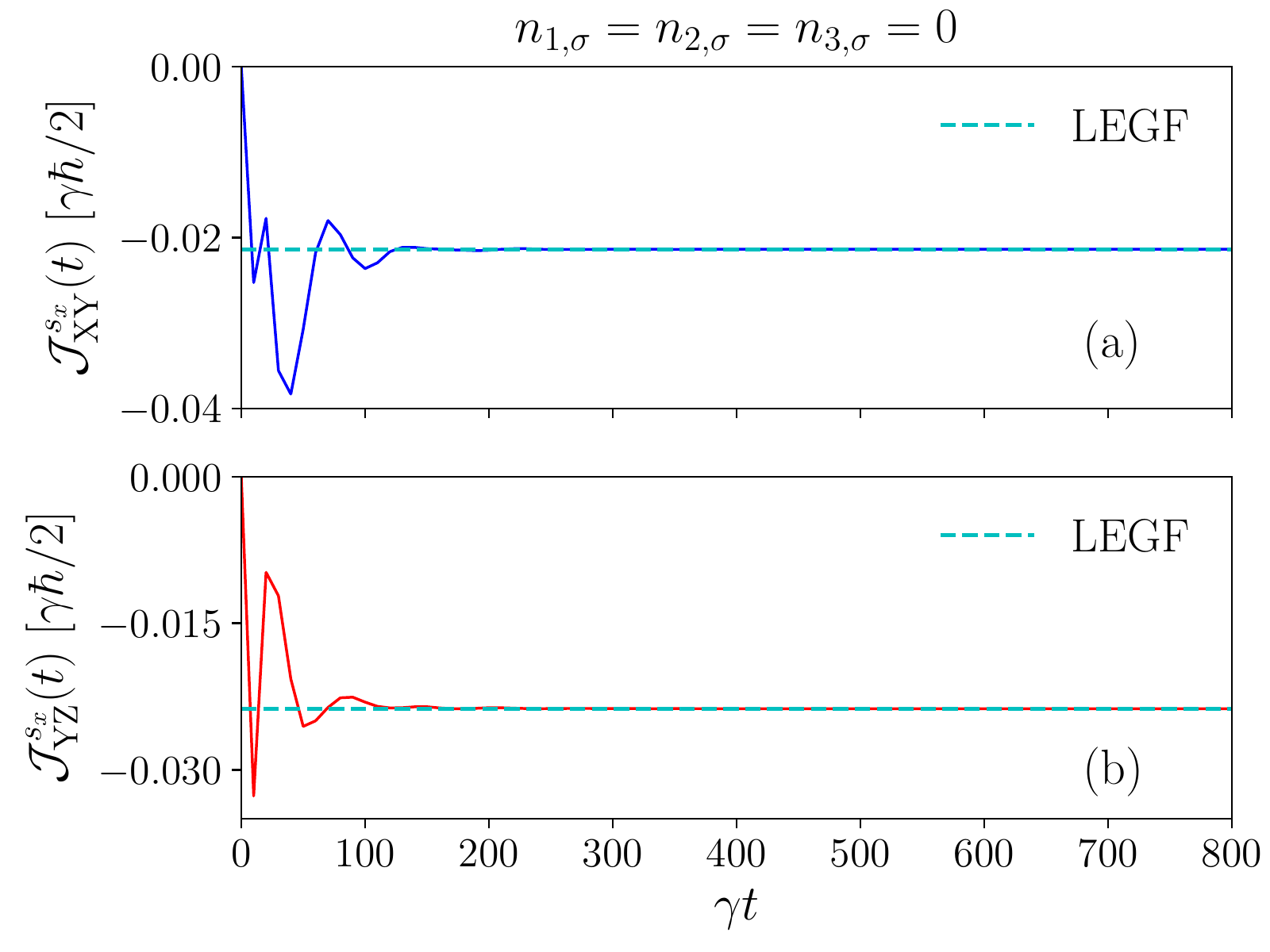}
\centering
\caption{Comparison of $x$-component of spin currents $\mathcal{J}^{s_x}_{\rm XY}(t)$ and $\mathcal{J}^{s_x}_{\rm YZ}(t)$ at both junctions obtained from the first-principle/direct time-evolution numerics (full lines) and the generalized LEGF method (dashed lines) for a TS-SM-SM device made of a Majorana wire. The initial density $n_{l',\sigma}=c^{\dg}_{l',\sigma}c_{l',\sigma}$ with $\sigma=\uparrow,\downarrow$ of the middle SM wire for the direct time-evolution numerics is indicated on the heading of the top panel. In both panels, $L_{\rm X}=L_{\rm Z}=900, L_{\rm Y}=3$, $\gamma_{\rm X}=\gamma_{\rm Z}=\gamma=1$, $\gamma_{\rm Y}=0.5$, $\Delta_{\rm X}=0.3,\Delta_{\rm Y}=\Delta_{\rm Z}=0, \zeta_{\rm X}=\zeta_{\rm Y}=\zeta_{\rm Z}=\zeta_{\rm XY}=\zeta_{\rm YZ}=0.2, B_{\rm X}=B_{\rm Y}=B_{\rm Z}=0.4$, $\epsilon_{\rm X}=\epsilon_{\rm Z}=0, \epsilon_{\rm Y}=0.05$, $\gamma_{\rm XY}=\gamma_{\rm YZ}=0.25$, $T_{\rm X}=T_{\rm Z}=0.02$ and $\mu_{\rm X}=0,\mu_{\rm Z}=0.5$. All above parameters except lengths are in units of $\gamma$.}
\label{spinCurr}
\end{figure}
 We now confirm the validity of steady-state spin current expression derived in the earlier section. For the Majorana wire, the total spin polarization along $x$-axis is not conserved in the presence of spin-orbit coupling (see Sec.~\ref{Majorana}). Therefore, the $x$-component of spin current does not remain the same at the left and right junctions of a semiconductor middle wire, i.e., $\mathcal{J}^{s_x}_{\rm XY} \ne \mathcal{J}^{s_x}_{\rm YZ}$. In Fig.~\ref{spinCurr}, we plot the $\mathcal{J}^{s_x}_{\rm XY}(t)$ and $\mathcal{J}^{s_x}_{\rm YZ}(t)$ calculated using the first-principle/direct time-evolution numerics and the generalized LEGF method in a TS-SM-SM device. We find good agreement between the steady-state values of $\mathcal{J}^{s_x}_{\rm XY}=-0.021369$ and $\mathcal{J}^{s_x}_{\rm YZ}= -0.023747$ (in units of $\gamma \hbar/2$) from the generalized LEGF method, and the long-time values of $\mathcal{J}^{s_x}_{\rm XY}(\gamma t=800)=-0.021369$ and $\mathcal{J}^{s_x}_{\rm YZ}(\gamma t=800)=-0.023746$ from the direct time-evolution numerics. We further investigate the steady-state spin current in an SM-TS-SM device, which was experimentally explored recently in \cite{YangNanoLetter2020}.  We get from the LEGF, $\mathcal{J}^{s_x}_{\rm XY}=0.004686$ and $\mathcal{J}^{s_x}_{\rm YZ}=-0.01072$ (in units of $\gamma \hbar/2$) for an SM-TS-SM device with $L_{\rm Y}=10$, $\gamma_{\rm Y}=1$, $\Delta_{\rm X}=0, \Delta_{\rm Y}=0.3$, and $\mu_{\rm X}=0.2,\mu_{\rm Z}=0.5$ (other parameters are the same as Fig.~\ref{spinCurr}). The long-time values from the direct time-evolution numerics are  $\mathcal{J}^{s_x}_{\rm XY}(\gamma t=1475)=0.004592$ and $\mathcal{J}^{s_x}_{\rm YZ}(\gamma t=1475)=-0.01047$, which show some deviations between the two methods for longer $L_{\rm Y}$. 
 
\begin{figure}
\includegraphics[width=0.99\linewidth]{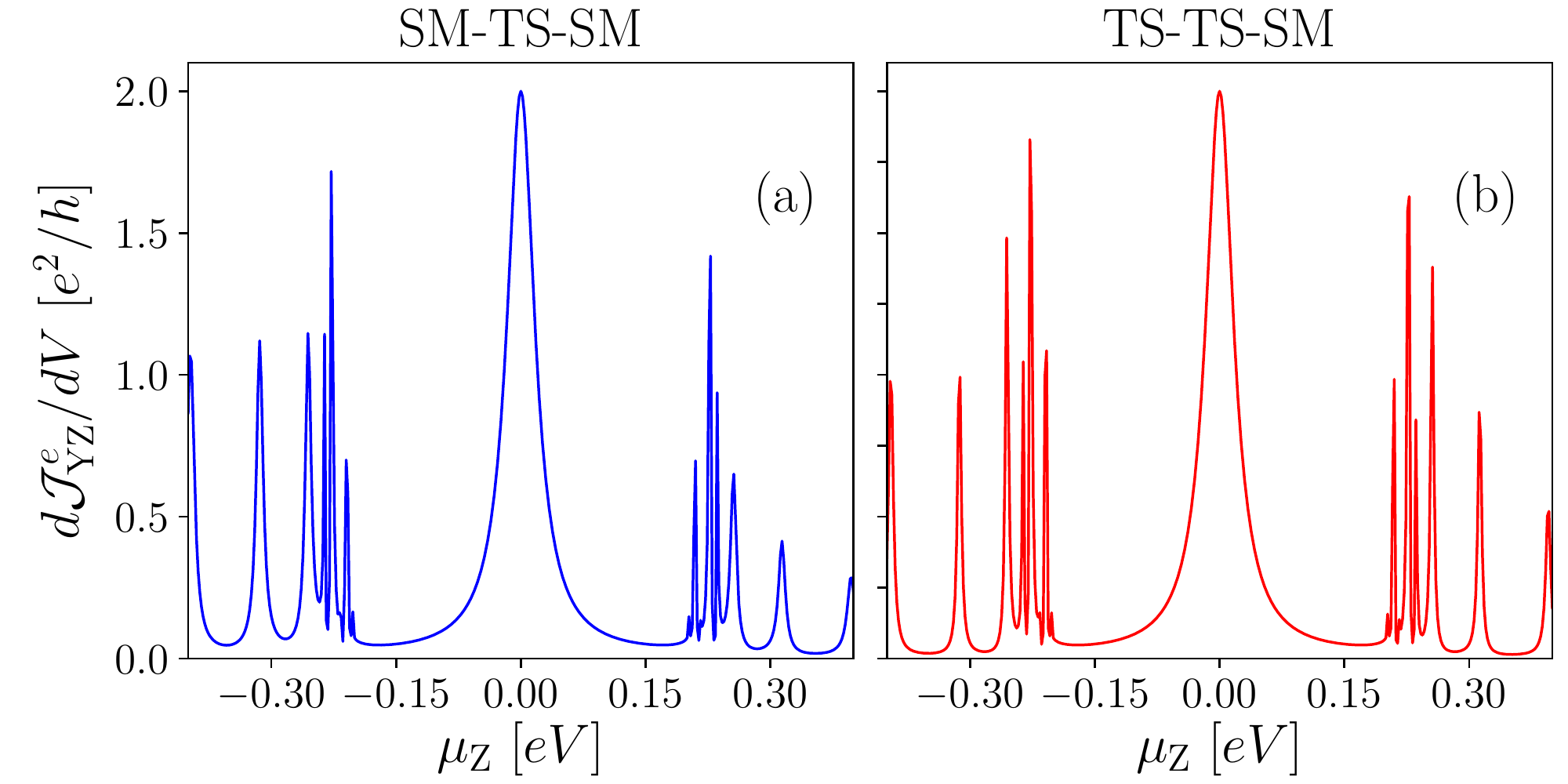}
\centering
\caption{Zero-temperature differential electrical conductance $d\mathcal{J}^e_{\rm YZ}/dV$ at the $Y$-$Z$ junction of an SM-TS-SM device (a), and a TS-TS-SM device (b), where the TS wires are the Majorana wire. In both panels, $L_{\rm Y}=40$, $\gamma_{\rm X}=\gamma_{\rm Z}=\gamma_{\rm Y}=1.0$, $\Delta_{\rm Y}=0.3, \Delta_{\rm Z}=0, \zeta_{\rm X}=\zeta_{\rm Y}=\zeta_{\rm Z}=\zeta_{\rm XY}=\zeta_{\rm YZ}=0.2, B_{\rm X}=B_{\rm Y}=B_{\rm Z}=0.4$, $\epsilon_{\rm Y}=0,\epsilon_{\rm Z}=1$, $\gamma_{\rm XY}=\gamma_{\rm YZ}=0.25$, and $\mu_{\rm X}=0$. Also, $\Delta_{\rm X}=0, \epsilon_{\rm X}=1$ in (a), and $\Delta_{\rm X}=0.3, \epsilon_{\rm X}=0$ in (b). All above parameters except lengths are in units of $\gamma$.}
\label{DECMajorana1}
\end{figure}

\begin{figure}
\includegraphics[width=0.99\linewidth]{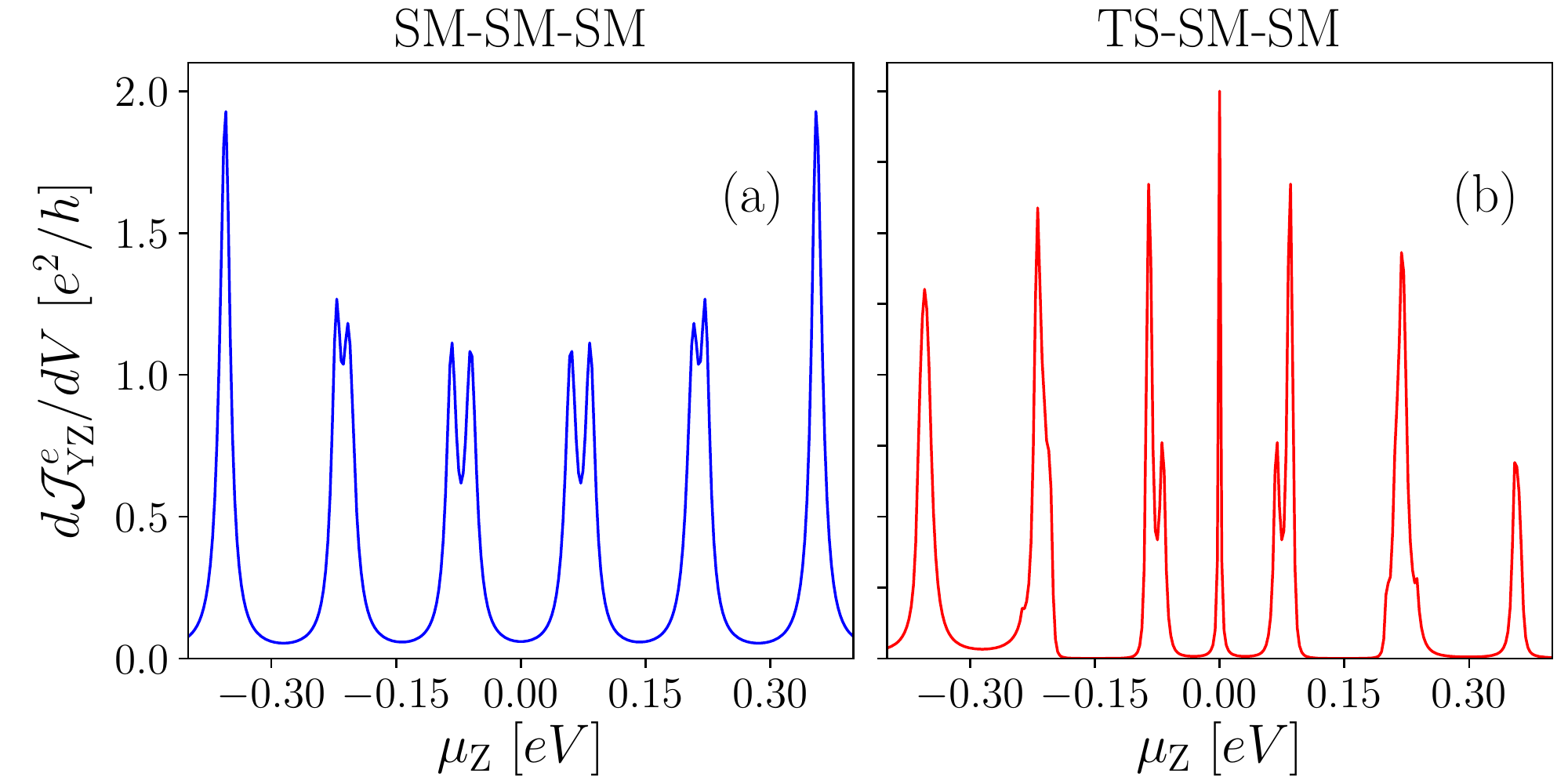}
\centering
\caption{Zero-temperature differential electrical conductance $d\mathcal{J}^e_{\rm YZ}/dV$ at the $Y$-$Z$ junction of an SM-SM-SM device (a), and a TS-SM-SM device (b), where the TS wires are the Majorana wire. In both panels, $L_{\rm Y}=40$, $\gamma_{\rm X}=\gamma_{\rm Z}=\gamma_{\rm Y}=1.0$, $\Delta_{\rm Y}=\Delta_{\rm Z}=0, \zeta_{\rm X}=\zeta_{\rm Y}=\zeta_{\rm Z}=\zeta_{\rm XY}=\zeta_{\rm YZ}=0.2, B_{\rm X}=B_{\rm Y}=B_{\rm Z}=0.4$, $\epsilon_{\rm Y}=\epsilon_{\rm Z}=1$, $\gamma_{\rm XY}=\gamma_{\rm YZ}=0.25$, and $\mu_{\rm X}=0$. Also, $\Delta_{\rm X}=0, \epsilon_{\rm X}=1$ in (a), and $\Delta_{\rm X}=0.3, \epsilon_{\rm X}=0$ in (b). All above parameters except lengths are in units of $\gamma$.}
\label{DECMajorana2}
\end{figure}

Next, we clarify the role of TS leads made of Majorana wires. For this, we again compare the zero-temperature DEC in a TS-SM-SM device to an SM-SM-SM device and a TS-TS-SM device to an SM-TS-SM device. In Fig.~\ref{DECMajorana1}, we show the zero-temperature DEC in TS-TS-SM and SM-TS-SM devices where the TS wires are kept in the topological phase. In contrast to the Kitaev chains, the difference in the finite bias DEC height above the bulk-gap between a TS-TS-SM device and an SM-TS-SM device is relatively less. It is probably due to the more structured density of states of the spin-orbit coupled wires. Nevertheless, there is a clear zero-bias DEC peak of height $2e^2/h$ in a TS-SM-SM device, which is absent in an SM-SM-SM device. We show it in Fig.~\ref{DECMajorana2}. Therefore, the TS leads in the topological phase do have a signature in transport even for the spinful model of TSs.

\begin{figure}[h]
\includegraphics[width=0.99\linewidth]{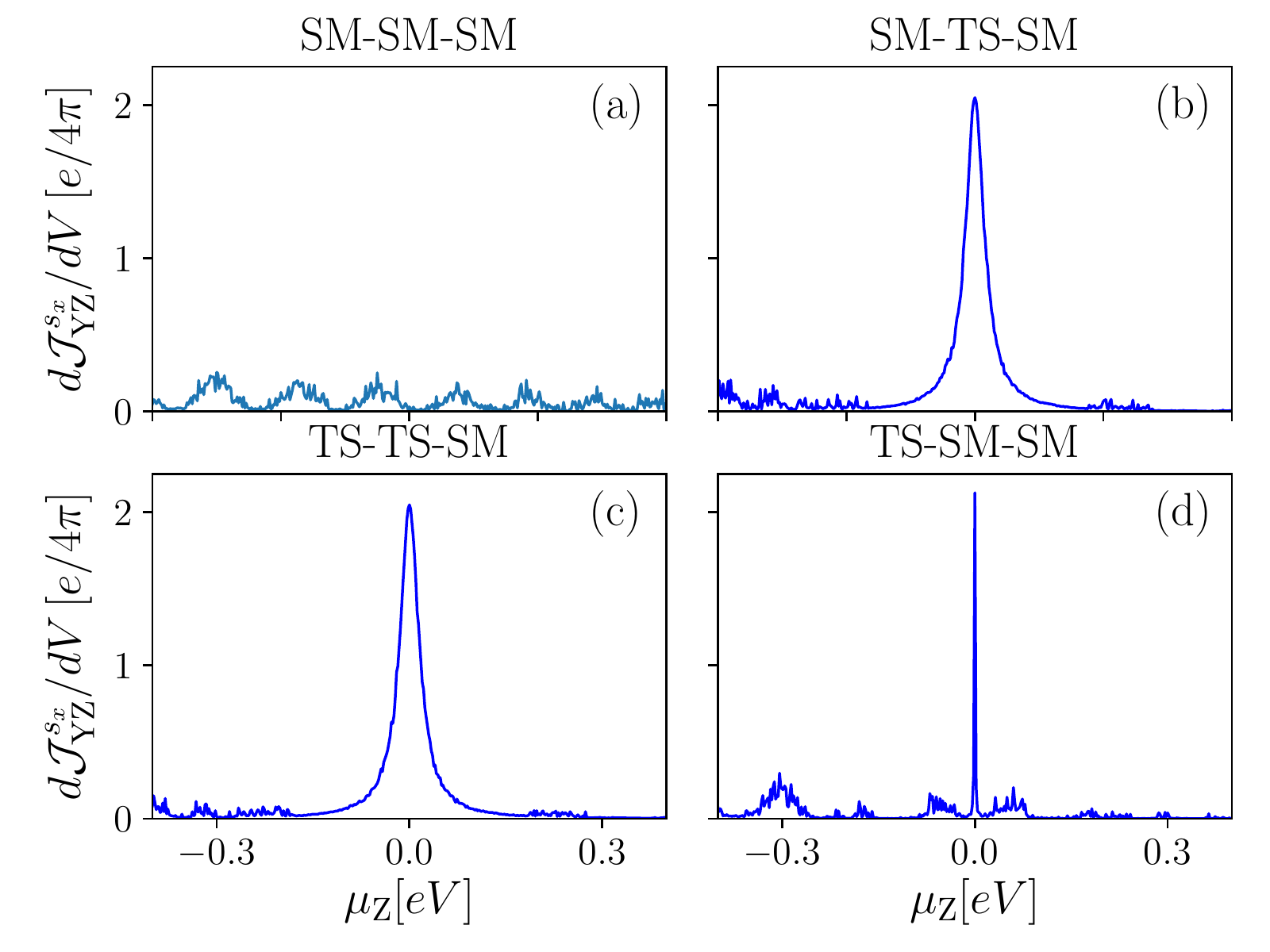}
\centering
\caption{Zero-temperature differential spin conductance $d\mathcal{J}^{s_x}_{\rm YZ}/dV$ at the $Y$-$Z$ junction of an SM-SM-SM device (a), an SM-TS-SM device (b), a TS-TS-SM device (c), and a TS-SM-SM device (d), where the TS wires are the Majorana wire. The middle $Y$ wire in all the devices has an onsite disorder with a uniform distribution of zero mean and variance $0.4$, and we show the DSC signals after disorder averaging over 20 realizations. In all panels, $L_{\rm Y}=40$, $\gamma_{\rm X}=\gamma_{\rm Z}=\gamma_{\rm Y}=1.0$, $\Delta_{\rm Z}=0, \zeta_{\rm X}=\zeta_{\rm Y}=\zeta_{\rm Z}=\zeta_{\rm XY}=\zeta_{\rm YZ}=0.2, B_{\rm X}=B_{\rm Y}=B_{\rm Z}=0.4$, $\epsilon_{\rm X}=\epsilon_{\rm Y}=\epsilon_{\rm Z}=0$, $\gamma_{\rm XY}=\gamma_{\rm YZ}=0.25$, and $\mu_{\rm X}=0$. Also, $\Delta_{\rm X}=\Delta_{\rm Y}=0$ in (a), $\Delta_{\rm X}=0,\Delta_{\rm Y}=0.3$ in (b), $\Delta_{\rm X}=\Delta_{\rm Y}=0.3, $ in (c), $\Delta_{\rm X}=0.3,\Delta_{\rm Y}=0$ in (d). All above parameters except lengths are in units of $\gamma$.}
\label{DSC}
\end{figure}

In Sec.~\ref{TransMajorana}, we introduce a definition of DSC for spinful TS wire junctions. We here discuss the features of zero-temperature DSC in SM-TS-SM, TS-TS-SM, and TS-SM-SM devices. We show in Fig.~\ref{DSC}(b,c) that the value of zero-bias DSC peak at zero temperature is also quantized in the above devices, while $Y$ wire is in the topological phase. The quantized value of $d\mathcal{J}^{s_x}_{\rm YZ}/dV$ at zero bias is $2$ in the unit of $(e/h)(\hbar/2) = (e/4 \pi)$. The quantized value of zero-bias DSC reminds us that of DEC in such devices. We note that the other components of DSC (e.g., $y$ and $z$ components) are not quantized in our devices. We observe the quantization of DSC is mainly along the direction of polarization of the electrons in the boundary wires (e.g., $Z$ wire for the DSC at $Y$-$Z$ junction) \cite{HePRL2014}. The quantization of zero-bias DSC at zero temperature is a consequence of Majorana zero modes that developed at the edges of the middle TS wire. It is a local phenomenon as it depends mainly on the localized Majorana zero mode and its tunnel coupling to the boundary wire. We have verified that the quantization of the zero-temperature zero-bias peak height of DSC is robust against disorder in onsite energies of the middle TS wire (check Fig.~\ref{DSC}) and the strength of the tunneling rates. It should be noted that neither the electrical charge nor the total spin components is conserved inside a topological superconductor like Majorana wire (TS with spin), however zero-bias peaks in both DSC and DEC signify the presence of Majorana zero modes at the edges of the Majorana wire. 


 For comparison, we further depict the zero-temperature DSC in an SM-SM-SM device in Fig.~\ref{DSC}(a), which does not show a quantized zero-bias peak of height $2$. Therefore, the quantized DSC along with the DEC can be employed to probe the emergence of MBSs in engineered semiconductor-superconductor heterostructures. In Fig.~\ref{DSC}(d), we also notice a sharp zero-bias peak in zero-temperature DSC from a TS-SM-SM device when the boundary $X$ wire is in the topological phase, and the height of the zero-bias peak is almost quantized to $2$. Nevertheless, there are finite DSC peaks within the superconducting gap of the boundary $X$ wire in Fig.~\ref{DSC}(d), and these peaks are due to the middle semiconductor wire as also seen for zero-temperature DEC in  Fig.~\ref{DECMajorana2}(b).


\section{Conclusion}
\label{conc}
This paper develops a unified open-quantum system description of nonequilibrium electrical, thermal, and spin transport in various devices whose systems and leads/baths are made of different TS wires. We demonstrate that the quantum LEGF method suits perfectly to derive neat expressions of different steady-state currents.  We here notably extend the applications of the LEGF method to the baths of TS wires. We mainly try to reveal several significant thermal and spin transport features in different junctions of TS, SM, and N wires, which are either less explored or have not been investigated. 
While the spin transport in the junctions of TS and SM wires are rarely investigated in the past, a systematic study of spin transport can be fruitful in disclosing necessary signatures of Majorana quasiparticles (such as spin polarizations of Majoranas \cite{Aligia2020}) and the topological nature of the systems \cite{YangNanoLetter2020,Tanaka2009,He2014}. The obtained expression of spin current here, is expected to be useful for further investigation of spin transport. We hope the quantized zero-bias peak in zero-temperature DSC would be tested in engineered spinful TS wire junctions \cite{YangNanoLetter2020} for experimentally detecting topological phases of Majorana wires.  Another interesting quantity that may be studied in this framework is the local compressibility which shows divergence at topological phase transition and can be an interesting probe to detect the same \cite{Nozad2016}. This compressibility while combined with thermal transport may reveal new Majorana quantization \cite{SmirnovPRB2020}.

The open-quantum system description of transport is more appropriate in determining the quantum materials' topological signatures as it incorporates the bath-induced dissipation in the topological materials (and generates an effective non-Hermitian notion) akin to the engineered TS devices in experiments \cite{MourikScience2012, DasNature2012, NadjPergeScience2014, DengScience2016, FornieriNature2019, YangNanoLetter2020}. Such a description is also required to explain the discrepancy between the experimentally measured and theoretically proposed (using transport in isolated systems) height of the zero-bias DEC manifesting emergence of the MBSs and the characteristics of the quantized peak in thermal conductance indicating the topological phase transition. For example, the position of the quantized peak in thermal conductance as a function of Fermi energy of the middle TS wires shifts with the system-bath coupling.

The expressions of electrical, thermal, and spin current, can be further explored to investigate thermoelectric, magnetoelectric, and thermomagnetic transport properties, especially in the linear response regime. Such studies may show further quantization of different unexplored transport quantities.

The LEGF method suits well to extend the above calculation to find current fluctuation across its average value due to thermal and quantum noises. For this, we need to obtain two-point correlators of current operators at the different instants of time, which can be analyzed to investigate correlations of the nonequilibrium current (shot noise) at the same or different junctions \cite{BolechPRL2007, SmirnovPRB2018, SmirnovPRB2019A} due to the quantization of the  charge and energy of the carriers.




\section*{Acknowledgements}
NB acknowledges funding from DST-FIST programme. DR acknowledges funding from the Department of Science and Technology, India via the Ramanujan Fellowship, and the Ministry of Electronics $\&$ Information Technology (MeitY), India under grant for ``Centre for Excellence in Quantum Technologies'' with Ref. No. 4(7)/2020-ITEA. 
\begin{appendices}
\setcounter{figure}{0}
\renewcommand\thefigure{A\arabic{figure}}
\section{Quantum Langevin equations for $Y$ wire connected to the Kitaev chain leads}
\label{App1}
The Eqs.~\ref{eombathl} and \ref{eombathr} in the main text can be solved using single-particle retarded Green's function of the isolated $\alpha$ lead/bath wire ($\alpha = X, Z$): 
\beq
G^+_\alpha(\tau)=-i e^{-i\mathcal{H}^{\alpha } \tau} \theta (\tau)
\label{rgf}
\eeq
where both $G^+_\alpha$ and $\mathcal{H}^{\alpha }$ are matrices, and $\theta (\tau)$ is the Heaviside step function. This matrix Green's function is related to the operator valued Green's function as $\mathcal{G}^+_\alpha(\tau)=\sum_{l,m}[{G}^+_\alpha(\tau)]_{l,m}a^{\dg}_la_m$, where $[{G}^+_\alpha (\tau)]_{l,m} \equiv [-i e^{-i\mathcal{H}^{\rm \alpha } \tau} ]_{lm} \theta (\tau)$. Here, $l,m= 1,\dots, 2L_{\rm X}$ for $\alpha=X$, and $l,m=2L_{\rm XY}+1, \dots, 2L$ for $\alpha=Z$. These bath Green's functions are the solution of the following equation:
 \beq
  \left( i \frac{\partial}{\partial t} -\mathcal{H}^{\alpha} \right) G_{\alpha}^+(t -t')= \delta(t-t')\one_{2L_{\alpha}}.
 \label{siGF}
 \eeq
Using these Green's functions (\ref{rgf}), we formally solve the Heisenberg equations of the boundary wires in Eqs.~\ref{eombathl}, \ref{eombathr}, and the solutions are for $t>t_0$:  
\bea
 a_{k}(t) &=& i \sum_{l=1}^{2L_{\rm X}} {[G_{\rm X}^+(t -t_0)]}_{k,l} \, a_{l}(t_0) -\gamma_{\rm XY}  \int_{t_0}^{t} dt'  \sum_{l=2L_{\rm X}-1}^{2L_{\rm X}} (-1)^{l+1} {[G_{\rm X}^+(t -t')]}_{k,l} a_{l+2}(t') \, , \nn \\
 a_{k'}(t) &=& i \sum_{l=2L_{\rm XY}+1}^{2L} {[G_{\rm Z}^+(t -t_0)]}_{k'-2L_{\rm XY},l-2L_{\rm XY}} \, a_{l}(t_0) \nn \\
&& ~~~~~~~~~~~~~~~ -\gamma_{\rm YZ} \int_{t_0}^{t} dt' \sum_{l=1}^2 (-1)^{l+1} {[G_{\rm Z}^+(t -t')]}_{k'-2L_{\rm XY},l} \,  a_{2 L_{\rm XY}-2+l}(t') \, ,\nn \\
 \label{solres} 
 \eea
where $ k=2L_{\rm X}-1,2L_{\rm X}$ and $k'=2L_{\rm XY}+1 ,2L_{\rm XY}+2$. $G_{\rm X}^+(t -t_0)$ and $G_{\rm Z}^+(t -t_0)$ are the single-particle retarded Green's functions of the $X$ and $Z$ baths, respectively. Now, plugging the formal solutions of the $X$ wire variables (\ref{solres}) into Eq.~\ref{eomwire}, we get the quantum Langevin equation of $Y$ wire variables at the $X$-$Y$ junction:
 \bea
 \dot{a}_{k}(t)= -i \sum_{l=2L_{\rm X}+1}^{2L_{\rm XY}} \mathcal{H}^{\rm Y}_{k,l} \, a_l(t)  -i \eta_{k-2L_{\rm X}}^{\rm X}(t)  - i \int_{t_0}^t dt'  \sum_{l=2L_{\rm X}+1}^{2L_{\rm X}+2}{[\Sigma_{\rm X}^+(t -t')]}_{k,l}\, a_{l}(t')\, ,\nn \\
\label{eom1}
 \eea
 for $k=2L_{\rm X}+1,2L_{\rm X}+2$, where 
\bea
\eta_1^{\rm X}(t)&=& -i\gamma_{\rm XY} \sum_{l=1}^{2L_{\rm X}}{[G_{\rm X}^+(t -t_0)]}_{2L_{\rm X}-1,l} \, a_{l}(t_0), \nn \\
 \eta_2^{\rm X}(t)& =& i\gamma_{\rm XY} \sum_{l=1}^{2L_{\rm X}}{[G_{\rm X}^+(t -t_0)]}_{2L_{\rm X},l} \, a_{l}(t_0),  
\label{Xnoise}
\eea
are the noise terms from the $X$ bath, and the dissipative terms generated in the $Y$ wire due to the coupling to the $X$ bath are:
\bea
 [\Sigma_{\rm X}^+(t)]_{l,m}&=&\gamma_{\rm XY}^2\sum_{k=2L_{\rm X}-1}^{2L_{\rm X}}[G_{\rm X}^+(t)]_{k,k} \, \delta_{l,k+2}\, \delta_{m,l} \nn \\
 && ~~~~~~~~~~~~~~~~~~~~~~~~~-\gamma_{\rm XY}^2  \sum_{k,k'=2L_{\rm X}-1 \atop k \neq k'}^{2L_{\rm X}}
[G_{\rm X}^+(t)]_{k,k'} \delta_{l,k+2}\delta_{m,k'+2} .\nn\\
\eea
Similarly, the quantum Langevin equations for the $Y$ wire variables at $Y$-$Z$ junction are:
\bea
 \dot{a}_{k}(t)&=&-i \sum_{j=2L_{\rm X}+1}^{2L_{\rm XY}} \mathcal{H}^{\rm Y}_{k,j} \, a_j(t) -i \eta_{k-2L_{\rm X}+2}^{\rm Z}(t) - i  \int_{t_0}^t dt' \sum_{l=2L_{\rm XY}-1}^{2L_{\rm XY}} {[\Sigma_{\rm Z}^+(t -t')]}_{k,l}\, a_{l}(t') \, ,\nn \\
 \label{eom2}
 \eea
for $k=2L_{\rm XY}-1,2L_{\rm XY}$, where
\bea
\eta_{1}^{\rm Z}(t)&=&-i\gamma_{\rm YZ} \sum_{l=2L_{\rm XY}+1}^{2L} {[G_{\rm Z}^+(t -t_0)]}_{1,l-2L_{\rm XY}} \, a_{l}(t_0), \nn \\
\eta_{2}^{\rm Z}(t)&=&i\gamma_{\rm YZ} \sum_{l=2L_{\rm XY}+1}^{2L} {[G_{\rm Z}^+(t -t_0)]}_{2,l-2L_{\rm XY}} \, a_{l}(t_0), \nn \\
\label{Znoise}
\eea
are the noises from the $Z$ bath, and 
\bea
  [\Sigma_{\rm Z}^+(t)]_{l,m}&=& \gamma_{\rm YZ}^2 \sum_{k=1}^{2}[G_{\rm Z}^+(t)]_{k,k} \, \delta_{l,2L_{\rm XY}-2+k}\delta_{m,l}  \nn \\
  &&~~~~~~~~~~~~~~~~~~ -\gamma_{\rm YZ}^2 \sum_{k,k'=1 \atop k \neq k'}^{2}[G_{\rm Z}^+(t)]_{k,k'} \, \delta_{l,2L_{\rm XY}-2+k}\delta_{m,2L_{\rm XY}-2+k'}, \nn \\
\eea
are the dissipative terms due to the $Z$ bath. For the internal sites of the $Y$ wire, the Heisenberg equations take the following form:
\bea
\dot{a}_{l}(t)=-i \sum_{j=2L_{\rm X}+1}^{2L_{\rm XY}} \mathcal{H}^{\rm Y}_{l,j} \, a_j(t), 
\label{eom3}
\eea
for $l=2L_{\rm X}+3,\dots,2L_{\rm XY}-2$. We finally rewrite Eqs.~\ref{eom1}, \ref{eom2}, and \ref{eom3} in a compact form:
\bea
 \dot{a}_{l}(t)&=&-i \sum_{m=2L_{\rm X}+1}^{2L_{\rm XY}} \mathcal{H}^{\rm Y}_{l,m} a_m(t) -i \sum_{k=1}^2 \left(  \delta_{l,2L_X+k}\, \eta_{k}^{\rm X}(t)+\delta_{l,2L_{\rm XY}-2+k}\, \eta_{k}^{\rm Z}(t) \right) \nn \\
&& -i(\delta_{l,2L_X+1}+\delta_{l,2L_X+2})\int_{t_0}^t dt' \sum_{m=2L_X+1}^{2L_X+2} {[\Sigma_{\rm X}^+(t -t')]}_{l,m}\, a_{m}(t')  \nn \\
&& -i(\delta_{l,2L_{\rm XY}-1}+\delta_{l,2L_{\rm XY}})\int_{t_0}^t dt' \sum_{m=2L_{\rm XY}-1}^{2L_{\rm XY}} {[\Sigma_{\rm Z}^+(t -t')]}_{l,m}\, a_{m}(t') , \nn \\
\label{eom4}
\eea
where, $l \in [2L_{\rm X}+1,2L_{\rm XY}]$. The Eq.~\ref{eom4} is in a form of generalized quantum Langevin equations \cite{DharPRB2006}. The Eq.~\ref{eom4} can be solved analytically by using the Fourier transformation for a particular class of systems, which have a unique NESS. Nevertheless, to apply the Fourier transformation on Eq.~\ref{eom4}, we first need to define the Fourier transformation of the noise, and the dissipative/self-energy terms: $\tilde{\eta}_{1,2}^{\alpha}(\omega)=\frac{1}{2\pi}\int_{-\infty}^{\infty} dt \,\eta_{1,2}^{\alpha}(t)\, e^{i \omega t} $ and $\tilde{\Sigma}_{\alpha}^+(\omega)=\int_{-\infty}^{\infty} dt \, \Sigma_{\alpha}^+(t)\, e^{i \omega t} $ for $\alpha=X,Z$. We note that the steady-state solutions for the $Y$ wire variables in frequency domain are given in Eq.~\ref{a1} of the main text.
\setcounter{equation}{0}
\setcounter{figure}{0}
\renewcommand\thefigure{B\arabic{figure}}
\section{Green's function of an isolated Kitaev chain and its properties}
\label{App2}
Let us consider an isolated $\alpha$ wire lead $(\alpha =X,Z)$ of the Kitaev chain represented by the Hamiltonian $H^{\alpha}$. The $r$-th eigenvalue of $H^{\alpha}$ and the corresponding eigenfunction are respectively denoted by $\hbar \omega_r^\alpha$ and $U_r^\alpha= [ \phi_r^\alpha(1), \psi_r^\alpha(1), \dots, \phi_r^\alpha(L ), \psi_r^\alpha(L )]^T$. As we have already discussed, the Hamiltonian matrix $\mathcal{H}^\alpha$ for the isolated $\alpha$ wire can be expressed as,
\beq
\mathcal{H}^\alpha= \sum_{r>0} \omega_r^\alpha (U^\alpha_r U_r^{\alpha \,\dagger}- V^\alpha_r V_r^{\alpha\, \dagger}).
\eeq
In Appendix \ref{App1}, we have introduced the single-particle retarded Green's function  of $\alpha$ wire using Eq.~\ref{siGF}. Next, we rewrite the bath Green's function by using the form of matrix $\mathcal{H}^\alpha$ as
\bea
 G^+_{\alpha }(t-t')=-i \theta(t-t') \sum_{r>0}\left(U^\alpha_r U_r^{\alpha \,\dagger} e^{-i \omega_r^\alpha (t-t')} 
+V^\alpha_r V_r^{\alpha\, \dagger} e^{i \omega_r^\alpha (t-t')} \right).
\label{GF}
\eea
It is easy to check the following properties of the Green's function:
\bea
&&{[G_\alpha^+(\tau)]}_{2i',2j'}=-{[G_\alpha^+(\tau)]}_{2i'-1,2j'-1}^*\, , \nn \\
&&{[G_\alpha^+(\tau)]}_{2i',2j'-1}=-{[G_\alpha^+(\tau)]}_{2i'-1,2j'}^* \, ,
\label{GFprop}
\eea 
for $ i',j'= 1,\dots, L_\alpha $. The Fourier transformation of Green's function is defined as: ${\tilde{G}^{\alpha +}_{l,m}(\omega)}=\frac{1}{2 \pi}\int_{-\infty}^{\infty} d\tau\, { [G_\alpha^+(\tau)]}_{l,m}\, e^{i \omega \tau}$. Hence, the retarded Green's function in the frequency domain satisfies the following relation:
\bea
&&{\tilde{G}^{\alpha +}_{2i',2j'}(\omega)}=-{[\tilde{G}^{\alpha +}_{2i'-1,2j'-1}(-\omega)]}^*, \nn \\
&&{\tilde{G}^{\alpha +}_{2i',2j'-1}(\omega)}=-{[\tilde{G}^{\alpha +}_{2i'-1,2j'}(-\omega)]}^*\, .
\label{GFprop1}
\eea
In the frequency domain, components of the bath Green's functions are expressed as
\bea
 {\tilde{G}^{\alpha +}_{2i'-1,2j'-1}(\omega)}&=& \sum_{r>0} \left[ \frac{\phi_r^\alpha(i')\phi^{\alpha *}_r(j')}{\omega -\omega_r^\alpha} +\frac{\psi^{\alpha *}_r(i')\psi_r^\alpha(j')}{\omega +\omega_r^\alpha} \right] \nn \\
&&~~~- i \pi  \sum_{r>0} \left[ \phi_r^\alpha(i')\phi^{\alpha *}_r(j')\delta( \omega -\omega_r^\alpha) + \psi^{\alpha *}_r(i')\psi_r^\alpha(j')
\delta( \omega +\omega_r^\alpha)\right], \nn \\
{\tilde{G}^{\alpha +}_{2i',2j'}(\omega)}&=& \sum_{r>0} \left[ \frac{\psi_r^\alpha(i')\psi^{\alpha *}_r(j')}{\omega -\omega_r^\alpha} 
+\frac{\phi^{\alpha *}_r(i')\phi_r^\alpha(j')}{\omega +\omega_r^\alpha} \right] \nn \\
&&~~~- i \pi  \sum_{r>0} \left[ \psi_r^\alpha(i')\psi^{\alpha *}_r(j')\delta( \omega -\omega_r^\alpha) + \phi^{\alpha *}_r(i')\phi_r^\alpha(j')
\delta( \omega +\omega_r^\alpha)\right], \nn \\
{\tilde{G}^{\alpha +}_{2i',2j'-1}(\omega)}&=& \sum_{r>0} \left[ \frac{\psi_r^\alpha(i')\phi^{\alpha *}_r(j')}{\omega -\omega_r^\alpha} +\frac{\phi^{\alpha *}_r(i')\psi_r^\alpha(j')}{\omega +\omega_r^\alpha} \right] \nn \\
&&~~~ - i \pi  \sum_{r>0} \left[ \psi_r^\alpha(i')\phi^{\alpha *}_r(j')\delta( \omega -\omega_r^\alpha) + \phi^{\alpha *}_r(i')\psi_r^\alpha(j')\delta( \omega +\omega_r^\alpha)\right], \nn \\
 {\tilde{G}^{\alpha +}_{2i'-1,2j'}(\omega)}&=& \sum_{r>0} \left[ \frac{\phi_r^\alpha(i')\psi^{\alpha *}_r(j')}{\omega -\omega_r^\alpha} +\frac{\psi^{\alpha *}_r(i')\phi_r^\alpha(j')}{\omega +\omega_r^\alpha} \right] \nn \\
&&~~~ - i \pi  \sum_{r>0} \left[ \phi_r^\alpha(i')\psi^{\alpha *}_r(j')\delta( \omega -\omega_r^\alpha) + \psi^{\alpha *}_r(i')\phi_r^\alpha(j')\delta( \omega +\omega_r^\alpha)\right].\nn \\
\label{GF2}
\eea
We observe that $\{ \phi_r^\alpha(i') \}$ and $ \{ \psi_r^\alpha(j') \} $ are real for a real superconducting gap $\Delta_\alpha$. Thus, we can write
\bea
  \text{Im}[ \tilde{G}^{\alpha +}_{2i'-1,2j'-1}(\omega)] &=& -\pi  \sum_{r>0} \left[ \phi_r^\alpha(i')\phi^{\alpha *}_r(j')\delta( \omega -\omega_r^\alpha) + \psi^{\alpha *}_r(i')\psi_r^\alpha(j') \delta( \omega +\omega_r^\alpha)\right],\nn \\
  \text{Im}[ \tilde{G}^{\alpha +}_{2i',2j'}(\omega)]&=& -\pi  \sum_{r>0} \left[ \psi_r^\alpha(i')\psi^{\alpha *}_r(j') \delta( \omega -\omega_r^\alpha) + \phi^{\alpha *}_r(i')\phi_r^\alpha(j')
\delta( \omega +\omega_r^\alpha)\right], \nn \\
  \text{Im}[\tilde{G}^{\alpha +}_{2i',2j'-1}(\omega) ]&=& -\pi  \sum_{r>0} \left[ \psi_r^\alpha(i')\phi^{\alpha *}_r(j')
\delta( \omega -\omega_r^\alpha) + \phi^{\alpha *}_r(i')\psi_r^\alpha(j')
\delta( \omega +\omega_r^\alpha)\right], \nn \\
 \text{Im}[\tilde{G}^{\alpha +}_{2i'-1,2j'}(\omega)] &=& - \pi  \sum_{r>0} \left[ \phi_r^\alpha(i')\psi^{\alpha *}_r(j')
\delta( \omega -\omega_r^\alpha) + \psi^{\alpha *}_r(i')\phi_r^\alpha(j')
\delta( \omega +\omega_r^\alpha)\right]. \nn \\
\label{GF2a}
\eea
The above relations (\ref{GF2a}) are of significant importance, and we subsequently use them in deriving the noise-noise correlations of superconducting leads with a real superconducting gap.
\setcounter{equation}{0}
\setcounter{figure}{0}
\renewcommand\thefigure{C\arabic{figure}}
\section{Noise-noise correlations for Kitaev chain leads}
\label{App3}
Employing the Eqs.~\ref{Klatticeop} and \ref{exp}, we calculate the equilibrium correlation functions for a Kitaev chain lead. For such an isolated $\alpha$ wire bath ($\alpha = X, Z$), the equilibrium correlations are given by, 
\bea
 \langle a_{2i'-1}^\dg \, a_{2j'-1} \rangle &=& \langle c_{i'}^{\dagger}\, c_{j'} \rangle = \sum_{r>0}\left[\phi^{\alpha *}_r(i') \phi^\alpha_r(j') f(\omega_r^\alpha, T_\alpha)+ \psi_r^\alpha(i') \psi_r^{\alpha *}(j')f(-\omega_r^\alpha, T_\alpha)\right], \nn \\
 \langle a_{2i'-1} \, a_{2j'-1}^\dg \rangle &=& \langle c_{i'}\, c_{j'}^\dagger  \rangle = \sum_{r>0}\left[ \psi_r^{\alpha *}(i') \psi_r^\alpha(j') f(\omega_r^\alpha, T_\alpha)+\phi_r^\alpha(i') \phi_r^{\alpha *}(j') f(-\omega_r^\alpha,T_\alpha)\right], \nn \\
 \langle a_{2i'-1} \, a_{2j'-1} \rangle &=& \langle c_{i'} \, c_{j'} \rangle = \sum_{r>0}\left[\psi_r^{\alpha *}(i') \phi_r^\alpha(j')f(\omega_r^\alpha,T_\alpha)+\phi_r^\alpha(i') \psi_r^{\alpha *}(j')f(-\omega_r^\alpha,T_\alpha)\right], \nn \\
 \langle a_{2i'-1}^\dg \, a_{2j'-1}^\dg \rangle &=& \langle c_{i'}^{ \dagger} \, c_{j'}^{ \dagger}  \rangle = \sum_{r>0}\left[\phi_r^{\alpha *}(i') \psi_r^\alpha(j')f(\omega_r^\alpha, T_\alpha)+\psi_r^\alpha(i') \phi_r^{\alpha *}(j')f(-\omega_r^\alpha, T_\alpha)\right],  \nn \\
\label{co1}
\eea
where $i',j'=1,\dots,L_{\alpha}$. Here, $T_\alpha$ is the temperature of the $\alpha$ wire bath, and $\hbar\omega_r^\alpha$ is the $r$-th eigenvalue, and $\{ \phi_r^{\alpha },\psi_r^{\alpha }\}$ are the components of corresponding eigenfunction. Moreover, these relations are valid only when the baths are kept at zero chemical potential. The other possible two-point correlations between $a_l$ variables can easily be derived employing the relation: $a_{2i'}=a_{2i'-1}^\dg$.  Using the definitions of noises (\ref{Xnoise}), we now find the noise-noise correlation in time domain:
\bea
 \langle \eta_1^{{\rm X} \dagger}(t) \eta_1^{\rm X}(t')\rangle & =&  \gamma_{\rm XY}^2 \theta(t-t_0)\theta(t'-t_0)
\sum_{r>0} \left( \phi_r^{\rm X}(L_{\rm X}) \phi_r^{\rm X *}(L_{\rm X}) e^{i \omega_r^{\rm X} (t-t')} f(\omega_r^{\rm X},T_{\rm X}) \right. \nn \\
 && ~~~~~~~~~~~~~~~~~~~~ \left. +\psi_r^{{\rm X} *}(L_{\rm X}) \psi_r^{\rm X}(L_{\rm X}) e^{-i \omega_r^{\rm X}(t-t')} f(-\omega_r^{\rm X},T_{\rm X}) \right). \nn \\
\label{cor}
\eea
The properties of the noises can be written in a convenient form in the frequency domain. Thus, we now convert the above noise-noise correlation to frequency domain, and we take the Fourier transformation (by letting $t_0 \to -\infty$)
\bea
 \langle \tilde{\eta}_1^{\rm X \dagger}(\omega) \tilde{\eta}_1^{\rm X}(\omega')\rangle &=&-{\left(\frac{1}{2 \pi}\right)}^2\int_{-\infty}^{\infty} dt e^{-i \omega t}
\int_{-\infty}^{\infty} dt' e^{i \omega' t'}\langle \eta_1^{\rm X \dagger}(t) \eta_1^{\rm X}(t')\rangle \nn \\
&= & \gamma_{\rm XY}^2 \, \delta(\omega-\omega')\,
\sum_{r>0} \left(\phi_r^{\rm X}(L_{\rm X}) \phi_r^{\rm X*}(L_{\rm X})\delta(\omega -\omega_r^{\rm X})f(\omega,T_{\rm X}) \right. \nn \\
& &  ~~~~~~~~~~~~~~~~~~~~~~\left. +\psi_r^{\rm X*}(L_{\rm X}) \psi_r^{\rm X}(L_{\rm X}) \delta(\omega +\omega_r^{\rm X})f(\omega,T_{\rm X}) \right)  \nn \\
&= & - \frac{\gamma_{\rm XY}^2}{\pi}   \text{Im}[\tilde{G}^{\rm X +}_{2L_{\rm X}-1,2L_{\rm X}-1}(\omega)] f(\omega,T_{\rm X}) \, \delta_{\omega,\omega'}\, ,
\label{NN1a}
\eea
where, $\delta_{\omega,\omega'} \equiv \delta(\omega-\omega')$. We have applied (\ref{GF2a}) in the last line of the above expression, and this substitution is valid when $\{ \phi_r^{\rm X}(i'),\psi_r^{\rm X}(j')\} $ are real, which is true for the N and the Kitaev chain Hamiltonian with a real superconducting gap ($\Delta _{\rm X}$). We similarly express all other noise-noise correlations from the $X$ bath (an N or a Kitaev chain with real $\Delta _{\rm X}$) in the compact forms by applying (\ref{GF2a}):
\bea
\langle \tilde{\eta}_2^{\rm X \dagger}(\omega) \tilde{\eta}_2^{\rm X}(\omega')\rangle &=& - \frac{\gamma_{\rm XY}^2}{\pi}  \text{Im}[\tilde{G}^{\rm X +}_{2L_{\rm X},2L_{\rm X}}(\omega)]f(\omega, T_{\rm X})\delta_{\omega,\omega'} \, , \nn \\
\langle \tilde{\eta}_1^{\rm X \dagger}(\omega) \tilde{\eta}_2^{\rm X}(\omega')\rangle &=&  \frac{\gamma_{\rm XY}^2}{\pi}  \text{Im}[\tilde{G}^{\rm X +}_{2L_{\rm X},2L_{\rm X}-1}(\omega)]f(\omega, T_{\rm X})\delta_{\omega,\omega'}\, , \nn \\
\langle \tilde{\eta}_2^{\rm X \dagger}(\omega) \tilde{\eta}_1^{\rm X}(\omega')\rangle &=&  \frac{\gamma_{\rm XY}^2}{\pi}  \text{Im}[\tilde{G}^{\rm X +}_{2L_{\rm X}-1,2L_{\rm X}}(\omega)]f(\omega, T_{\rm X})\delta_{\omega,\omega'}. \nn \\
\label{NN1b}
\eea
The noise-noise correlations for the $Z$ bath can also be written in the following compact forms provided the bath is an N or a Kitaev chain with a real $\Delta _{\rm Z}$:
\bea
\langle \tilde{\eta}_1^{\rm Z \dagger}(\omega) \tilde{\eta}_1^{\rm Z}(\omega')\rangle &=& - \frac{ \gamma_{\rm YZ}^2}{\pi} \text{Im}[\tilde{G}^{\rm Z +}_{1,1}(\omega)]f(\omega, T_{\rm Z})\delta_{\omega,\omega'} \, , \nn \\
\langle \tilde{\eta}_2^{\rm Z \dagger}(\omega) \tilde{\eta}_2^{\rm Z}(\omega')\rangle &=& - \frac{ \gamma_{\rm YZ}^2}{\pi}  \text{Im}[\tilde{G}^{\rm Z +}_{2,2}(\omega)]f(\omega, T_{\rm Z})\delta_{\omega,\omega'} \, , \nn \\
\langle \tilde{\eta}_1^{\rm Z \dagger}(\omega) \tilde{\eta}_2^{\rm Z}(\omega')\rangle &=&  \frac{ \gamma_{\rm YZ}^2}{\pi}  \text{Im}[\tilde{G}^{\rm Z +}_{2,1}(\omega)]f(\omega, T_{\rm Z})\delta_{\omega,\omega'} \, , \nn \\
\langle \tilde{\eta}_2^{\rm Z \dagger}(\omega) \tilde{\eta}_1^{\rm Z}(\omega')\rangle &=&  \frac{ \gamma_{\rm YZ}^2}{\pi}  \text{Im}[\tilde{G}^{\rm Z +}_{1,2}(\omega)]f(\omega, T_{\rm Z})\delta_{\omega,\omega'}.
\label{NN2}
\eea
The above noise-noise correlations (\ref{NN1a}-\ref{NN2}) are in the form of fluctuation-dissipation relations. It is clear from these relations that we need to find the boundary Green's function of the leads to evaluate the noise-noise correlations. We know an exact analytical form of these boundary Green's functions for an N bath \cite{DharPRB2006,RoyDharPRB2007}. However, we use numerical methods to find the boundary Green's function of semiconducting and superconducting leads. In this paper, we apply the highly-convergent iterative method of Lopez Sancho {\it et al}. \cite{LopezSancho1985} to calculate the boundary Green's function.

We note that the above noise-noise correlations, which are proportional to the off-diagonal terms of boundary Green's function, are identically zero for an N lead. These off-diagonal terms only exist for the TS leads in the presence of superconducting pairing. Since we here often take a non-zero chemical potential for the N baths, the Fermi functions used in the aforementioned noise-noise correlations need some modifications. When the metallic  $\alpha$ wire bath is kept at a chemical potential $\mu_\alpha$, the Fermi functions in the $\langle \tilde{\eta}_1^{ \alpha \dagger}(\omega) \tilde{\eta}_1^{\alpha}(\omega')\rangle$ and $\langle \tilde{\eta}_2^{\alpha \dagger}(\omega) \tilde{\eta}_2^{\rm  \alpha}(\omega')\rangle$ should be modified as $f(\omega-\frac{\mu_\alpha}{\hbar}, T_{\alpha})$ and $ f(\omega+\frac{\mu_\alpha}{\hbar}, T_{\alpha})$, respectively. Thus, the expressions of the junction currents would also be modified accordingly. We further clarify that we use an exact analytical expression for the boundary Green's function for an N lead \cite{DharPRB2006, RoyDharPRB2007}.

\setcounter{equation}{0}
\setcounter{figure}{0}
\renewcommand\thefigure{D\arabic{figure}}
\section{Application of LEGF to a device made of Majorana wire }
\label{App4a}

Here, for a $X$-$Y$-$Z$ device with $L$ lattice sites, we use the following generalized basis: ${\bf b} \equiv  [ b_1, b_2, b_3, b_4, \dots, b_{4L-3}, b_{4L-2}, b_{4L-1}, b_{4L}]^T $. For time $t > t_0$, the Heisenberg equations of motion of the $Y$ wire variables are 
 \bea
  \dot{b}_l &=& -i \sum_{m=4L_{\rm X}+1}^{4L_{\rm XY}} \mathcal{H}^{\rm Y}_{lm} b_m  +i\gamma_{\rm XY}\sum_{m=4L_{\rm X}+1}^{4L_{\rm X}+2} b_{m-4} \, \delta_{l,m}-i\gamma_{\rm XY}\sum_{m=4L_{\rm X}+3}^{4L_{\rm X}+4} \, b_{m-4} \delta_{l,m} \nn \\
  &&+i\gamma_{\rm YZ}\sum_{m=4L_{\rm XY}-3}^{4L_{\rm XY}-2} b_{m+4} \, \delta_{l,m}  -i\gamma_{\rm YZ}\sum_{m=4L_{\rm XY}-1}^{4L_{\rm XY}} \, b_{m+4} \delta_{l,m} \nn \\
  && - i\zeta_{\rm XY} \big( b_{4L_{\rm X}-2}\, \delta_{l,4L_{\rm X}+1}  -b_{4L_{\rm X}-3}\, \delta_{l,4L_{\rm X}+2}-b_{4L_{\rm X}}\, \delta_{l,4L_{\rm X}+3}+b_{4L_{\rm X}-1}\, \delta_{l,4L_{\rm X}+4}\big) \nn \\
 && +i\zeta_{\rm YZ}\big(b_{4L_{\rm XY}+2}\, \delta_{l,4L_{\rm XY}-3} -b_{4L_{\rm XY}+1}\, \delta_{l,4L_{\rm XY}-2} \nn \\
 &&~~~~~~~~~~~~~~~~~~~~~~~~~~~~~~~~~~~~~~~- b_{4L_{\rm XY}+4}\, \delta_{l,4L_{XY}-1}+ b_{4L_{\rm XY}+3}\, \delta_{l,4L_{XY}}\big), \nn \\
 \label{Meomwire}
 \eea
 where, $l = 4L_{\rm X}+1,\dots,4L_{\rm XY} $. The equations of motion for the bath ($X$ and $Z$ wires) variables read as
  \bea
  \dot{b}_l &=& -i \sum_{m=1}^{4L_{\rm X}} \mathcal{H}^{\rm X}_{lm} b_m +i\gamma_{\rm XY}\sum_{m=4L_{\rm X}-3}^{4L_{\rm X}-2} b_{m+4} \, \delta_{l,m}-i\gamma_{\rm XY}\sum_{m=4L_{\rm X}-1}^{4L_{\rm X}} \, b_{m+4} \delta_{l,m} \nn \\
 && + i\zeta_{\rm XY}\left( b_{4L_{\rm X}+2}\, \delta_{l,4L_{\rm X}-3}-b_{4L_{\rm X}+1}\, \delta_{l,4L_{\rm X}-2}-b_{4L_{\rm X}+4}\, \delta_{l,4L_{\rm X}-1}+b_{4L_{\rm X}+3}\, \delta_{l,4L_{\rm X}}\right), 
 \qquad \label{Meombathl}
 \eea
 for $l = 1, \dots, 4L_{\rm X}$, and
\bea
  \dot{b}_l &= & -i \sum_{m=4L_{\rm XY}+1}^{4L} \mathcal{H}^{\rm Z}_{lm} b_m +i\gamma_{\rm YZ}\sum_{m=4L_{\rm XY}+1}^{4L_{\rm XY}+2} b_{m-4} \, \delta_{l,m}-i\gamma_{\rm YZ}\sum_{m=4L_{\rm XY}+3}^{4L_{\rm XY}+4} \, b_{m-4} \delta_{l,m} \nn \\
  && -i\zeta_{\rm YZ}\left( b_{4L_{\rm XY}-2}\, \delta_{l,4L_{XY}+1}- b_{4L_{\rm XY}-3}\, \delta_{l,4L_{XY}+2} \right. \nn \\
  &&\left. ~~~~~~~~~~~~~~~~~~~~~~~~~~~~~~~~~~~~~~-b_{4L_{\rm XY}}\, \delta_{l,4L_{\rm XY}+3}+b_{4L_{\rm XY}-1}\, \delta_{l,4L_{\rm XY}+4}\right), 
\qquad \label{Meombathr}
 \eea
 for $l = 4L_{\rm XY}+1,\dots,4L$. Like the Kitaev wire in the previous subsection \ref{kita}, Eq.~\ref{Meombathl} and Eq.~\ref{Meombathr} can be solved using the single-particle retarded Green's function of the isolated bath wire. In the first step, the above equations for the variables of $X$ and $Z$ near the junctions are solved using the Green's function to find formal solutions of $b_l(t)$ in the integral form in time domain. In the next step, these formal solutions of bath variables are Fourier transformed assuming steady state at long time, and the bath modes, $ \tilde{b}_l(\omega) $ are written in terms of their noises and $Y$ wire variables. 
 
It is noteworthy that the matrix Green's function for $ \alpha $ Majorana/semiconductor wire with $L_\alpha$ sites is represented by a square matrix of dimension $4L_\alpha \times 4L_\alpha$ in ${\bf{b}}$ basis as $[{G}^+_\alpha (\tau)]_{l,m} \equiv [-i e^{-i\mathcal{H}^{ \alpha } \tau} ]_{lm} \theta (\tau) $, where $l,m=1,\dots,4L_\alpha$ and $\mathcal{H}^{ \alpha }$ is the corresponding wire Hamiltonian. Substituting the formal solutions of Eq.~\ref{Meombathl} and Eq.~\ref{Meombathr} into Eq.~\ref{Meomwire}, we get a set of generalized quantum Langevin equations for $Y$ wire variables in which the contributions of the baths (boundary wires) enters into the equations via the self-energy corrections and the noise terms. Due to the presence of spin-orbit coupling in the SM junctions, both the self-energy and noise terms also depend on the spin-orbit coupling in addition to its dependency on the tunnel coupling. The steady-state solutions of the aforesaid $Y$ wire equations are possible through the Fourier transformation method when the middle wire attains a unique NESS. In such cases, the solutions of the variables of $Y$ wire, $\tilde{b}_l(\omega)$ read (for $l=4L_{\rm X}+1,\dots,4L_{\rm XY}$)
\bea
 \tilde{b}_l(\omega)&=& \sum_{m=4L_{\rm X}+1}^{4L_{\rm XY}} \tilde{G}^+_{l,m}(\omega)\left( \tilde{\xi}_{1 \uparrow}^{\rm X} (\omega)\, \delta_{m,4L_{\rm X}+1}+\tilde{\xi}_{1 \downarrow}^{\rm X} (\omega)\, \delta_{m,4L_{\rm X}+2} \right. \nn \\
&& +\tilde{\xi}_{2 \uparrow }^{\rm X} (\omega) \, \delta_{m,4L_{\rm X}+3} +\tilde{\xi}_{2 \downarrow }^{\rm X} (\omega) \, \delta_{m,4L_{\rm X}+4}  +\tilde{\xi }_{1 \uparrow }^{\rm Z} (\omega)\, \delta_{m,4L_{\rm XY} -3} \nn \\
 &&\left. +\tilde{\xi}_{1  \downarrow }^{\rm Z} (\omega)\, \delta_{m,4L_{\rm XY} -2} +\tilde{\xi}_{2 \uparrow}^{\rm Z} (\omega) \, \delta_{m,4L_{\rm XY}-1} +\tilde{\xi}_{2 \downarrow}^{\rm Z} (\omega) \, \delta_{m,4L_{\rm XY}}\right), 
\label{b1}
\eea 
where $\tilde{\xi}^{\alpha}_{1 \sigma}(\omega)$ and  $\tilde{\xi}^{\alpha}_{2 \sigma}(\omega)$ ($\sigma= \uparrow,\downarrow$) are the Fourier transform of the noise terms associated with $\alpha$= $X$, $Z$ bath (see \ref{noiseM} for definition), and $\tilde{G}^+_{l,m}(\omega)$ is the Fourier transform of the retarded Green's function of the full system which, in this case, is a square matrix of dimension $4L_{\rm Y} \times 4L_{\rm Y}$. This retarded Green's function is defined in a similar fashion to the relation (\ref{FG0}), and the effective non-Hermitian Hamiltonian of $Y$ wire reads $\tilde{\mathcal{H}}^{\rm Y}=\mathcal{H}^{\rm Y}+ \tilde{\Sigma}^+_{\rm X}(\omega) +\tilde{\Sigma}^+_{\rm Z}(\omega)$. Here, $\mathcal{H}^{\rm Y}$ is the Hamiltonian of the finite Majorana/SM wire. The structure of $\tilde{\Sigma}^+_{\rm X/Z}(\omega)$ is determined by the tunnelling Hamiltonian (\ref{tunham}) and the Hamiltonian of the lead made of either a Majorana or an SM wire (\ref{Mham}). The self-energy correction term associated with the $X$ lead can be written as:
\bea
[\tilde{\Sigma}_{\rm X}^+(\omega)]_{lm} &=& \sum_{k=4L_{\rm X}-3}^{4L_{\rm X}}\left[ \delta_{l,4L_{\rm X}+1} \left( \gamma_{\rm XY}   \, {\tilde{G}^{\rm  X +}_{4L_{\rm X}-3,k}(\omega)}-\zeta_{\rm XY}  {\tilde{G}^{\rm  X +}_{4L_{\rm X}-2,k}(\omega)}\right) \right. \nn \\
&&+\delta_{l,4L_{\rm X}+2} \left( \gamma_{\rm XY}   \, {\tilde{G}^{\rm  X +}_{4L_{\rm X}-2,k}(\omega)}+\zeta_{\rm XY}  {\tilde{G}^{\rm  X +}_{4L_{\rm X}-3,k}(\omega)}\right)  \nn \\
&&-\delta_{l,4L_{\rm X}+3} \left( \gamma_{\rm XY}   \, {\tilde{G}^{\rm  X +}_{4L_{\rm X}-1,k}(\omega)}-\zeta_{\rm XY}  {\tilde{G}^{\rm  X +}_{4L_{\rm X},k}(\omega)}\right) \nn \\
&&\left. -\delta_{l,4L_{\rm X}+4} \left( \gamma_{\rm XY}   \, {\tilde{G}^{\rm  X +}_{4L_{\rm X},k}(\omega)}+\zeta_{\rm XY}  {\tilde{G}^{\rm  X +}_{4L_{\rm X}-1,k}(\omega)}\right)\right] \Gamma^{\rm X}_{km}\, ,\nn \\
\\
{\rm where}, \, \Gamma^{\rm X}_{km} &=& \left( \mathcal{X}_{k m}^{\rm X} \gamma_{\rm XY}+\mathcal{Y}_{k m}^{\rm X}\zeta_{\rm XY}\right),\nn \\
\mathcal{X}_{k m}^{\rm X} &=& \sum_{s=4L_{\rm X}-3}^{4L_{\rm X}-2} \delta_{k,s} \,\delta_{m,s+4}-\sum_{s=4L_{\rm X}-1}^{4L_{\rm X}} \delta_{k,s} \,\delta_{m,s+4}\, ,
\nn \\
 \text{and},  \,\mathcal{Y}_{k m}^{\rm X} &=& \delta_{k,4L_{\rm X}-3} \,\delta_{m,4L_{\rm X}+2}-\delta_{k,4L_{\rm X}-2} \, \delta_{m,4L_{\rm X}+1} \nn \\
&&~~~~~~~~~~~~~~~~~~~~~~~~~~~ -\delta_{k,4L_{\rm X}-1}\, \delta_{m,4L_{\rm X}+4}+\delta_{k,4L_{\rm X}} \, \delta_{m,4L_{\rm X}+3}\, . \nn 
\eea
\noindent
Similarly, the self-energy correction term for the $Z$ bath reads
\bea
 [\tilde{\Sigma}_{\rm Z}^+(\omega)]_{lm} &=& \sum_{k=1}^{4} \left[ \delta_{l,4L_{\rm XY}-3} \left( \gamma_{\rm YZ}   \, {\tilde{G}^{\rm  Z +}_{1,k}(\omega)}+\zeta_{\rm YZ}  {\tilde{G}^{\rm  Z +}_{2,k}(\omega)}\right) \right.\nn \\
 && +\delta_{l,4L_{\rm XY}-2} \left( \gamma_{\rm YZ}   \, {\tilde{G}^{\rm  Z +}_{2,k}(\omega)}-\zeta_{\rm YZ}  {\tilde{G}^{\rm  Z +}_{1,k}(\omega)}\right)\nn \\
&& -\delta_{l,4L_{\rm XY}-1} \left( \gamma_{\rm YZ}   \, {\tilde{G}^{\rm  Z +}_{3,k}(\omega)}+\zeta_{\rm YZ}  {\tilde{G}^{\rm  Z +}_{4,k}(\omega)}\right) \nn \\
&& \left. -\delta_{l,4L_{\rm XY}} \left( \gamma_{\rm YZ}   \, {\tilde{G}^{\rm  Z +}_{4,k}(\omega)}-\zeta_{\rm YZ}  {\tilde{G}^{\rm  Z +}_{3,k}(\omega)}\right) \right]\Gamma^{\rm Z}_{km}\, , \nn\\ 
\\
 {\rm where},\, \Gamma^{\rm Z}_{km}&=&\left( \mathcal{X}_{k m}^{\rm Z }\gamma_{\rm YZ}+  \mathcal{Y}_{k m}^{\rm Z}\zeta_{\rm YZ}\right), \nn \\
~\mathcal{X}_{k m}^{\rm Z}& =& \sum_{s=1}^{2} \delta_{k,s} \,\delta_{m,4L_{\rm XY}-4+s}-\sum_{s=3}^{4} \delta_{k,s} \,\delta_{m,4L_{\rm XY}-4+s}\, , \nn\\
 {\rm and}, \, \mathcal{Y}_{k m}^{\rm Z}&=&-\delta_{k,1} \,\delta_{m,4L_{\rm XY}-2}+\delta_{k,2} \, \delta_{m,4L_{\rm XY}-3}+\delta_{k,3}\, \delta_{m,4L_{\rm XY}}-\delta_{k,4} \, \delta_{m,4L_{\rm XY}-1}\, . \nn 
\eea
It is worth-noting that for $\tilde{G}^{\rm X +}_{l,m}(\omega)$, $l,m =4L_{\rm X}-3,\dots,4L_{\rm X}$ correspond to the right most site (i.e. $L_{\rm X}$-th site) of the $X$ reservoir, whereas in case of $\tilde{G}^{\rm Z +}_{l,m}(\omega)$, $l,m=1,\dots,4$ correspond to the left most site (i.e. $L_{\rm XY}+1$-th site) of the $Z$ reservoir in the full $X$-$Y$-$Z$ device. As we have observed earlier, the noise-noise correlations are essential for finding the currents at the junctions. We outline the derivations of such correlations for the Majorana wire bath in Appendix \ref{App4b}.

\setcounter{equation}{0}
\setcounter{figure}{0}
\renewcommand\thefigure{E\arabic{figure}}
\section{Noise-noise correlations for Majorana wire leads}
\label{App4b}

The noise-noise correlations for the Majorana wire leads have several new types of terms in comparison to the Kitaev chain in the presence of two types of spin degrees of freedom in the former. To this end, we first calculate the equilibrium correlation functions by applying Eq.~\ref{Mlatticeop}. For an isolated $\alpha$ bath wire ($\alpha=X,Z$), whose  $r$-th eigenvalue is $\hbar\omega_r^\alpha$, and $\{ \phi_{r \uparrow}^{\alpha },\phi_{r \downarrow}^{\alpha },\psi_{r \uparrow}^{\alpha },\psi_{r \downarrow}^{\alpha }\}$ are the components of the corresponding eigenfunction, the equilibrium correlation functions at temperature $T_\alpha$ are:
\bea
\langle c_{i' \sigma }^{\dagger}\, c_{j' \rho} \rangle &=& \sum_{r>0}\left[ \phi^{\alpha * }_{r \sigma}(i')\phi^\alpha_{r \rho}(j')f(\omega_r^\alpha, T_\alpha)+\psi_{r \sigma}^\alpha(i') \psi_{r \rho}^{\alpha *}(j') f(-\omega_r^\alpha, T_\alpha)\right], \nn \\
\langle c_{i' \sigma }\, c_{j' \rho}^\dagger  \rangle &=& \sum_{r>0}\left[ \psi_{r \sigma}^{\alpha *}(i') \psi_{r \rho }^\alpha(j') f(\omega_r^\alpha, T_\alpha)+\phi_{r \sigma }^\alpha(i') \phi_{r \rho}^{\alpha *}(j') f(-\omega_r^\alpha,T_\alpha)\right], \nn \\
\langle c_{i' \sigma } \, c_{j' \rho} \rangle &=& \sum_{r>0}\left[\psi_{r \sigma }^{\alpha *}(i') \phi_{r \rho}^\alpha(j')f(\omega_r^\alpha,T_\alpha)+\phi_{r \sigma}^\alpha(i') \psi_{r \rho}^{\alpha *}(j')f(-\omega_r^\alpha,T_\alpha)\right], \nn \\
\langle c_{i' \sigma}^{ \dagger} \, c_{j'\rho}^{ \dagger}  \rangle &=& \sum_{r>0}\left[\phi_{r \sigma}^{\alpha *}(i') \psi_{r \rho}^\alpha(j')f(\omega_r^\alpha, T_\alpha)+\psi_{r \sigma}^\alpha(i') \phi_{r \rho}^{\alpha *}(j')f(-\omega_r^\alpha, T_\alpha)\right], \qquad
\label{coM}
\eea
where $i',j'=1,\dots,L_{\alpha}$ and $\sigma ,\rho =\uparrow, \downarrow $. The noise terms related to the $X$ and $Z$ baths are defined as follows:
\bea
 \xi^{\rm X}_{k \uparrow}(t)= \eta^{\rm X}_{k \uparrow}(t)-\bar{\zeta}_{XY}\eta^{\rm X}_{k \downarrow}(t)  , && \,
\xi^{\rm X}_{k \downarrow}(t)= \eta^{\rm X}_{k \downarrow}(t)+\bar{\zeta}_{XY}\eta^{\rm X}_{k \uparrow}(t) \, , \nn \\
\xi^{\rm Z}_{k \uparrow}(t)= \eta^{\rm Z}_{k \uparrow}(t)+\bar{\zeta}_{YZ}\eta^{\rm Z}_{k \downarrow}(t) , && \,  
\xi^{\rm Z}_{k \downarrow}(t)= \eta^{\rm Z}_{k \downarrow}(t)-\bar{\zeta}_{YZ}\eta^{\rm Z}_{k \uparrow}(t) \, , \nn \\
\label{noiseM}
\eea
where $k=1, 2$ and $\bar{\zeta}_{\alpha \beta}=\zeta_{\alpha \beta}/\gamma_{\alpha \beta}$. These $\eta^\alpha_{k \sigma}$ terms are quite similar to the noise terms of the Kitaev chain leads, and they read as
\bea
\eta_{k \uparrow}^{\rm X}(t)&=&i\gamma_{\rm XY} (-1)^k  \sum_{l=1}^{4L_{\rm X}}{[G_{\rm X}^+(t -t_0)]}_{4L_{\rm X}+2k-5,l} \, b_{l}(t_0)\, ,\nn \\
\eta_{k \downarrow}^{\rm X}(t)&=&i \gamma_{\rm XY} (-1)^k  \sum_{l=1}^{4L_{\rm X}}{[G_{\rm X}^+(t -t_0)]}_{4L_{\rm X}+2k-4,l} \, b_{l}(t_0) \, ,\nn \\
\eta_{k \uparrow}^{\rm Z}(t)&=&i \gamma_{\rm YZ} (-1)^k  \sum_{l=4L_{\rm XY}+1}^{4L}{[G_{\rm Z}^+(t -t_0)]}_{2k-1,l-4L_{\rm XY}} \, b_{l}(t_0)\, ,\nn \\
\eta_{k \downarrow}^{\rm Z}(t)&=&i\gamma_{\rm YZ} (-1)^k   \sum_{l=4L_{\rm XY}+1}^{4L}{[G_{\rm Z}^+(t -t_0)]}_{2k,l-4L_{\rm XY}} \, b_{l}(t_0)\, , \nn \\
\label{pnoise}
\eea
where ${[G_{\alpha}^+(t -t_0)]}_{l,m}$ are the components of bath Green's functions, and $b_l$'s are operators in the generalized basis ${\bf b}$ (\ref{bas2}). These bath Green's functions can be expanded in terms of eigenfunctions of the Hamiltonian of the bath wire. For example, some components of ${[G_{\alpha}^+(t -t_0)]}_{l,m}$ are given below, and others can be derived identically.
\bea
 {[G_\alpha^+(\tau)]}_{4i'-3,4j'-3} &=& - i \theta(\tau) \sum_{r>0} \left( \phi_{r\uparrow}^\alpha(i' ) \phi_{r \uparrow}^{\alpha *}(j') e^{-i\,  \omega^\alpha_r \tau}+\psi^{\alpha *}_{r \uparrow}(i') \psi_{r \uparrow}^\alpha(j') e^{i\, \omega_r^\alpha\tau }\right), \nn \\
 {[G_\alpha^+(\tau)]}_{4i'-2,4j'-2} &=& - i \theta(\tau) \sum_{r>0} \left( \phi_{r\downarrow}^\alpha(i' ) \phi_{r \downarrow}^{\alpha *}(j') e^{-i\,  \omega^\alpha_r \tau}+\psi^{\alpha *}_{r \downarrow}(i') \psi_{r \downarrow}^\alpha(j') e^{i\, \omega_r^\alpha \tau}\right), \nn \\
 {[G_\alpha^+(\tau)]}_{4i'-1,4j'-1} &=& - i \theta(\tau) \sum_{r>0} \left( \psi_{r \uparrow}^\alpha(i') \psi_{r \uparrow}^{\alpha *}(j') e^{-i \, \omega_r^\alpha \tau}+\phi^{\alpha *}_{r \uparrow}(i') \phi_{r \uparrow}^\alpha(j') e^{i\, \omega_r^\alpha \tau} \right), \nn \\
 {[G_\alpha^+(\tau)]}_{4i',4j'} &=& - i \theta(\tau) \sum_{r>0} \left( \psi_{r \downarrow}^\alpha(i') \psi_{r \downarrow}^{\alpha *}(j') e^{-i \, \omega_r^\alpha \tau}+\phi^{\alpha *}_{r \downarrow}(i') \phi_{r \downarrow}^\alpha(j') e^{i\, \omega_r^\alpha \tau} \right), \nn \\
 {[G_\alpha^+(\tau)]}_{4i'-3,4j'-2} &=& - i \theta(\tau) \sum_{r>0} \left( \phi_{r\uparrow}^\alpha(i' ) \phi_{r \downarrow}^{\alpha *}(j') e^{-i\,  \omega^\alpha_r \tau}+\psi^{\alpha *}_{r \uparrow}(i') \psi_{r \downarrow}^\alpha(j') e^{i \, \omega_r^\alpha \tau}\right), \nn \\
 {[G_\alpha^+(\tau)]}_{4i'-3,4j'-1} &=& - i \theta(\tau) \sum_{r>0} \left( \phi_{r \uparrow}^\alpha(i') \psi_{r \uparrow}^{\alpha *}(j') e^{-i \, \omega_{r \uparrow}^\alpha \tau}+\psi^{\alpha *}_{r \uparrow}(i') \phi_{r  \uparrow}^\alpha(j') e^{i \, \omega_r^\alpha \tau}\right). \nn 
\eea
In our steady-state transport analysis, we do not apply ${[G_{\alpha}^+(t -t_0)]}_{l,m}$ directly, we rather take the Fourier transformation of these bath Green's functions using 
\beq
{\tilde{G}^{\alpha +}_{l,m}(\omega)}=\frac{1}{2 \pi}\int_{-\infty}^{\infty} d\tau\, { [G_\alpha^+(\tau)]}_{l,m}\, e^{i \omega \tau}.\nn 
\eeq
Since the steady-state current calculation requires the noise-noise correlations in the Fourier domain, we further define the Fourier transform of the aforementioned functions as $\tilde{\eta}_{k \sigma }^{\alpha}(\omega)=\frac{1}{2\pi}\int_{-\infty}^{\infty} dt \,\eta_{k \sigma}^{\alpha}(t)\, e^{i \omega t} $ and $\tilde{\xi}_{k \sigma }^{\alpha}(\omega)=\frac{1}{2\pi}\int_{-\infty}^{\infty} dt \,\xi_{k \sigma}^{\alpha}(t)\, e^{i \omega t} $. 

Like the Kitaev chain leads in Appendix \ref{App3}, the noise-noise correlations for the Majorana wires are initially calculated in time domain using \ref{noiseM}, \ref{pnoise}, \ref{coM}, and the orthonormality relations between the eigenvectors of the bath's Hamiltonian. In the next step, we perform the Fourier transformation of the noise-noise correlations. Finally, we substitute ${\tilde{G}^{\alpha +}_{l,m}(\omega)}$ into the Fourier transformed noise-noise correlation to rewrite them in a simplified form. This calculation is tedious but straightforward. Hence, we here only mention the important results, which are the correlations between $ \tilde{\eta}_{k \sigma }^{\alpha}(\omega)$'s. These last correlations, in turn, can be applied to derive the required noise-noise correlations $ \langle \tilde{\xi}_{k \sigma }^{\alpha \dg}(\omega) \tilde{\xi}_{m \sigma }^{\alpha}(\omega') \rangle$. The correlations, $\langle \tilde{\eta}_{k \sigma }^{\alpha \dg}(\omega) \tilde{\eta}_{m \sigma }^{\alpha}(\omega') \rangle$, are expressed in a compact notation as follows:
\bea
 \langle \tilde{\eta}_{k \sigma}^{\rm X \dagger}(\omega) \tilde{\eta}_{m \sigma'}^{\rm X}(\omega')\rangle &=& (-1)^{k+m+1} \frac{ \gamma_{\rm XY}^2}{\pi} \text{Im}[\tilde{G}^{\rm X +}_{4L_{\rm X}-6+2m+r(\sigma'),4L_{\rm X}-6+2k+r(\sigma)}(\omega)]\nn \\
 && ~~~~~~~~~~~~~~~~~~~~~~~~~~~~~~~~~~~~~~~~~~~~~~~~~~~\times f(\omega, T_{\rm X})\delta_{\omega,\omega'} \, , \nn \\
 \langle \tilde{\eta}_{k \sigma}^{\rm Z \dagger}(\omega) \tilde{\eta}_{m \sigma'}^{\rm Z}(\omega')\rangle &=& (-1)^{k+m+1} \frac{ \gamma_{\rm YZ}^2}{\pi} \text{Im}[\tilde{G}^{\rm Z +}_{2m-2+r(\sigma'),2k-2+r(\sigma)}(\omega)]f(\omega, T_{\rm Z})\delta_{\omega,\omega'} \, , \nn \\
\label{noiseFM}
\eea
where $\sigma, \sigma' =\uparrow, \downarrow$ and $r(\uparrow)(r(\downarrow))=1(2)$. It is easy to find all the noise-noise correlations using \ref{noiseM} and \ref{noiseFM}. The noise-noise correlations depend linearly on the boundary Green's functions of the baths. We apply the same iterative method of Ref. \cite{LopezSancho1985} to numerically calculate these boundary Green's functions for the Majorana wire and the SM lead. 

For an SM lead, all correlations of type $\langle \tilde{\eta}_{k \sigma}^{ \alpha \dagger}(\omega) \tilde{\eta}_{m \tau}^{\alpha}(\omega')\rangle$ are zero when $k \neq m$. Like an N bath, an SM bath can also be kept at a non-zero chemical potential. When an SM lead is kept at chemical potential $\mu_\alpha$, the Eq. \ref{noiseFM} needs modification, and the Fermi functions in the $\langle \tilde{\eta}_{1 \sigma} ^{ \alpha \dagger}(\omega) \tilde{\eta}_{1 \tau}^{\alpha}(\omega')\rangle$ and $\langle \tilde{\eta}_{2 \sigma} ^{\alpha \dagger}(\omega) \tilde{\eta}_{2 \tau}^{\rm  \alpha}(\omega')\rangle$ should be replaced by $ f(\omega-\frac{\mu_\alpha}{\hbar}, T_{\alpha})$ and $ f(\omega+\frac{\mu_\alpha}{\hbar}, T_{\alpha})$, respectively.

\setcounter{equation}{0}
\setcounter{figure}{0}
\renewcommand\thefigure{F\arabic{figure}}
\section{Expressions of electrical currents $J^{e}_{\rm XY}$ and $J^{e}_{\rm YZ}$}
\label{App5}
The full expression of the electrical current from the $X$ bath to the $Y$ wire can be separated in three different parts. When the chemical potentials of the boundary leads are zero, only the first part survives in the presence of a temperature bias from the boundary leads. We also need to set the chemical potential $(\mu_{\rm X},\mu_{\rm Z})$ to zero for the TS bath(s) of various devices in the following current expressions. 
\bea
 J^{e}_{\rm XY} &=& \int_{- \infty}^{ \infty}d \omega \:e\left( \mathcal{T}^1_{\rm XY}(\omega)+\mathcal{T}^2_{\rm XY}(\omega) \right)(f(\omega-\frac{\mu_{\rm X}}{\hbar},T_{\rm X}) -f(\omega-\frac{\mu_{\rm Z}}{\hbar},T_{\rm Z})) \nn \\
&& ~~~~~~~~~~~~~~~~+\int_{- \infty}^{ \infty}d \omega \,e\, \mathcal{T}^2_{\rm XY}(\omega) \,(f(\omega+\frac{\mu_{\rm X}}{\hbar},T_{\rm X}) -f(\omega-\frac{\mu_{\rm X}}{\hbar},T_{\rm X}))  \nn \\
&& ~~~~~~~~~~~~~~~~+\int_{- \infty}^{ \infty}d \omega \,e\, \mathcal{T}^3_{\rm XY}(\omega)\, (f(\omega+\frac{\mu_{\rm Z}}{\hbar},T_{\rm Z}) -f(\omega-\frac{\mu_{\rm Z}}{\hbar},T_{\rm Z})),  \nn \\
\label{cuL}
\eea
where,
\bea
\mathcal{T}^1_{\rm XY}(\omega) &=& (-\frac{2 \gamma_{\rm XY}^2}{\pi}) \sum_{l=2L_{\rm X}-1}^{2L_{\rm X}}
(-1)^{l+1}  \text{Im} \left[[\tilde{G}^+(\omega)]_{2L_{\rm X}+1,l}^* \, \text{Im}[\tilde{G}^{\rm X +}_{2L_{\rm X}-1,l-2}(\omega)] \right. \nn \\
&+& [\tilde{\Sigma}_{\rm X}^+(\omega)]_{2L_{\rm X}+1,2L_{\rm X}+1}[\tilde{G}^+(\omega)]_{2L_{\rm X}+1,2L_{\rm X}+1} [\tilde{G}^+(\omega)]_{2L_{\rm X}+1,l}^* \text{Im}[\tilde{G}^{\rm X +}_{2L_{\rm X}-1,l-2}(\omega)] \nn \\
&+& \left. [\tilde{\Sigma}_{\rm X}^+(\omega)]_{2L_{\rm X}+1,2L_{\rm X}+2}[\tilde{G}^+(\omega)]_{2L_{\rm X}+1,2L_{\rm X}+1}^*  [\tilde{G}^+(\omega)]_{2L_{\rm X}+2,l} \text{Im}[\tilde{G}^{\rm X +}_{l-2,2L_{\rm X}-1}(\omega)] \right] \nn ,
\eea
\bea
\mathcal{T}^2_{\rm XY}(\omega)&=& (\frac{2 \gamma_{\rm XY}^2}{\pi}) \sum_{l=2L_{\rm X}-1}^{2L_{\rm X}}
(-1)^{l+1}  \text{Im} \left[[\tilde{\Sigma}_{\rm X}^+(\omega)]_{2L_{\rm X}+1,2L_{\rm X}+1}[\tilde{G}^+(\omega)]_{2L_{\rm X}+1,2L_{\rm X}+2}\right.\nn \\
&&~~~~~~~~~~~~~~~~~~~~~~~~~~~~~~~~~~~~~~~~~~~~ \times[\tilde{G}^+(\omega)]_{2L_{\rm X}+1,l}^* \text{Im}[\tilde{G}^{\rm X +}_{2L_{\rm X},l-2}(\omega)]\nn \\
&+&\left. [\tilde{\Sigma}_{\rm X}^+(\omega)]_{2L_{\rm X}+1,2L_{\rm X}+2}[\tilde{G}^+(\omega)]_{2L_{\rm X}+1,2L_{\rm X}+2}^*
[\tilde{G}^+(\omega)]_{2L_{\rm X}+2,l} \text{Im}[\tilde{G}^{\rm X +}_{l-2,2L_{\rm X}}(\omega)] \right] , \nn  
\eea
and, 
\bea
\mathcal{T}^3_{\rm XY}(\omega) &=& (\frac{2 \gamma_{\rm YZ}^2}{\pi})\sum_{l=2L_{\rm XY}-1}^{2L_{\rm XY}}
(-1)^{l+1} \text{Im} \left[ [\tilde{\Sigma}_{\rm X}^+(\omega)]_{2L_{\rm X}+1,2L_{\rm X}+1}[\tilde{G}^+(\omega)]^*_{2L_{\rm X}+1,2L_{\rm XY}} \right. \nn \\
&& ~~~~~~~~~~~~~~~~~~~~~~~~~~~~~~~~~~~~~~~\times [\tilde{G}^+(\omega)]_{2L_{\rm X}+1,l} \text{Im}[\tilde{G}^{\rm Z +}_{l-2L_{\rm XY}+2,2}(\omega)] \nn \\
&+& \left. [\tilde{\Sigma}_{\rm X}^+(\omega)]_{2L_{\rm X}+1,2L_{\rm X}+2}[\tilde{G}^+(\omega)]_{2L_{\rm X}+1,2L_{\rm XY}}^* [\tilde{G}^+(\omega)]_{2L_{\rm X}+2,l} \text{Im}[\tilde{G}^{\rm Z +}_{l-2L_{\rm XY}+2,2}(\omega)] \right] . \nn
\eea
We can similarly simplify the electrical current from the $Y$ wire to the $Z$ bath:
\bea
 J^{e}_{\rm YZ }&=& \int_{- \infty}^{ \infty}d \omega \,e\left( \mathcal{T}^1_{\rm YZ}(\omega)+\mathcal{T}^2_{\rm YZ}(\omega) \right)(f(\omega-\frac{\mu_{\rm X}}{\hbar},T_{\rm X}) -f(\omega-\frac{\mu_{\rm Z}}{\hbar},T_{\rm Z})) \nn \\
&&+\int_{- \infty}^{ \infty}d \omega \,e\, \mathcal{T}^2_{\rm YZ}(\omega) \,(f(\omega+\frac{\mu_{\rm X}}{\hbar},T_{\rm X}) -f(\omega-\frac{\mu_{\rm X}}{\hbar},T_{\rm X})) \nn \\
&& +\int_{- \infty}^{ \infty}d \omega \,e\, \mathcal{T}^3_{\rm YZ}(\omega)\, (f(\omega+\frac{\mu_{\rm Z}}{\hbar},T_{\rm Z}) -f(\omega-\frac{\mu_{\rm Z}}{\hbar},T_{\rm Z})), 
\label{cuR}
\eea
\noindent
where, 
\bea
\mathcal{T}^1_{\rm YZ}(\omega) &=& (\frac{2 \gamma_{\rm XY}^2}{\pi}) \sum_{l=2L_{\rm XY}-1}^{2L_{\rm XY}}
(-1)^{l+1} \text{Im} \left[[\tilde{\Sigma}_{\rm Z}^+(\omega)]_{1,1} [\tilde{G}^+(\omega)]_{2L_{\rm XY}-1,2L_{\rm X}+1}\right. \nn \\
&& ~~~~~~~~~~~~~~~~~~~~~~~~~~~~~~~~~~~~~\times  [\tilde{G}^+(\omega)]_{2L_{\rm XY}-1,l}^* \text{Im}[\tilde{G}^{\rm X +}_{2L_{\rm X}-1,l-2}(\omega)] \nn \\ 
&& \left. + [\tilde{\Sigma}_{\rm Z}^+(\omega)]_{1,2}[\tilde{G}^+(\omega)]_{2L_{\rm XY}-1,2L_{\rm X}+1}^* [\tilde{G}^+(\omega)]_{2L_{\rm XY},l} \text{Im}[\tilde{G}^{\rm X +}_{l-2,2L_{\rm X}-1}(\omega)] \right] , \nn 
\eea
\bea
\mathcal{T}^2_{\rm YZ}(\omega) &=& (-\frac{2 \gamma_{\rm XY}^2}{\pi}) \sum_{l=2L_{\rm XY}-1}^{2L_{\rm XY}}
(-1)^{l+1}  \text{Im} \left[[\tilde{\Sigma}_{\rm Z}^+(\omega)]_{1,1}[\tilde{G}^+(\omega)]_{2L_{\rm XY}-1,2L_{\rm X}+2} \right. \nn \\
&&~~~~~~~~~~~~~~~~~~~~~~~~~~~~~~~~~~~~~~~~\times [\tilde{G}^+(\omega)]_{2L_{\rm XY}-1,l}^* \text{Im}[\tilde{G}^{\rm X +}_{2L_{\rm X},l-2}(\omega)] \nn \\
&&~~~ +[\tilde{\Sigma}_{\rm Z}^+(\omega)]_{1,2}[\tilde{G}^+(\omega)]_{2L_{\rm XY}-1,2L_{\rm X}+2}^* [\tilde{G}^+(\omega)]_{2L_{\rm XY},l} \text{Im}[\tilde{G}^{\rm X +}_{l-2,2L_{\rm X}}(\omega)] , \nn 
\eea
and, 
\bea
\mathcal{T}^3_{\rm YZ}(\omega)&=& (-\frac{2 \gamma_{\rm YZ}^2}{\pi})\sum_{l=2L_{\rm XY}-1}^{2L_{\rm XY}}
(-1)^{l+1}  \text{Im} \left[[\tilde{\Sigma}_{\rm Z}^+(\omega)]_{1,1}[\tilde{G}^+(\omega)]^*_{2L_{\rm XY}-1,2L_{\rm XY}} \right. \nn \\
&& ~~~~~~~~~~~~~~~~~~~~~~~~~~~~~~~~~~~~~~~ \times [\tilde{G}^+(\omega)]_{2L_{\rm XY}-1,l} \text{Im}[\tilde{G}^{\rm Z +}_{l-2L_{\rm XY}+2,2}(\omega)] \nn \\
&&~~ \left. + [\tilde{\Sigma}_{\rm Z}^+(\omega)]_{1,2}[\tilde{G}^+(\omega)]_{2L_{\rm XY}-1,2L_{\rm XY}}^*[\tilde{G}^+(\omega)]_{2L_{\rm XY},l} \text{Im}[\tilde{G}^{\rm Z +}_{l-2L_{\rm XY}+2,2}(\omega)] \right] .\nn 
\eea
We note that for TS-TS-N/TS-N-N  devices with bias $\mu_{\rm X}=0, \mu_{\rm Z}=\mu$, the above expression becomes independent of $\mathcal{T}^2_{\rm YZ}(\omega)$ whose coefficient returns zero value for such a biasing.

We can further simplify the aforesaid expression (\ref{cuR}) for an N-TS-N device by considering $J^{e}_{\rm XY } = J^{e}_{\rm YZ }$ for either $\mu_{\rm X}=\mu_{\rm Z}=0$ or $\mu_{\rm X}=-\mu_{\rm Z}$ and $T_{\rm X}=T_{\rm Z}$. We thus obtain the following expression:
\bea
 J^{e}_{\rm YZ } &= & \int_{- \infty}^{ \infty}d \omega \,e\left( \mathcal{T}^1_{\rm XY}(\omega)+\mathcal{T}^2_{\rm XY}(\omega) \right)f_{\rm XZ} +\int_{- \infty}^{ \infty}d \omega \,e\, ( \mathcal{T}^2_{\rm XY}(\omega)-\mathcal{T}^3_{\rm XY}(\omega)+\mathcal{T}^3_{\rm YZ}(\omega))f_{\rm \bar{X}X} \nn \\
&&~~~~~~~~~~~~~~~~+\int_{- \infty}^{ \infty}d \omega \,e\, \mathcal{T}^3_{\rm YZ}(\omega)f_{\rm \bar{Z}Z}\, ,\label{cuR1}
\eea
where
\bea
 f_{\rm XZ} &=& f(\omega-\frac{\mu_{\rm X}}{\hbar},T_{\rm X}) -f(\omega-\frac{\mu_{\rm Z}}{\hbar},T_{\rm Z}) ,\, 
\,f_{\rm \bar{X}X}=f(\omega+\frac{\mu_{\rm X}}{\hbar},T_{\rm X}) -f(\omega-\frac{\mu_{\rm X}}{\hbar},T_{\rm X}) , \nn  \\
&& ~~~~~~~~~~~~~~~~~~~~~~~~ f_{\rm \bar{Z} Z}=f(\omega+\frac{\mu_{\rm Z}}{\hbar},T_{\rm Z}) -f(\omega-\frac{\mu_{\rm Z}}{\hbar},T_{\rm Z}). \nn 
\eea
\setcounter{equation}{0}
\setcounter{figure}{0}
\renewcommand\thefigure{G\arabic{figure}}
\section{Expressions of energy currents $J^{u}_{\rm XY}$ and $J^{u}_{\rm YZ}$}
\label{App6}
The energy currents $J^{u}_{\rm XY}$ (\ref{ecuurent1}) and $J^{u}_{\rm YZ}$ (\ref{ecuurent2}) are written in the generalized basis ${\bf a}$ (\ref{bas1}) as 
\bea
 J^{u}_{\rm XY} &=& 2\hbar \gamma_{\rm XY} \left( \gamma_{\rm Y} {\rm Im}[\langle a^\dg_{2L_{\rm X}+3}(t) a_{2L_{\rm X}-1} (t)\rangle ]+  \epsilon_{\rm Y}{\rm Im}[\langle a^\dg_{2L_{\rm X}+1}(t) a_{2L_{\rm X}-1} (t)\rangle ] \right) \nn \\
&&~~~~~~~~~~~~~~~~~~~~~~~~~~~ +  2\gamma_{\rm XY} \Delta_{\rm Y}{\rm Im}[\langle a^\dg_{2L_{\rm X}+3}(t) a_{2L_{\rm X}} (t)\rangle ], \\
 J^{u}_{\rm YZ} &=& 2\hbar \gamma_{\rm YZ} \left( \gamma_{\rm Y} {\rm Im}[\langle a^\dg_{2L_{\rm XY}+1}(t) a_{2L_{\rm XY}-3} (t)\rangle ]+\epsilon_{\rm Y}{\rm Im}[\langle a^\dg_{2L_{\rm XY}+1}(t) a_{2L_{\rm XY}-1} (t)\rangle ]\right) \nn \\
&&~~~~~~~~~~~~~~~~~~~~~~~~~~~~~ - 2 \gamma_{\rm YZ} \Delta_{\rm Y} {\rm Im}[\langle a^\dg_{2L_{\rm XY}+1}(t) a_{2L_{\rm XY}-2} (t)\rangle ]. 
\eea
Since the total energy is conserved in the middle $Y$ wire, the values of $J^{u}_{\rm XY}$ and $J^{u}_{\rm YZ}$ are equal. These energy currents can also be separated in three parts like the electrical currents in Appendix \ref{App5}. We again need to set the chemical potential $(\mu_{\rm X},\mu_{\rm Z})$ to zero for the TS bath(s) of different devices in the following expressions of $ J^{u}_{\rm XY}$ and $J^{u}_{\rm YZ}$.  
\bea
  J^{u}_{\rm XY} =J^{u}_{\rm YZ}&=& \int_{- \infty}^{ \infty} d\omega \: \hbar\left( \mathcal{A}^1(\omega)+\mathcal{A}^2(\omega) \right) f_{\rm XZ}  +\int_{- \infty}^{ \infty} d \omega \: \hbar\, \mathcal{A}^2(\omega) f_{\rm \bar{X}X}  \nn \\
  && +\int_{- \infty}^{ \infty} d \omega\: \hbar \, \mathcal{A}^3(\omega)\, f_{\rm \bar{Z}Z}, \nn \\
\label{encfull}
\eea
where,
\bea
\mathcal{A}^1(\omega)& =& \left(\frac{2 \gamma_{\rm XY}^2}{\pi}\right) \sum_{l=2L_{\rm X}+1}^{2L_{\rm X}+2} (-1)^{l+1}\text{Im} \left[\gamma_{\rm Y}
[\tilde{G}^+(\omega)]_{2L_{\rm X}+3,l}^*\text{Im}[\tilde{G}^{\rm X +}_{2L_{\rm X}-1,l-2}(\omega)] \right. \nn \\
&& ~~~~~~~~~~~~~~~~~~~~~~~~~~~~~~~~~~~~~+\epsilon_{\rm Y} [\tilde{G}^+(\omega)]_{2L_{\rm X}+1,l}^*\text{Im}[\tilde{G}^{\rm X +}_{2L_{\rm X}-1,l-2}(\omega)]   \nn \\
&& +  \left( \gamma_{\rm Y}[\tilde{\Sigma}_{\rm X }^+(\omega)]_{2L_{\rm X}+1,2L_{\rm X}+1}[\tilde{G}^+(\omega)]_{2L_{\rm X}+1,2L_{\rm X}+1}[\tilde{G}^+(\omega)]_{2L_{\rm X}+3,l}^* \right. \nn \\
&& - \Delta_{\rm Y}[\tilde{\Sigma}_{\rm X }^+(\omega)]_{2L_{\rm X}+2,2L_{\rm X}+1} [\tilde{G}^+(\omega)]_{2L_{\rm X}+1,2L_{\rm X}+1}[\tilde{G}^+(\omega)]_{2L_{\rm X}+3,l}^* \nn \\
&& + \left. \epsilon_{\rm Y}  [\tilde{\Sigma}_{\rm X }^+(\omega)]_{2L_{\rm X}+1,2L_{\rm X}+1} [\tilde{G}^+(\omega)]_{2L_{\rm X}+1,2L_{\rm X}+1}[\tilde{G}^+(\omega)]_{2L_{\rm X}+1,l}^* \right) \nn \\
&& ~~~~~~~~~~~~~~~~~~~~~~~~~~~~~~~~~~~~~~~~~~~~~~~~~~~~~~~~~~~~~~~~\times \text{Im}[\tilde{G}^{\rm X +}_{2L_{\rm X}-1,l-2}(\omega)] \nn \\
&& + \left( \gamma_{\rm Y}[\tilde{\Sigma}_{\rm X }^+(\omega)]_{2L_{\rm X}+1,2L_{\rm X}+2}[\tilde{G}^+(\omega)]_{2L_{\rm X}+3,2L_{\rm X}+1}[\tilde{G}^+(\omega)]_{2L_{\rm X}+2,l}^* \right. \nn \\
&& - \Delta_{\rm Y}[\tilde{\Sigma}_{\rm X }^+(\omega)]_{2L_{\rm X}+2,2L_{\rm X}+2} [\tilde{G}^+(\omega)]_{2L_{\rm X}+3,2L_{\rm X}+1}[\tilde{G}^+(\omega)]_{2L_{\rm X}+2,l}^* \nn \\
&& +  \left. \epsilon_{\rm Y}  [\tilde{\Sigma}_{\rm X }^+(\omega)]_{2L_{\rm X}+1,2L_{\rm X}+2} [\tilde{G}^+(\omega)]_{2L_{\rm X}+1,2L_{\rm X}+1}[\tilde{G}^+(\omega)]_{2L_{\rm X}+2,l}^* \right) \nn \\
&&\left.  ~~~~~~~~~~~~~~~~~~~~~~~~~~~~~~~~~~~~~~~~~~~~~~~~~~~~~~~~~~~~~~~\times \text{Im}[\tilde{G}^{\rm X +}_{l-2,2L_{\rm X}-1}(\omega)] \right], \nn 
\eea
\bea
\mathcal{A}^2(\omega)&=& \left(\frac{2 \gamma_{\rm XY}^2}{\pi}\right) \sum_{l=2L_{\rm X}+1}^{2L_{\rm X}+2} (-1)^{l+1} \text{Im} \left[\Delta_{\rm Y}[\tilde{G}^+(\omega)]_{2L_{\rm X}+3,l}^*\text{Im}[\tilde{G}^{\rm X +}_{2L_{\rm X},l-2}(\omega)] \right. \nn \\
&& -\left( \gamma_{\rm Y}[\tilde{\Sigma}_{\rm X }^+(\omega)]_{2L_{\rm X}+1,2L_{\rm X}+1} [\tilde{G}^+(\omega)]_{2L_{\rm X}+1,2L_{\rm X}+2} [\tilde{G}^+(\omega)]_{2L_{\rm X}+3,l}^* \right. \nn \\
&& -\Delta_{\rm Y}[\tilde{\Sigma}_{\rm X }^+(\omega)]_{2L_{\rm X}+2,2L_{\rm X}+1} [\tilde{G}^+(\omega)]_{2L_{\rm X}+1,2L_{\rm X}+2} [\tilde{G}^+(\omega)]_{2L_{\rm X}+3,l}^*\nn \\
&& \left. +\epsilon_{\rm Y}[\tilde{\Sigma}_{\rm X }^+(\omega)]_{2L_{\rm X}+1,2L_{\rm X}+1} [\tilde{G}^+(\omega)]_{2L_{\rm X}+1,2L_{\rm X}+1} [\tilde{G}^+(\omega)]_{2L_{\rm X}+1,l}^*\right) \nn \\
&& ~~~~~~~~~~~~~~~~~~~~~~~~~~~~~~~~~~~~~~~~~~~~~~~~~~~~~~~~~~~~~~~~\times\text{Im}[\tilde{G}^{\rm X +}_{2L_{\rm X},l-2}(\omega)] \nn \\
&& -\left( \gamma_{\rm Y}[\tilde{\Sigma}_{\rm X }^+(\omega)]_{2L_{\rm X}+1,2L_{\rm X}+2} [\tilde{G}^+(\omega)]_{2L_{\rm X}+3,2L_{\rm X}+1} [\tilde{G}^+(\omega)]_{2L_{\rm X}+1,l}^* \right. \nn \\
&& -\Delta_{\rm Y}[\tilde{\Sigma}_{\rm X }^+(\omega)]_{2L_{\rm X}+2,2L_{\rm X}+2} [\tilde{G}^+(\omega)]_{2L_{\rm X}+3,2L_{\rm X}+1} [\tilde{G}^+(\omega)]_{2L_{\rm X}+2,l}^*\nn \\
&& \left. +\epsilon_{\rm Y}[\tilde{\Sigma}_{\rm X }^+(\omega)]_{2L_{\rm X}+1,2L_{\rm X}+2} [\tilde{G}^+(\omega)]_{2L_{\rm X}+1,2L_{\rm X}+1} [\tilde{G}^+(\omega)]_{2L_{\rm X}+2,l}^*\right) \nn \\
&& ~~~~~~~~~~~~~~~~~~~~~~~~~~~~~~~~~~~~~~~~~~~~~~~~~~~~~~~~~~~~~~~~\times \text{Im}[\tilde{G}^{\rm X +}_{l-2,2L_{\rm X}}(\omega)] , \nn 
\eea
and,
\bea
 \mathcal{A}^3(\omega)&=&  \left(-\frac{2 \gamma_{\rm YZ}^2}{\pi}\right)\sum_{l=2L_{\rm XY}-1}^{2L_{\rm XY}} (-1)^{l+1} \text{Im} \left[\left(\gamma_{\rm Y} [\tilde{\Sigma}_{\rm X }^+(\omega)]_{2L_{\rm X}+1,2L_{\rm X}+1} [\tilde{G}^+(\omega)]_{2L_{\rm X}+3,2L_{\rm XY}}^* \right. \right. \nn \\
&&~~~~~~~-\Delta_{\rm Y} [\tilde{\Sigma}_{\rm X }^+(\omega)]_{2L_{\rm X}+2,2L_{\rm X}+1} [\tilde{G}^+(\omega)]_{2L_{\rm X}+3,2L_{\rm XY}}^* \nn \\
 &&~~~~~~~\left. +\epsilon_{\rm Y} [\tilde{\Sigma}_{\rm X }^+(\omega)]_{2L_{\rm X}+1,2L_{\rm X}+1} [\tilde{G}^+(\omega)]_{2L_{\rm X}+1,2L_{\rm XY}}^* \right) \nn \\
 && ~~~~~~~~~~~~~~~~~~~~~~~~~~~~~~~~~~~~~~~~~~~\times [\tilde{G}^+(\omega)]_{2L_{\rm X}+1,l} \text{Im}[\tilde{G}^{\rm Z +}_{l-2L_{\rm XY}+2,2}(\omega)] \nn \\
 && ~~~~~~~+\left(\gamma_{\rm Y} [\tilde{\Sigma}_{\rm X }^+(\omega)]_{2L_{\rm X}+1,2L_{\rm X}+2} [\tilde{G}^+(\omega)]_{2L_{\rm X}+3,2L_{\rm XY}}^* \right. \nn \\
&&~~~~~~~-\Delta_{\rm Y} [\tilde{\Sigma}_{\rm X }^+(\omega)]_{2L_{\rm X}+2,2L_{\rm X}+2} [\tilde{G}^+(\omega)]_{2L_{\rm X}+3,2L_{\rm XY}}^* \nn \\
 &&~~~~~~~\left. +\epsilon_{\rm Y} [\tilde{\Sigma}_{\rm X }^+(\omega)]_{2L_{\rm X}+1,2L_{\rm X}+2} [\tilde{G}^+(\omega)]_{2L_{\rm X}+1,2L_{\rm XY}}^* \right) \nn \\
 && ~~~~~~~~~~~~~~~~~~~~~~~~~~~~~~~~~~~~~~~~~~~\times [\tilde{G}^+(\omega)]_{2L_{\rm X}+2,l} \text{Im}[\tilde{G}^{\rm Z +}_{l-2L_{\rm XY}+2,2}(\omega)] \nn .
 \eea

\setcounter{equation}{0}
\setcounter{figure}{0}
\renewcommand\thefigure{H\arabic{figure}}
\section{Numerical time-evolution of density matrix}
\label{App7}
To study the time-evolution dynamics of the devices with the Kitaev chains, we first consider three disjoint wires $X$, $Y$ and $Z$, where $L_{\rm X}, ~L_{\rm Z} >> L_{\rm Y}$, so that we can treat $X$ and $Z$ as baths. At $t=t_0$, two opposite ends of $Y$ wire are connected to the $X$ and $Z$ bath. The equilibrium density matrix for sites on the isolated boundary wires kept at zero chemical potential reads
\bea
\langle a^{\dagger}_{l}(t_0) a_{m}(t_0) \rangle = \sum_r {\psi_r^{\alpha}}^*(l)\psi_r^{\alpha}(m) f(\omega^{\alpha}_r,T_{\alpha}),\label{indm1}
\eea
with $l,m=1,\dots,2L_{\rm X}$ for $\alpha={\rm X}$ and $l,m=2L_{ \rm XY}+1,\dots,2L$ for $\alpha={\rm Z}$. Here, $f(\omega,T)=1/(\exp[\hbar\omega/k_BT]+1)$ is the Fermi function. However, we need to modify the aforesaid expression in case of an N bath kept at a chemical potential $\mu_\alpha$. The Fermi function in the $\langle a^{\dagger}_{2l'-1}(t_0) a_{2m'-1}(t_0) \rangle$ and $\langle a^{\dagger}_{2l'}(t_0) a_{2m'}(t_0) \rangle$ should be modified as $f(\omega^{\alpha}_r-\mu_\alpha/ \hbar,T_{\alpha})$  and $f(\omega^{\alpha}_r+\mu_\alpha/\hbar,T_{\alpha})$, respectively. All other correlations involving odd-even and even-odd indices are also zero for an N bath.

Since $X$, $Y$, and $Z$ wires are disconnected at $t \leq t_0$, their uncorrelated operators at $t=t_0$ satisfy 
\bea
&& \langle a^{\dagger}_{l}(t_0) a_{p}(t_0) \rangle =\langle a_{l}(t_0) a^{\dagger}_{p}(t_0)\rangle=0 \, ,\nn \\
&& \langle a_{l}(t_0) a_{p}(t_0) \rangle=\langle a_{l}^{\dagger}(t_0) a^{\dagger}_{p}(t_0)\rangle=0 \,,\label{indm2}
\eea
where $l$ and $p$ represent indices corresponding to two different wires of the hybrid device. We further choose some arbitrary initial density matrix for the $Y$ wire, such as, 
\bea \langle a_{2l'}(t_0) a_{2m'-1}(t_0) \rangle = \left \{
\begin{array}{l l} n_{l'} & \quad \mbox{when}~l'=m' \\ 0 & \quad \mbox{when}~ l' \ne m' \end{array} \right. \label{indm3}
\eea
for physical sites: $l',m' \in \{L_{\rm X}+1,\dots, L_{\rm XY}\}$ and $n_{l'}$ denotes the number of fermions at a site $l'$. We then connect the three wires through the tunneling Hamiltonians at time $t_0$ and investigate the time-evolution of the full device using the Heisenberg equations of motion. The solution of the equations of motion for $t \geq t_0$ is given by
\bea
{\bf a}(t)=i \mathcal{G}^+(t-t_0){\bf a}(t_0),
\eea
where $\mathcal{G}^+(\tau)=-i e^{-i H^{\rm F} \tau/ \hbar } \theta (\tau)=\sum_{l,m}\mathcal{G}_{lm}^+(\tau)a^{\dg}_la_m$ is the retarded Green's function of the full device and the matrix elements are $\mathcal{G}_{lm}^+ (\tau) \equiv [-i e^{-i \mathcal{H}^{\rm F} \tau} ]_{lm} \theta (\tau) $, where $\mathcal{H}^{\rm F}_{lm}$ is the matrix Hamiltonian associated with the full Hamiltonian $H^{\rm F}$ (\ref{H1}). Suppose, $\Psi_q(m)$ and $\Lambda_q$ denote the eigenvectors and eigenvalues of the full Hamiltonian matrix, $\mathcal{H}^{\rm F}_{lm}$, hence they satisfy
\beq
\sum_{m=1}^{2L} \mathcal{H}^{\rm F}_{lm}\Psi_q(m)=\Lambda_q \Psi_q(l),~~l=1,2,\dots,2L.
\eeq
Further, the matrix elements of full Green's function can be expanded in the following form for $t >t_0$:
\beq
\mathcal{G}^+_{rs}(t-t_0)=- i \sum_{q=1}^{2L} \Psi_{q}(r) \Psi_{q}^*(s) e^{- i \Lambda_q (t-t_0)} \, ,
\label{EGF}
\eeq
where $r,s \in \{1,\dots, 2L\}$. Finally, the time-evolved density matrix of the full device is expressed as
\bea
\langle a^\dagger_l(t) a_m(t) \rangle= \sum_{r,s=1}^{2L} \mathcal{G}^+_{ms}(t-t_0) \langle a^\dagger_{r}(t_0) a_{s}(t_0) \rangle [\mathcal{G}^+_{lr}(t-t_0)]^{\dagger},\nn\\ \label{tedm}
\eea
where we plug the initial density matrix $\langle a^\dagger_{r}(t_0) a_{s}(t_0) \rangle$ from Eqs.~\ref{indm1}, \ref{indm2}, \ref{indm3}. Using the above expression for $\langle a^\dagger_l(t) a_m(t) \rangle$, we can numerically calculate time-evolution of both electrical and energy currents at the junctions. An extension of this direct time-evolution method for the devices with the Majorana wire leads is straightforward.

\begin{figure}
\includegraphics[width=0.99\linewidth]{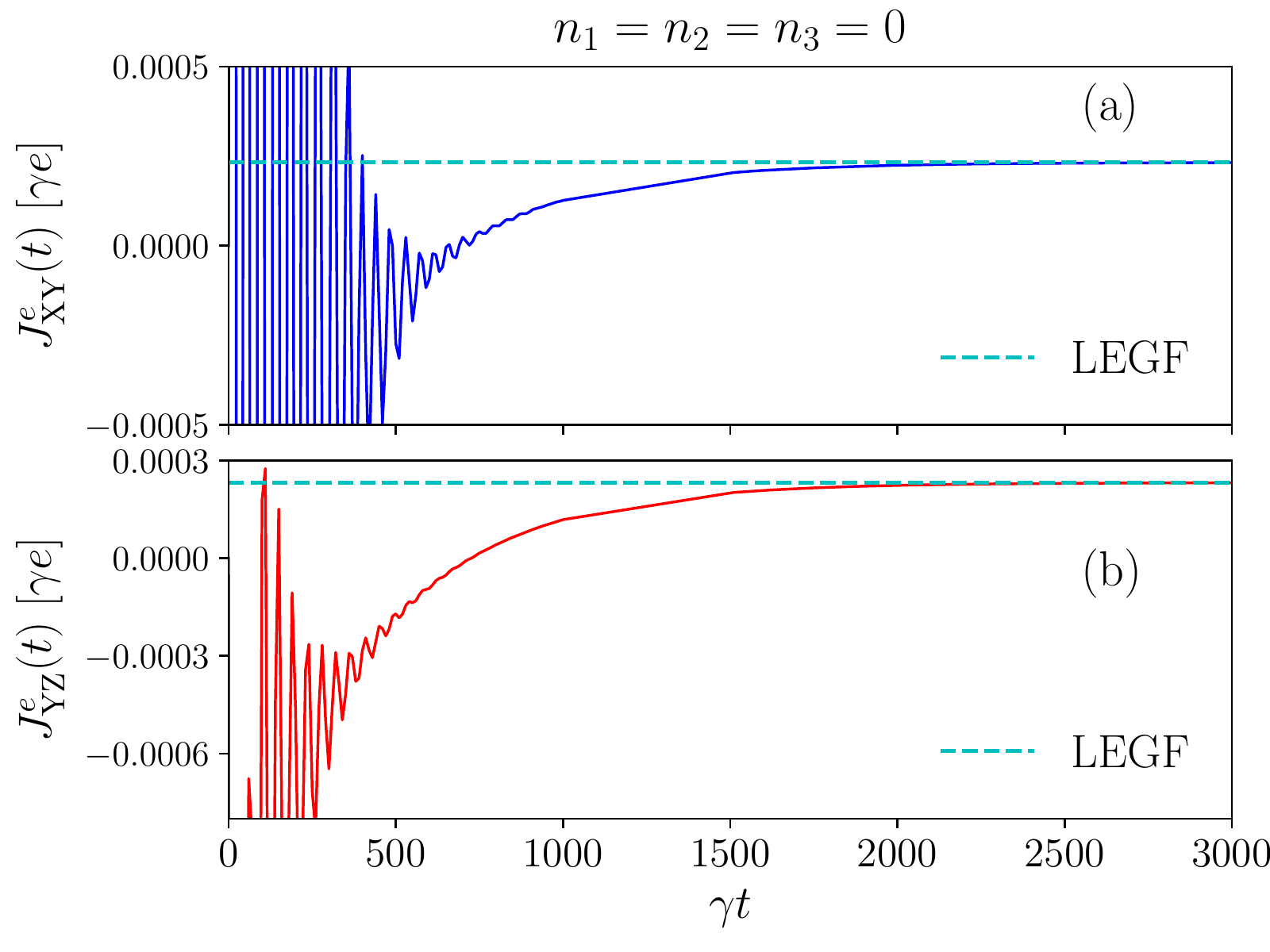}
\centering
\caption{Comparison of electrical currents at both junctions obtained from the first-principle/direct time-evolution numerics (full lines) and the generalized LEGF method (dashed lines) for a TS-N-TS at TP, where the TS wires are made of Kitaev chains. The initial numbers of spinless electrons at the middle N wire for the direct time-evolution numerics are provided on the heading of the top panel. In both panels, $L_{\rm X}=L_{\rm Z}=3000, L_{\rm Y}=3$, $\gamma_{\rm X}=\gamma_{\rm Z}=1$, $\gamma_{\rm Y}=0.5$, $\Delta_{\rm X}=\Delta_{\rm Z}=0.3,\Delta_{\rm Y}=0$, $\epsilon_{\rm X}=\epsilon_{\rm Z}=0$, $\epsilon_{\rm Y}=0.05$, $\gamma_{\rm XY}=\gamma_{\rm YZ}=0.25$, $T_{\rm X}=0.02, T_{\rm Z}=0.2$ and $\mu_{\rm X}=\mu_{\rm Z}=0$. All above parameters except lengths are in units of $\gamma$.}
\label{TSNTPT_n3}
\end{figure}
We next demonstrate the correctness of our steady-state electrical and energy current formulas in different devices by comparing them with the long-time values of the time-dependent currents obtained using the above first-principle numerics.  We show results for steady-state electrical transport in TS-N-Z and TS-TS-Z with Z=N and TS at TP. First, we consider TS-N-TS at TP device in Fig.~\ref{TSNTPT_n3} to verify the validity of the generalized LEGF for such devices. In Fig.~\ref{TSNTPT_n3}(a,b), we respectively depict $J^e_{\rm XY}(t)$ and $J^e_{\rm YZ}(t)$ calculated from the direct time-evolution numerics and the generalized LEGF. These results show an excellent agreement between the two different methods at a long time. However, it takes very long time (equivalently, very long leads) in the direct numerics to achieve the steady state in such TS-N-TS at TP device in comparison to TS-TS-N, TS-N-N. It even requires much longer time to reach the steady state in a TS-TS-TS at TP device in the direct time-dependent numerics. The relaxation time scales for attaining steady-state (as observed in the direct time-evolution numerics) depend on the initial densities of the middle wire, the strength of the tunnel couplings, nature of the sub-gap states, and the bulk-gap of the superconductors \cite{Bondyopadhaya2019}. In comparison, it takes much shorter time to numerically evaluate the steady-state values of $J^e_{\rm XY}(t)$ and $J^e_{\rm YZ}(t)$ in all devices using the LEGF. This indicates an advantage of using the LEGF method over direct/first-principle numerics for calculating steady-state currents in a hybrid device with relatively longer middle wire.  Since the middle wire is made of an N in a TS-N-TS device, we have $J^e_{\rm XY}= J^e_{\rm YZ}$, and the steady-state value of the currents for this set of parameters is  $0.0002325$ (in units of $\gamma e$).

\begin{figure}
\includegraphics[width=0.99\linewidth]{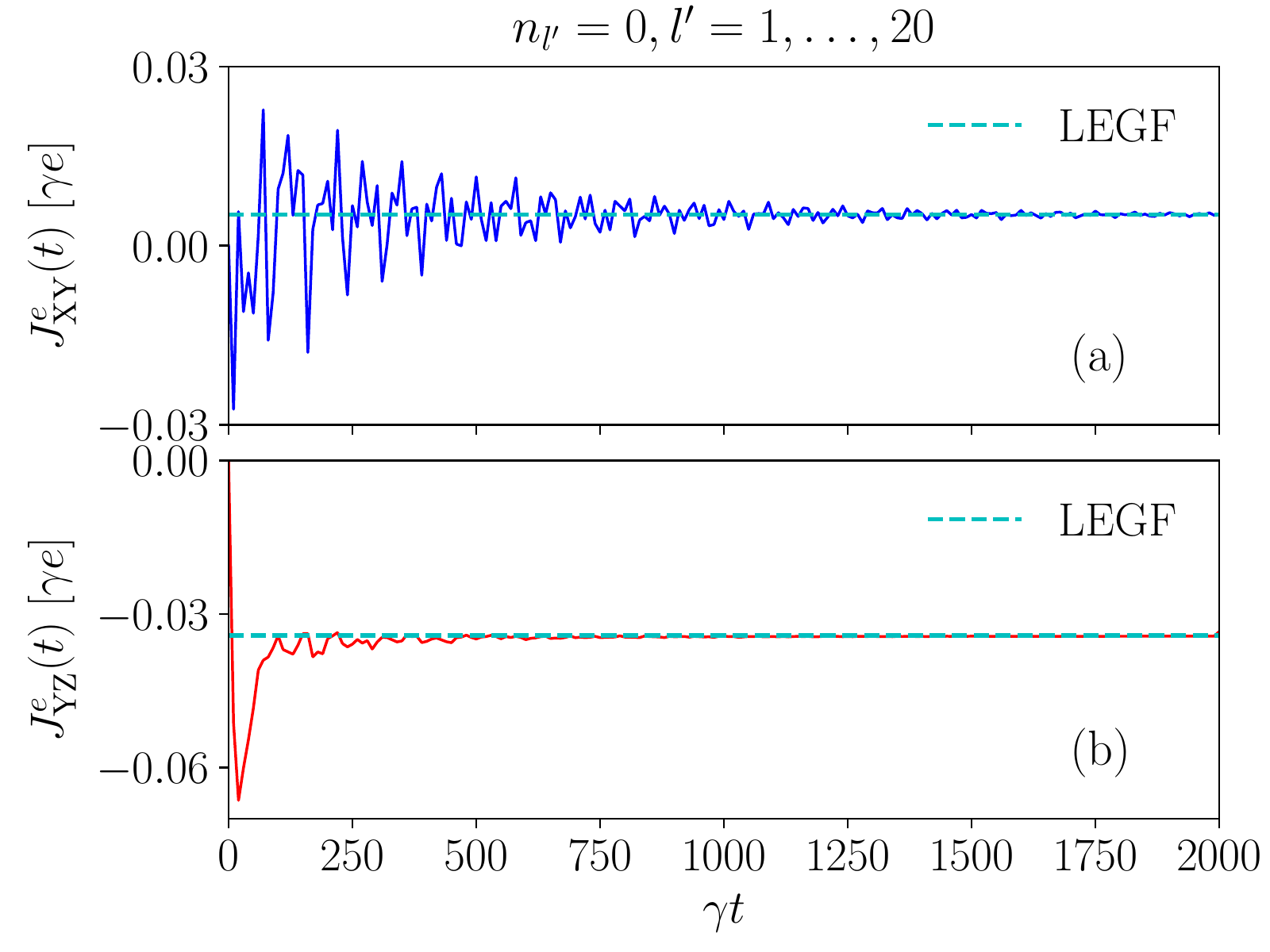}
\centering
\caption{Comparison of electrical currents at both junctions obtained from the first-principle/direct time-evolution numerics (full lines) and the generalized LEGF method (dashed lines) for a TS-TS-N device made of  Kitaev chains. The initial numbers of spinless fermions at the middle TS wire for the direct time-evolution numerics are given on the heading of the top panel. In both panels, $L_{\rm X}=L_{\rm Z}=2000, L_{\rm Y}=20$, $\gamma_{\rm X}=\gamma_{\rm Z}=1$, $\gamma_{\rm Y}=0.5$, $\Delta_{\rm X}=0.3,\Delta_{\rm Y}=0.1,\Delta_{\rm Z}=0$, $\epsilon_{\rm X}=\epsilon_{\rm Y}=\epsilon_{\rm Z}=0$, $\gamma_{\rm XY}=\gamma_{\rm YZ}=0.25$, $T_{\rm X}=T_{\rm Z}=0.02$ and $\mu_{\rm X}=0,\mu_{\rm Z}=0.5$. All above parameters except lengths are in units of $\gamma$.}
\label{TSTSN_n20}
\end{figure}

We further show the validity of the generalized LEGF for a TS-TS-N device in Fig.~\ref{TSTSN_n20} by comparing the steady-state currents with the time-evolution numerics. In the upper (lower) panel, we plot the value of $J^e_{\rm XY}$ ($J^e_{\rm YZ}$) found from the generalized LEGF and the direct time-evolution numerics. Again, the LEGF results match accurately with those obtained from the direct time-evolution numerics at a long time. We notice here that $J^e_{\rm XY} \ne J^e_{\rm YZ}$ as the middle wire is a TS. We also get such excellent agreements between the LEGF and the direct time-evolution numerics for steady-state electrical currents in a TS-N-N and a TS-TS-TS at TP.  

\begin{figure}
\includegraphics[width=0.99\linewidth]{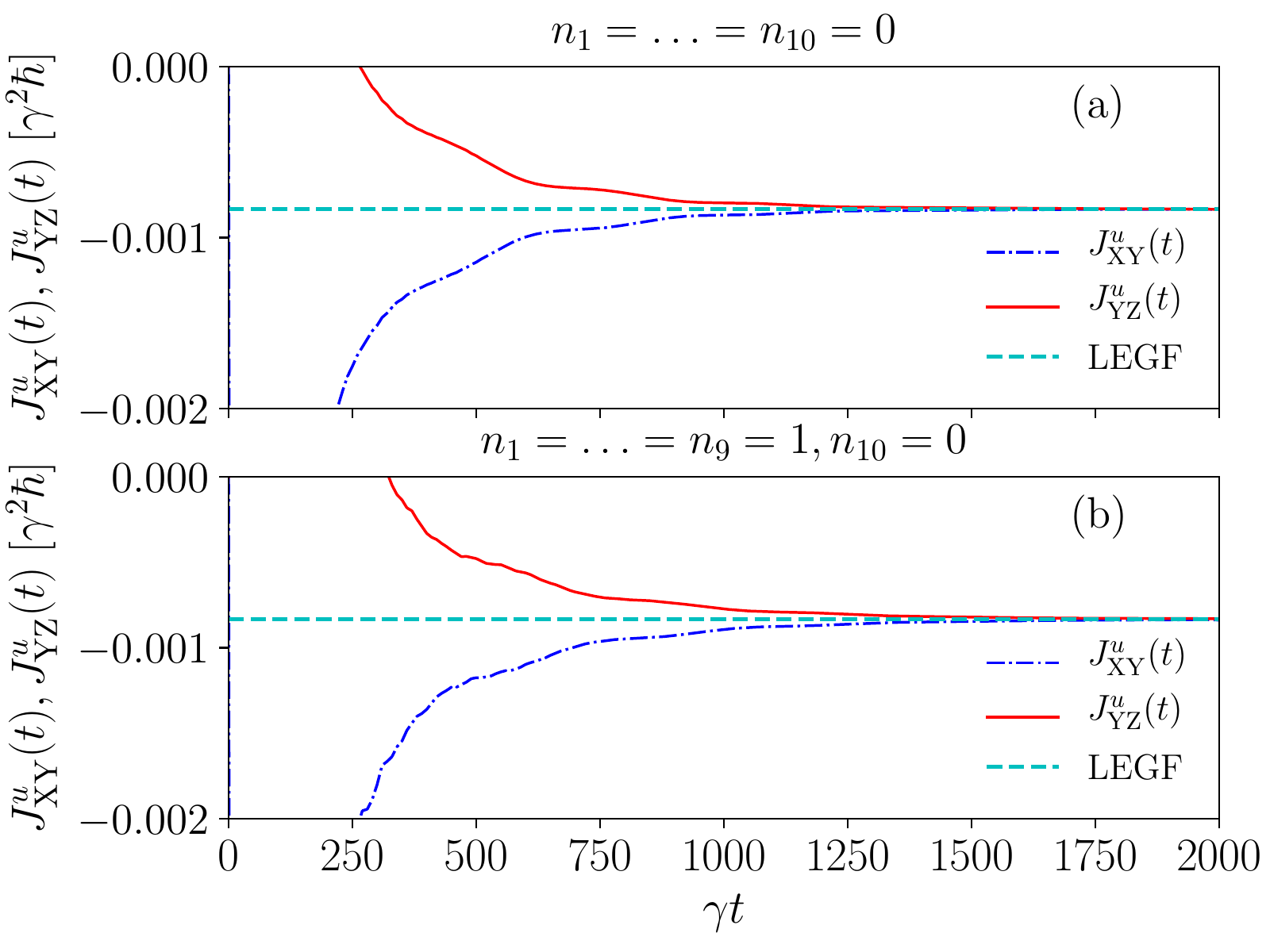}
\centering
\caption{Comparison of energy currents at both junctions obtained from the first-principle/direct time-evolution numerics and the generalized LEGF method for an N-TS-N device made of a Kitaev chain. The initial numbers of spinless electrons ($n_{l'}$) at the middle TS wire used for the time-evolution numerics are indicated on the headings. In both panels, $L_{\rm X}=L_{\rm Z}=2000, L_{\rm Y}=10$, $\gamma_{\rm X}=\gamma_{\rm Z}=\gamma=1$, $\gamma_{\rm Y}=0.5$, $\Delta_{\rm X}=\Delta_{\rm Z}=0,\Delta_{\rm Y}=0.15$, $\epsilon_{\rm X}=\epsilon_{\rm Z}=0$, $\epsilon_{\rm Y}=0.01$, $\gamma_{\rm XY}=\gamma_{\rm YZ}=0.25$, $T_{\rm X}=0.02,T_{\rm Z}=0.02$ and $\mu_{\rm X}=0.2,\mu_{\rm Z}=-0.4$. All above parameters except lengths are in units of $\gamma$.}
\label{EcNTSN}
\end{figure}

In Fig.~\ref{EcNTSN}, we compare the steady-state energy current $J^{u}_{\rm XY}$ and $J^{u}_{\rm YZ}$ in (\ref{encfull}) with the long-time values of the time-dependent currents obtained using the first-principle numerics. The values of $J^{u}_{\rm XY}(t)$ and $J^{u}_{\rm YZ}(t)$ in Fig.~\ref{EcNTSN} due to a voltage bias are the same at long times for any initialization of the middle wire. 

\end{appendices}

\bibliography{bibliography2}

\end{document}